\documentclass[journal,10pt]{IEEEtran}
\usepackage[a4paper,margin={1.61 cm,1.61 cm,1.61 cm,1.61 cm}]{geometry}
\usepackage{graphicx}
\usepackage[cmex10]{amsmath}
\usepackage{stfloats}
\usepackage{epstopdf}
\usepackage{algorithm}
\usepackage{algorithmic}

\usepackage{amsmath}
\usepackage{amssymb}
\usepackage{color}
\usepackage{xcolor}
\usepackage[
            CJKbookmarks=true,
            bookmarksnumbered=true,
            bookmarksopen=true,
            colorlinks=true,
            citecolor=blue,
            linkcolor=red,
            anchorcolor=green,
            urlcolor=blue
            ]{hyperref}
\newtheorem{proof}{Proof}

\title{Delay-Dependent Distributed Kalman Fusion Estimation with Dimensionality Reduction\\ in Cyber-Physical Systems}
\author{Bo Chen,~\IEEEmembership{Member,~IEEE},\;Daniel W. C. Ho,~\IEEEmembership{Fellow,~IEEE}\\Guoqiang Hu,~\IEEEmembership{Senior Member,~IEEE},
\;Li Yu,~\IEEEmembership{Member,~IEEE}
\thanks{B. Chen and L. Yu are with Department of Automation, Zhejiang University of Technology, Hangzhou 310023, China, and also with Institute of Cyberspace Security, Zhejiang University of Technology, Hangzhou 310023, China (email: bchen@aliyun.com; lyu@zjut.edu.cn).}%
\thanks{D. W. C. Ho is with the Department of Mathematics, City University of Hong Kong, Hong Kong, 999077.
(email: madaniel@cityu.edu.hk).}%
\thanks{G. Hu is with the School of Electrical and Electronic Engineering, Nanyang Technological University, 639798 Singapore
(email: gqhu@ntu.edu.sg).}}

\begin{document}

\markboth{}
{Shell \MakeLowercase{\textit{et al.}}: Bare Demo of IEEEtran.cls for Journals}
\maketitle

\begin{abstract}
This paper studies the distributed dimensionality reduction fusion estimation problem with communication delays for a class of cyber-physical systems (CPSs). The raw measurements are preprocessed in each sink node to obtain the local optimal estimate (LOE) of a CPS, and the compressed LOE under dimensionality reduction encounters with communication delays during the transmission. Under this case, a mathematical model with compensation strategy is proposed to characterize the dimensionality reduction and communication delays. This model also has the property to reduce the information loss caused by the dimensionality reduction and delays. Based on this model, a recursive distributed Kalman fusion estimator (DKFE) is derived by optimal weighted fusion criterion in the linear minimum variance sense. A stability condition for the DKFE, which can be easily verified by the exiting software, is derived. In addition, this condition can guarantee that estimation error covariance matrix of the DKFE converges to the unique steady-state matrix for any initial values, and thus the steady-state DKFE (SDKFE) is given.  Notice that the computational complexity of the SDKFE is much lower than that of the DKFE. Moreover, a probability selection criterion for determining the dimensionality reduction strategy is also presented to guarantee the stability of the DKFE. Two illustrative examples are given to show the advantage and effectiveness of the proposed methods.
\end{abstract}


\section{Introduction}
Information fusion has attracted considerable research interest during the past decades, and has found applications in a variety of areas, including internet of things \cite{c1}, sensor networks \cite{c90,c91} and cyber-physical systems (CPSs) \cite{c2}. Particularly, multi-sensor fusion estimation utilizes useful information contained in multiple sets of data for the purpose of estimating a quantity or parameter in a process \cite{c6}. It is widely used in practical applications because it can potentially improve estimation accuracy and enhance reliability and robustness against faults \cite{c6,c3,c5}. Many fusion estimation approaches have been presented in the literature (see \cite{c7,c8,c9,c10,c13,c12}, and the references therein). At the same time, advances in embedded computing, communication, and related hardware technologies have recently brought the paradigm of CPSs to a new research frontier \cite{c14}. Moreover, CPSs have found applications in a broad range of areas such as intelligent transportation systems \cite{c19}, multi-robot systems \cite{c21}, and smart grid systems \cite{c16}. As one of important issues in CPSs, real-time state estimation based on sensor measurements has recently attracted considerable research interests because state estimate can provide a CPS with the real-time monitoring and control capability \cite{c17,c25}. For example, estimating the real voltage from sensor information must be completed before taking certain actions to regulate the voltage into some desired range in a power grid \cite{c16}. It is noted that the accuracy of state estimation has an important impact on computing control commands for safe and efficient operation of a CPS \cite{c17,c25,c22,ac22}. Therefore, it is of theoretical significance and practical relevance to investigate the problem of information fusion estimation for the CPSs \cite{c15,ac15}.
\begin{figure}[thpb]
\centering
\includegraphics[height=4.5cm, width=8.0cm]{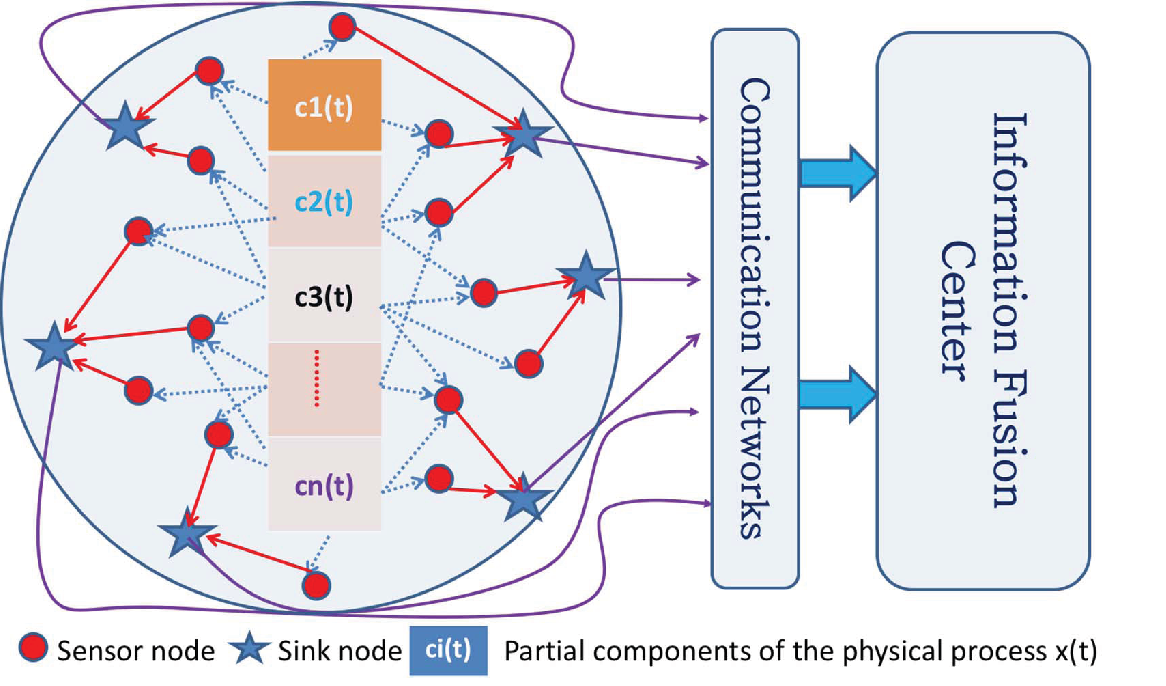}
\caption{Information fusion estimation for a class of spatially distributed physical systems over communication networks: i) x(t) is the state of the physical process, where x(t) is composed of c1(t), c2(t), c3(t),..., cn(t); ii) sensor node only measures the target information; iii) sink node is a gateway, which is responsible for receiving measurements, computing the local optimal estimate (LOE) and sending the LOE to an information fusion center via communication networks.}
\label{fig1}
\end{figure}

There mainly exist two kinds of fusion architectures: centralized fusion structure and distributed fusion structure.
However, the distributed fusion structure is generally more robust and fault-tolerant as compared with the centralized fusion structure \cite{c3,c5,c7,c8,c9}. This motivates us to consider the~distributed fusion estimation problem in this paper for a class of CPS architecture (see Fig.\ref{fig1}), where system state is spatially distributed in the physical space. When the local estimates are transmitted to the fusion center (FC) via communication channels, bandwidth constrains and communication delays are unavoidable in communication networks \cite{c23}. Moreover, the above two factors can degrade the fusion estimation performance because of the information loss caused by bandwidth and delay constrains \cite{c24,o24,ad35}. Thus, how to design distributed fusion methods in the presence of bandwidth and delay constraints is essential for real-time state estimate of CPSs.

\subsection{Related Work}
When considering the problem of bandwidth constraints in multi-sensor systems, as pointed out in \cite{c35}, there are mainly two approaches to reduce the communication traffic: the quantization method (see \cite{c40,c36,c37}, and references therein) and the dimensionality reduction method (see \cite{c32,c33,c34}, and the references therein). Particularly, by analyzing the statistical property of measurement information and resorting to the principal component analysis method, the dimensionality reduction strategy has been designed in \cite{ad32} to compress the measurement signals, while the dimensionality reduction strategy with the quantization error was developed in \cite{c26} to deal with stable multi-sensor fusion systems. Notice that the dimensionality reduction method in \cite{ad32} requires to know the global measurement matrix that may be difficult to be satisfied in distributed systems, and solving non-convex nonlinear optimization problem under this strategy may also add the computational cost and design difficulty. Under the distributed fusion structure, when the physical state x(t) as shown in Fig.\ref{fig1} is multi-dimensional (or even high-dimensional) in a CPS, it is unrealistic to completely send the local estimate of the state x(t) to the FC via a bandwidth-constrained communication channel. In this sense, bandwidth constraint in the CPSs is the primary consideration when designing a distributed state fusion estimator. Notice that, to reduce the communication traffic, the idea of the dimensionality reduction method is that a multi-dimensional signal is directly converted into a low-dimensional signal, while the idea of the quantization method is that the number of coding bits for each component of a multidimensional signal is reduced before being transmitted. Meanwhile, the quantization usually results in nonlinear dynamics, and it is difficult to find a data compression operator analytically, particularly, for the multidimensional signals. Therefore, the dimensionality reduction method can provide an attractive alternative to solve the distributed fusion estimation problem with bandwidth constraints in the CPSs.

Though the dimensionality reduction fusion estimation algorithms have been proposed in \cite{c32, c33, c34, ad32, c26} to reduce the communication traffic, the communication delays, which occurs during the transmission, were not taken into account. With the communication delays, the dimensionality reduction fusion estimation must solve two challenging issues: one is how to compensate the information loss caused by the communication delays and bandwidth constraints under a unified mathematical model; The other one is how to fuse the asynchronous local compressed estimates because of communication delays. Notice that the centralized and distributed fusion estimation algorithms have been proposed in \cite{c24,c27,c28,c29,c30,o30,e30} based on different communication delay models, however, the main results in \cite{c24,c27,c28,c29,c30,o30,e30} cannot be extended to the case of the dimensionality reduction estimation with communication delays. The reason is that the data compression and information compensation in dimensionality reduction may change the property of the original measurements (e.g., the statistical correlation in \cite{c32,c34} has been changed under the Kalman fusion structure). Under this case, we have studied the information fusion estimation problem in \cite{c15,o24} for the CPSs with bandwidth constraints and communication delays. It should be pointed out that the steady-state fusion estimator with simple calculation cannot be obtained based on the proposed communication model in \cite{c15}, while the covariance intersection (CI) fusion strategy in \cite{o24} was suboptimal because fusion estimator was determined by minimizing an upper bound of estimation error covariance.
\vspace{-6pt}
\subsection{Contributions}
Motivated by the aforementioned analysis, we study the distributed stochastic dimensionality reduction fusion estimation problem with communication delays for the CPSs. Notice that the information loss is inevitable because of the dimensionality reduction and communication delays, and such a fusion estimation with incomplete information will degrade the estimation performance. Since the delays are caused by communication channels, the key issue is how to design an efficient dimensionality reduction strategy to guarantee the stability of the distributed fusion estimator. Although our previous works in \cite{c15,o24,c32} have studied the related stochastic dimensionality fusion estimation problems, there are still fundamental problems that cannot be solved up to now. In detail,
\begin{itemize}
\item When only considering stochastic dimensionality reduction strategy, the stable probability selection criterion in \cite{c32} was derived from the inequality relaxation of the matrix trace. However, the inequality relaxation will lead to certain conservatism, thus how to find a new derivation idea to reduce the conservatism is very important for the application of the proposed dimensionality reduction strategy. Notice that the stability conditions in \cite{c15} were directly derived from the similar derivation in \cite{c32}, and thus the corresponding conservatism also cannot be avoided in \cite{c15}. Moreover, the stability conditions in \cite{o24} also have certain conservatism due to the inequality relaxation of 1-norm and $\infty$-norm.
\item When considering stochastic dimensionality reduction strategy under communication delays, the distributed CI fusion estimator in \cite{o24} was suboptimal because the corresponding optimization objective was an upper bound of the estimation error covariance matrix.  Particularly, the CI fusion results in \cite{o24} required to solve non-convex nonlinear optimization problems online at each time, which may lead to a large number of calculation. Though the distributed fusion estimator in \cite{c15} was optimal based on the optimal weighted fusion criterion, the model of communication delays cannot be applicable to the case of time-varying delays. More importantly, the computational complexity of the fusion estimator in \cite{c15} was also high. Obviously, the common disadvantage of the results in \cite{c15} and \cite{o24} is the high computation cost, and the optimal weighed fusion criterion can provide the optimal and analytic solutions. Therefore, based on the optimal weighted fusion criterion, how to design steady-state dimensionality reduction fusion estimators with simple calculation is of great significance in the presence of communication delays.
\end{itemize}
We shall solve the above two problems, and the main contributions of this paper can be summarized as follows:
\begin{itemize}
\item By constructing a new common orthogonal space, the cross-covariance matrix is calculated by the recursive form, and then an optimal distributed Kalman fusion estimator (DKFE) is derived in the linear minimum variance sense when there are bandwidth and communication delay constraints in CPSs. Notice that each weighting fusion matrix is calculated by the analytic form.

\item A delay-dependent and probability-dependent stability condition is derived such that the fusion estimation error covariance matrix of the DKFE converges to a unique steady-state matrix for any initial values. Under this condition, the steady-state DKFE, which has much lower computational complexity as compared with the DKFE, is given. Moreover, when each communication delay is known, the probability selection criterion for determining dimensionality reduction strategy is presented to guarantee the stability of the DKFE.

\item Compared with the fusion estimation method in \cite{c15}, the model of communication delays in this paper does not require that each sink node knows the communication delay in advance, and the steady-state DKFE with simple calculation is derived (see Remark 1). Since the covariance intersection fusion criterion in \cite{o24} is suboptimal, the estimation performance of the designed DKFE must be better than that of the fusion estimator in \cite{o24} when each communication delay is constant. Moreover, the computation cost of the steady-state DKFE must be lower than that of the CI fusion estimator in \cite{o24} and the conservatism of delay-dependent stability conditions is less than that of the conditions in \cite{o24} (see Remark 2 and Remark 7).

\item When there is no communication delay for the scenario described in Fig.\ref{fig1}, it is shown that the stability condition in this paper has less conservatism than the result in \cite{c32}. This is because a new derivation idea without any inequality relaxation is proposed to design the stochastic dimensionality reduction strategy. Moreover, when considering communication delays, the corresponding stability analysis is also based on this new derivation idea. Notice that it is difficult to obtain the stability condition by using the derivation idea in \cite{c32} when the communication delay is modeled in this paper (see Remarks 7-8).
\end{itemize}

The rest of this paper is organized as follows. Section II presents the problem formulation. The finite-horizon DKFE is designed in Section III. In Section IV, the stability condition and the steady-state DKFE are derived, and the probability selection criteria are given to determine satisfactory compression operators. Two illustrative examples are presented in Section V to show the advantage and effectiveness of the proposed approaches, and then the conclusions are drawn in Section VI.

\emph{Notations:} The notations used throughout the paper are fairly standard. The superscript $'{\rm{T'}}$ represents the transpose, and ${\rm E}\{\cdot\}$ is the mathematical expectation. ${I_m}$ represents the identity matrix of size $m \times m$, while ${\rm{diag}\{\cdot\}}$ stands for a block diagonal matrix. ${\rm{Prob\{}}A{\rm{\}}}$ means the occurrence probability of the event $A$, while ${\rm{Tr}}(B)$ denotes the trace of the matrix $B$. $||A|{|_2}$ represent the 2-norm of the matrix $A$. $x \bot y$ denotes that $x$ and $y$ are orthogonal vectors, and ${\rm{col}}\{ {a_1}, \cdots ,{a_L}\}$ represents the column vector that is composed of the elements ${a_1}, \cdots ,{a_L}$. The symbol ${\rm{lcm}}(a,b)$ is the least common multiple of $a$ and $b$, while ${\rm{rank(}} A {\rm{)}}$ denotes the rank of the matrix $A$. The function $f_ \circ ^\hbar (t)$ is defined by $f_ \circ ^\hbar (t) \buildrel \Delta \over = \underbrace {f(f( \cdots (f}_{\hbar \;{\rm{times}}}(t)) \cdots ))$, and $X>(<)0$~denotes a positive-definite (negative-definite) matrix.

\section{Problem Formulation}
\subsection{Dimensionality Reduction and Communication Delays}
Consider the physical process in Fig.1 described by the following discrete state-space model:
\begin{eqnarray}
x(t + 1) = Ax(t) + w(t),
\label {eq:1}
\end{eqnarray}
where $x(t) \in {{\rm{R}}^n}\;(n>1)$ is the state of the process, $w(t)$ is the system noise, and $A$ is a constant matrix with appropriate dimension. As pointed out in \cite{c25}, the model (\ref{eq:1}) is widely adopted for describing state dynamics of CPSs including power systems, smart grid infrastructures, and building automation systems, etc. When the measurements from each sensor are sent to sink nodes, the $i{\rm{th}}$ sink node's measurement ${y_i}(t) \in {{\rm{R}}^{{q_i}}}$ is modeled by:
\begin{eqnarray}
{y_i}(t) = {C_i}x(t) + {v_i}(t)(i = 1,2, \cdots ,L),
\label {eq:2}
\end{eqnarray}
where ${C_i}$ is the measurement matrix with appropriate dimension, and $v_i(t)$ is the measurement noise. Moreover, $w(t)$ and ${v_i}(t)$ are uncorrelated zero-mean Gaussian white noises satisfying
\begin{eqnarray}
\begin{array}{l}
 {\rm E}\{ {[{w^{\rm{T}}}(t)\;v_i^{\rm{T}}(t)]^{\rm{T}}}[{w^{\rm{T}}}(t_1)\;v_j^{\rm{T}}(t_1)]\}\\
  \;\;\;\;\;\;\;\;\;\;\;\;\;\;\;\;\;\;\;\;\;\;\;\;\;\;\;\;\;\;\;\;\;\;\;\;\;\;\;\;\;= {\delta _{t,{t_1}}}{\rm{diag\{ }}{Q_w},{\delta _{i,j}}{Q_{{v_i}}}{\rm{\} }} \\
 \end{array},
\label {eq:3}
\end{eqnarray}
where ${\delta _{t,{t_1}}}$ is defined by:
\begin{eqnarray}
{\delta _{t,{t_1}}} = \left\{ \begin{array}{l}
 1\;\;\;\;\;\;{\rm{if}}\;\;t = {t_1} \\
 0\;\;\;\;\;\;{\rm{if}}\;\;t \ne {t_1} \\
 \end{array}. \right.\
\label {eq:4}
\end{eqnarray}
Then, based on the measurements $\{ {y_i}(1),\cdots ,{y_i}(t)\}$, the local optimal estimate (LOE) ${\hat x_i}(t)$ is given by the Kalman filter:
\begin{eqnarray}
{{\hat x}_i}(t) = {{\rm{G}}_{{{\rm K}_i}}}(t)A{{\hat x}_i}(t - 1) + {{\rm K}_i}(t){y_i}(t),
\label {eq:5}
\end{eqnarray}
where
\vspace{-8pt}
\begin{eqnarray}
{{\rm{G}}_{{{\rm K}_i}}}(t) \buildrel \Delta \over = {I_n} - {{\rm K}_i}(t){C_i}.
\label {eq:6}
\end{eqnarray}
Define ${{\tilde x}_i}(t) \buildrel \Delta \over = x(t) - {{\hat x}_i}(t)$. Then, the optimal gain matrix~${{\rm K}_i}(t)$~and the local estimation error covariance matrix~${P_{ii}}(t) \buildrel \Delta \over = {\rm E}\{ {{\tilde x}_i}(t)\tilde x_i^{\rm{T}}(t)\}$~are calculated by
\begin{eqnarray}
\left\{ \begin{array}{l}
 {{\rm K}_i}(t) = P_{ii}^ * (t)C_i^{\rm{T}}{[{C_i}P_{ii}^ * (t)C_i^{\rm{T}} + {Q_{{v_i}}}]^{ - 1}} \\
 {P_{ii}}(t) = {{\rm{G}}_{{{\rm K}_i}}}(t)P_{ii}^ * (t) \\
 P_{ii}^ * (t) = A{P_{ii}}(t - 1){A^{\rm{T}}} +  {Q_w} \\
 \end{array}, \right.
\label {eq:7}
\end{eqnarray}
where $P_{ii}^*(t)$ denotes the error covariance matrix of one-step prediction.
Moreover, it follows from (\ref{eq:1}), (\ref{eq:5}) and (\ref{eq:7}) that the local estimation error cross-covariance matrix ${P_{ij}}(t) \buildrel \Delta \over = {\rm E}\{ {{\tilde x}_i}(t)\tilde x_j^{\rm{T}}(t)\} (i \ne j)$ is calculated by:
\begin{eqnarray}
{P_{ij}}(t) = {{\rm{G}}_{{{\rm K}_i}}}(t)[{Q_w} + A{P_{ij}}(t - 1){A^{\rm{T}}}{\rm{]G}}_{{{\rm K}_j}}^{\rm{T}}(t).
\label {eq:8}
\end{eqnarray}

\begin{figure}[thpb]
\centering
\includegraphics[height=5.0cm, width=9.0cm]{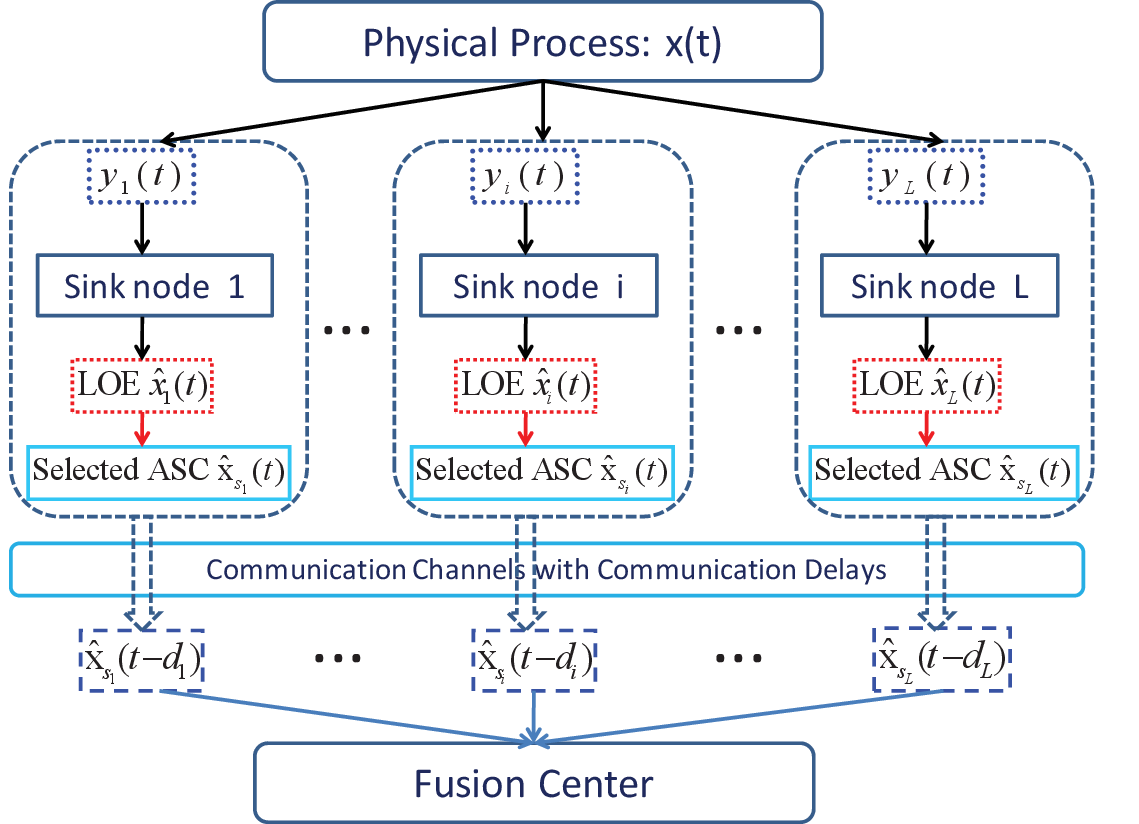}
\caption{Distributed dimensionality reduction fusion estimation with communication delays in CPSs}
\label{fig8}
\end{figure}

Under the distributed fusion structure, each LOE $\hat x_i(t)$ must be sent to the FC to design an optimal fusion estimator. However, it is unrealistic to send the complete information included in $\hat x_i(t)(\in {{\rm{R}}^n})$ to the FC over communication networks because almost all communication network can only carry a finite amount of information per unit time. This problem is especially prominent in the fusion estimation for the large-scale CPSs integrated by wireless sensor networks. To reduce communication traffic, only ${r_i}(1 \le r_i < n)$ components of the $i{\rm{th}}$ LOE $\hat x_i(t)$ are allowed to be transmitted to the FC at each time, and other components are discarded. Compared with the original LOE $\hat x_i(t)$, the dimension of the transmitted signal is reduced. In this sense, the above method can be viewed as one of the dimensionality reduction strategies. According to this dimensionality reduction strategy, the allowed sending components (ASC) of~$\hat x_i(t)$~has~$\Delta_i$~possible cases, where ${\Delta _i} = {{\prod\limits_{\ell _i^n = 0}^{{r_i} - 1} {(n - \ell _i^n)} } \mathord{\left/
{\vphantom {{\prod\limits_{\ell _i^n = 0}^{{r_i} - 1} {(n - \ell _i^n)} } {\prod\limits_{\ell _i^r = 1}^{{r_i}} {\ell _i^r} }}} \right.
\kern-\nulldelimiterspace} {\prod\limits_{\ell _i^r = 1}^{{r_i}} {\ell _i^r} }}$. Then, at a particular time, only one vector signal, which is taken from one group of the above ${\Delta _i}$ cases, is selected and transmitted to the FC, and this selected signal is denoted by ${{\rm{\hat x}}_{{s_i}}}(t)\in {{\rm{R}}^{{r_i}}}$. When ${{\rm{\hat x}}_{{s_i}}}(t)$ is sent to the FC by the sink node, the FC will receive the data packet containing ${{\rm{\hat x}}_{{s_i}}}(t)$ at time $t+d_i$ because of communication delay. Let ${{\rm{\bar x}}_{{s_i}}}(t)$ denote the local estimation information received by the FC at time $t$. Then, ${{\rm{\bar x}}_{{s_i}}}(t)$ in the FC is given by:
\begin{eqnarray}
{{{\rm{\bar x}}}_{{s_i}}}(t) = {{{\rm{\hat x}}}_{{s_i}}}(t - {d_i}).
\label {eq:9}
\end{eqnarray}
It should be pointed out that the communication delays for different sink nodes are not the same (i.e., ${d_i} \ne {d_j}$), which means that the addressed delays are not constant from the perspective of the whole fusion systems. Moreover, the constant communication delay for each sink node is mainly determined by signal transmission power, and is difficult to be avoid due to resource constraints. Particularly, when the fusion center received the signals from multiple sink nodes, there must exist the problem of resource scheduling. Though time-varying delays may occur for different sink nodes, they must lead to the information disorder that directly add the design difficulties of dimensionality reduction fusion estimators. In this case, to avoid the information disorder in multi-sensor fusion, resource scheduling can be controlled by designing the physical mechanism for guaranteeing constant delays (see an example in Remark 1). Therefore, it is of significance in studying the dimensionality reduction fusion estimation problem with constant delays in this paper. Up to now, the problem of dimensionality reduction and communication delays has been presented, and the process diagram is shown in Fig.\ref{fig8}.

It is noted that the signal ${{\rm{\hat x}}_{{s_i}}}(t)$ only takes one element from the following finite set:
\begin{eqnarray}
{S_i}(t) = \{ {\rm{\hat x}}_{{s_i}}^{{\hbar _i}}(t)|{\hbar _i} = 1,2, \cdots ,{\Delta _i}\},
\label {eq:10}
\end{eqnarray}
where $\hat x_{si}^{{\hbar _i}}(t) \in {{\rm{R}}^{{r_i}}}$ represents one group of ASCs. To characterize the determining process of ${{\rm{\hat x}}_{{s_i}}}(t)$, we introduce the following indicator functions:
\begin{eqnarray}
\sigma _{{\hbar _i}}^i(t) = \left\{ {\begin{array}{*{20}{c}}
   {1\:\:\:\:\:{\rm{if}}\:\:{{{\rm{\hat x}}}_{{s_i}}}(t) = {\rm{\hat x}}_{{s_i}}^{{\hbar _i}}(t)}  \\
   {0\:\:\:\:\:{\rm{if}}\:\:{{{\rm{\hat x}}}_{{s_i}}}(t) \ne {\rm{\hat x}}_{{s_i}}^{{\hbar _i}}(t)}  \\
\end{array}}, \right.
\label {eq:11}
\end{eqnarray}
where $\sigma _{{\hbar _i}}^i(t)({\hbar _i} = 1,2, \cdots ,{\Delta _i})$ are required to satisfy
\begin{eqnarray}
\sigma _{{\hbar _i}}^i(t)\sigma _{\hbar _i^0}^i(t) = 0({\hbar _i} \ne \hbar _i^0)\:,\sum\nolimits_{{\hbar _i} = 1}^{{\Delta _i}} {\sigma _{{\hbar _i}}^i(t)}  = 1
\label {eq:12}
\end{eqnarray}
such that ${{{{\rm{\hat x}}}_{{s_i}}}(t)}$ only takes one ASC from the set (\ref{eq:10}) at time~$t$, i.e.,
\vspace{-8pt}
\begin{eqnarray}
{{{\rm{\hat x}}}_{{s_i}}}(t) = \sum\nolimits_{{\hbar _i} = 1}^{{\Delta _i}} {\sigma _{{\hbar _i}}^i(t){\rm{\hat x}}_{{s_i}}^{{\hbar _i}}(t)}.
\label {eq:13}
\end{eqnarray}
Then, it is derived from (\ref{eq:9}) and (\ref{eq:13}) that
\begin{eqnarray}
{{{\rm{\bar x}}}_{{s_i}}}(t) = \sum\nolimits_{{\hbar _i} = 1}^{{\Delta _i}} {\sigma _{{\hbar _i}}^i(t - {d_i}){\rm{\hat x}}_{{s_i}}^{{\hbar _i}}(t - {d_i})}.
\label {eq:14}
\end{eqnarray}
At time $t$, if the fusion estimate of $x(t)$ is directly designed based on ${{{\rm{\bar x}}}_{{s_i}}}(t)$, the fusion estimation performance must be poor because of the communication delays and the un-transmitted component of ${{{\rm{\hat x}}}_i}(t)$. In this case, the compensating state estimate (CSE) of $x(t)$, denoted by ${\rm{\hat x}}_i^{\rm{c}}(t)$, can be modeled as follows:
\begin{eqnarray}
\begin{array}{l}
 {\rm{\hat x}}_i^{\rm{c}}(t) = {A^{{d_i}}}{H_i}(t - {d_i}){{\hat x}_i}(t - {d_i})\:\:\: \\
 \;\;\;\;\;\;\;\;\;\;\;\; + {A^{{d_i}}}[{I_n} - {H_i}(t - {d_i})]A{\rm{\hat x}}_i^{\rm{c}}(t - {d_i} - 1) \\
 \end{array},
\label {eq:15}
\end{eqnarray}
where $H_i(t-d_i)$ is determined by
\begin{eqnarray}
{H_i}(t) = \sum\nolimits_{{\hbar _i} = 1}^{{\Delta _i}} {\sigma _{{\hbar _i}}^i(t)H_{{\hbar _i}}^i}  = {\rm{diag}}\{ \gamma _1^i(t), \cdots ,\gamma _n^i(t)\}.
\label {eq:16}
\end{eqnarray}
Here, $H_{{\hbar _i}}^i$ represents a diagonal matrix that contains~$r_i$~diagonal elements ``1'' and~$n-r_i$~diagonal elements ``0''. Then it follows from (\ref{eq:11}) and (\ref{eq:12}) that
\begin{eqnarray}
\gamma _\ell ^i(t) \in \{ 0,1\} ,\sum\nolimits_{\ell  = 1}^n {\gamma _\ell ^i(t)}  = {r_i}(i = 1, \cdots ,L),
\label {eq:17}
\end{eqnarray}
where $\gamma _\ell ^i(t) = 1$ means that the $\ell {\rm{th}}$ component of ${{\hat x}_i}(t)$ is selected and sent to the FC, while $\gamma _\ell ^i(t) = 0$ means that the $\ell {\rm{th}}$ component of ${{\hat x}_i}(t)$ is discarded. Particularly, at time $t$, the compensation strategy in the CSE model (\ref{eq:15}) is reflected by the following aspects:
\begin{itemize}
\item The un-transmitted components of $\hat x_i(t)$ are compensated by the one-step prediction based on ${\rm{\hat x}}_i^{\rm{c}}(t - {d_i} - 1)$.
\item The delayed information ${{{\rm{\bar x}}}_{{s_i}}}(t)$ is compensated by the $d_i$-step prediction based on ${\rm{\hat x}}_i^{\rm{r}}(t - {d_i})$, where $\hat x_i^{\rm{r}}(t - {d_i}) = {H_i}(t - {d_i}){{\hat x}_i}(t - {d_i}) + [I - {H_i}(t - {d_i})]A{\rm{\hat x}}_i^{\rm{c}}(t - {d_i} - 1)$.
\end{itemize}

\textbf{Remark 1}. In \cite{c15}, at the $i{\rm{th}}$ sink node, the $d_i$-step prediction based on the local estimate $\hat x_i(t)$ was given by $\left( {\hat x_i^{{d_i}}(t) = {A^{{d_i}}}{{\hat x}_i}(t)} \right)$. Due to the bandwidth constraints, only ${r_i}$ components of $\hat x_i^{{d_i}}(t)$ were allowed to be sent. Then, the CSE of $x(t)$, denoted as $\hat x_{{d_i}}^c(t)$, was given by (i.e., the model (18) in \cite{c15}):
\begin{eqnarray}
\begin{array}{l}
 \hat x_{{d_i}}^c(t) = H_i^c(t - {d_i}){A^{{d_i}}}{{\hat x}_i}(t - {d_i}) \\
 \;\;\;\;\;\;\;\;\;\;\;\;\;\;\;\;\;\;\;\; + [I - H_i^c(t - {d_i})]A\hat x^f(t - 1), \\
 \end{array}
\label {eq:a7}
\end{eqnarray}
where the definition of $H_i^c(t - {d_i})$ is the same as that of $H_i(t - {d_i})$, and ${{\hat x}^f}(t - 1)$ denotes the fusion estimate designed by \cite{c15}. For the CSE model (\ref{eq:a7}), the $d_i$-step prediction $\hat x_i^{{d_i}}(t)$ must be completed at the sink node, which implies that each sink node must know the communication delay from the sink node to the FC \emph{in advance}. Under this case, when the communication delay is unknown for the sink node or time-varying, the model (\ref{eq:a7}) will be \emph{invalid}. Different from the modeling method in \cite{c15}, the CSE model (\ref{eq:15}) does not require that each sink node knows the communication delay in advance, and thus the model (\ref{eq:15}) can be more easily implemented in a practical system. Particularly, when considering the time-varying communication delay $d_i(t)$, the local estimation information received by the FC, denoted as ${\rm{\bar x}}_i^{{d_i}}(t)$, is given by:
\begin{eqnarray}
{\rm{\bar x}}_i^{{d_i}}(t) = {\rm{\hat x}}_i^{{s_i}}(t - {d_i}(t)),
\label {eq:c7}
\end{eqnarray}
where ${\rm{\hat x}}_i^{{s_i}}(t)$ denotes the selected ASC at the sink node. Meanwhile, it is reasonable to consider that the time-varying delay $d_i(t)$ is bounded in practical applications, and satisfies ${d_i}(t) \le d_i^u$. Then, by resorting to the buffers at the FC, each time-varying delay can be prolonged to its upper bound $d_i^u$ at each time, i.e., the model (\ref{eq:c7}) is reduced to:
\begin{eqnarray}
{\rm{\bar x}}_i^{{d_i}}(t) = {\rm{\hat x}}_i^{{s_i}}(t - d_i^u)
\label {eq:d7}
\end{eqnarray}
Since the structure of (\ref{eq:d7}) is the same as that of (\ref{eq:9}), the case of time-varying delays can still be modeled by (\ref{eq:15}). Notice that the CSE model (\ref{eq:a7}) in \cite{c15} will not be applicable to this case, because the time-varying communication delays are only known to the FC, and each sink node impossibly know the time-varying delays a priori. On the other hand, the stability condition in \cite{c15} could only guarantee the MSE of the fusion estimator converged to a steady-state value. It should be pointed out that the computational complexity of the fusion estimator in \cite{c15} is a slightly high, yet the corresponding steady-state fusion estimator cannot be derived from the stability condition in \cite{c15}. In contrast, the steady-state DKFE with simple calculation can be designed based on the stability condition in Theorem 3.

\textbf{Remark 2}. For the case of time-varying delays, the estimation error cross-covariance matrices cannot be obtained under the dimensionality reduction strategy in this paper. Fortunately, the CI fusion criterion does not need the cross-covariance matrices. Therefore, the distributed CI fusion estimation algorithm was developed in \cite{o24} to deal with the time-varying delays. Notice that the CI fusion criterion is not optimal because the optimization objective is an upper bound of estimation error covariance matrix, and each weighting matrix is obtained by solving non-convex nonlinear optimization problems \emph{at each time}.
Different from the fusion criterion in \cite{o24}, the \emph{optimal} weighted fusion criterion with analytic solutions is used to design the DKFE in this paper. Thus, when considering the constant communication delays, the estimation performance of the DKFE is better than that of the fusion estimator in \cite{o24}. On the other hand, as pointed out in Remark 1, the designed fusion estimation algorithms in this paper can be also applicable to the case of time-varying communication delays. However, it is difficult to show whose estimation performance is better, the DKFE in this paper or the fusion estimator in \cite{o24}, when dealing with time-varying delays. This is because the performance loss in this paper is introduced from the delay model (i.e., prolonging the time-varying delay to its upper bound at each time), while the performance loss in \cite{o24} is introduced from the CI fusion criterion (i.e., minimizing an upper bound of the fusion estimation error covariance). However, from the perspective of computational complexity, the steady-state DKFE in this paper is better than the CI fusion estimator in \cite{o24} whenever considering the constant delays or time-varying delays.

\subsection{Problem of Interest}
It is concluded from (\ref{eq:13}) that the selected ASC ${{{\rm{\hat x}}}_{{{\rm{s}}_i}}}(t)( \in {{\rm{R}}^{{r_i}}})$ at the sink node is determined by the binary variables $\sigma _{{\hbar _i}}^i(t)({\hbar _i} = 1,2, \cdots ,{\Delta _i})$. On the other hand, it is known from (\ref{eq:15}) that the design of optimal $\sigma _{{\hbar _i}}^i(t)({\hbar _i} = 1,2, \cdots ,{\Delta _i})$ must be completed at the FC, because the communication delay (from the sink node to the FC) and each CSE ${\rm{\hat x}}_i^{\rm{c}}(t)$ are only obtained by the FC, but these information are unknown to each sink node. Therefore, an optimal ${{{\rm{\hat x}}}_{{{\rm{s}}_i}}}(t)$ may be difficult to be designed at the sink node. Based on the above consideration, let each binary variable $\sigma _{{\hbar_i}}^i(t)$ be generated in a random way at the sink node, and let random variables $\{ \sigma _1^i(t),\sigma _2^i(t), \cdots ,\sigma _{{\Delta _i}}^i(t)\}$ obey the categorical distribution satisfying
\begin{eqnarray}
\begin{array}{l}
 {\rm{E}}\{ \sigma _{{\hbar _i}}^i(t)\sigma _{\hbar _j^0}^j(t)\}  \\
 \;\;\;\;\;\;\;\;\;\;\; = \left\{ \begin{array}{l}
 {\delta _{{\hbar _i},\hbar _i^0}}{\rm{E}}\{ \sigma _{{\hbar _i}}^i(t)\} \;\;\;\;\;\;\;\;\;\;\;{\rm{if}}\;\;i = j,t = {t_1} \\
 {\rm{E}}\{ \sigma _{{\hbar _i}}^i(t)\} {\rm{E}}\{ \sigma _{\hbar _i^0}^i({t_1})\} \;\;{\rm{if}}\;i = j,t \ne {t_1} \\
 {\rm{E}}\{ \sigma _{{\hbar _i}}^i(t)\} {\rm{E}}\{ \sigma _{\hbar _j^0}^j({t_1})\} \;\;{\rm{if}}\;i \ne j \\
 \end{array} \right. \\
 \end{array}.
\label {eq:19}
\end{eqnarray}
Under this case, a group of ASC ${\rm{\hat x}}_{{s_i}}^{{\hbar _i}}(t)({\hbar _i} \in \{ 1,2, \cdots ,{\Delta _i}\})$ in the set (\ref{eq:10}) is randomly selected as the ${{{\rm{\hat x}}}_{{s_i}}}(t)$ at time $t$. Moreover, the occurrence probabilities of the cases $\sigma _{{\hbar _i}}^i(t)=1$ and $\sigma _{{\hbar _i}}^i(t)=0$ are given by ${\rm{Prob}}\{ \sigma _{{\hbar _i}}^i(t) = 1\}  = \pi _{{\hbar _i}}^i$ and ${\rm{Prob}}\{ \sigma _{{\hbar _i}}^i(t) = 0\}  = 1 - \pi _{{\hbar _i}}^i$, where the selection probability $\pi _{{\hbar _i}}^i \ge 0$ satisfies:
\vspace{-8pt}
\begin{eqnarray}
\sum\nolimits_{{\hbar _i} = 1}^{{\Delta _i}} {\pi _{{\hbar _i}}^i}  = 1\;(i \in \{ 1,2, \cdots ,L\} ).
\label {eq:20}
\end{eqnarray}
Then, it is concluded from (\ref{eq:16}) and (\ref{eq:19}) that the binary variables $\gamma _\ell ^i(t)(\ell  = 1, \cdots ,n)$ in (\ref{eq:17}) are independent Bernoulli distributed white noise sequences with ${\rm{Prob}}\{ \gamma _\ell ^i(t) = 1\}  \buildrel \Delta \over = \gamma _\ell ^i$ and ${\rm{Prob}}\{ \gamma _\ell ^i(t) = 0\}  \buildrel \Delta \over = 1 - \gamma _\ell ^i$, which yields
\begin{eqnarray}
{H_i} \buildrel \Delta \over = {\rm E}\{ {H_i}(t-d_i)\}  = {\rm{diag}}\{ \gamma _1^i,\gamma _2^i, \cdots ,\gamma _n^i\}.
\label {eq:21}
\end{eqnarray}
From (\ref{eq:16}) and (\ref{eq:21}), there must exist a constant matrix~$U_\ell ^i \in {{\rm{R}}^{1 \times {\Delta _i}}}$ such that
\begin{eqnarray}
\gamma _\ell ^i = U_\ell ^i{\varsigma _i}(\ell  = 1, \cdots ,n;i = 1, \cdots ,L),
\label {eq:22}
\end{eqnarray}
where ${\varsigma _i} \buildrel \Delta \over = {\rm{col}}\{ \pi _1^i,\pi _2^i, \cdots ,\pi _{{\Delta _i}}^i\}$. This means that when each selection probability $\pi _{{\hbar _i}}^i$ is given by (\ref{eq:19}--\ref{eq:20}), $\gamma _\ell ^i$ in (\ref{eq:21}) will be determined by (\ref{eq:22}). Notice that the selection probabilities ${\varsigma _i}(i = 1,2, \cdots ,L)$ are to be designed in this paper for guaranteeing the stability of the DKFE.

Let~${\rm{\tilde x}}_i^{\rm{c}}(t) \buildrel \Delta \over = x(t) - {\rm{\hat x}}_i^{\rm{c}}(t)$~denote the estimation error of each CSE. Then, it follows from~(\ref{eq:1})~and~(\ref{eq:15})~that
\begin{eqnarray}
\begin{array}{l}
 {\rm{\tilde x}}_i^{\rm{c}}(t) = {A^{{d_i}}}{H_i}(t - {d_i}){{{\rm{\tilde x}}}_i}(t - {d_i}) \\
 \;\;\;\;\;\; + {A^{{d_i}}}[{I_n} - {H_i}(t - {d_i})]A{\rm{\tilde x}}_i^{\rm{c}}(t - {d_i} - 1) \\
 \;\;\;\;\;\; + {A^{{d_i}}}[{I_n} - {H_i}(t - {d_i})]w(t - {d_i} - 1) + {{\rm{F}}_w}({d_i},t), \\
 \end{array}
\label {eq:23}
\end{eqnarray}
where~${\rm{F}}_w(d_i,t)$~is determined by the following function:
\begin{eqnarray}
{{\rm{F}}_w}(g,t) \buildrel \Delta \over = \sum\nolimits_{\theta  = 1}^{g} {{A^{\theta  - 1}} w(t - \theta )}.
\label {eq:24}
\end{eqnarray}
When ${\rm E}\{ {\rm{\hat x}}_i^{\rm{c}}({-d_{{n _i}}})\}  = {\rm E}\{ x({-d_{{n _i}}})\} ({d_{{n _i}}} = 0,1,\cdots ,d_i)$, it is concluded from (\ref{eq:3}), (\ref{eq:23}) and the fact ${\rm E}\{ x(t)\}  = {\rm E}\{ {\hat x_i}(t)\}$ that each CSE ${\rm{\hat x}}_i^{\rm{c}}(t)$ is unbiased, i.e.,
\begin{eqnarray}
{\rm E}\{ {\rm{\hat x}}_i^{\rm{c}}(t)\}  = {\rm E}\{ x(t)\} (i = 1,2, \cdots ,L).
\label {eq:25}
\end{eqnarray}
According to the CSEs~${\rm{\hat x}}_i^{\rm{c}}(t)(i = 1,2, \cdots ,L)$~in the FC, the DKFE for the addressed CPSs is given by:
\begin{eqnarray}
{\rm{\hat x}}(t) = \sum\nolimits_{i = 1}^L {{\Omega _i}(t){\rm{\hat x}}_i^{\rm{c}}(t)},
\label {eq:26}
\end{eqnarray}
where $\sum\nolimits_{i = 1}^L {{\Omega _i}(t)}  = {I_n}$, and combining (\ref{eq:25}) yields that the DKFE~${\rm{\hat x}}(t)$ is unbiased if ${\rm E}\{ {\rm{\hat x}}_i^{\rm{c}}({d_{{n _i}}})\}  = {\rm E}\{ x({d_{{n _i}}})\} ({d_{{n _i}}} = 0,1,\cdots ,d_i)$.

Consequently, the problems to be solved in this paper are described as follows:

1) When the selection probabilities $\pi _{{\hbar _i}}^i({\hbar _i} = 1,\cdots ,{\Delta _i};i = 1,\cdots ,L)$ satisfying (\ref{eq:20}) are given in advance, the aim is to design optimal weighting matrices ${\Omega _1}(t), \cdots ,{\Omega _L}(t)$ such that the MSE of the DKFE ${\rm{\hat x}}(t)$ is minimal at each time step, i.e.,
\begin{eqnarray}
\begin{array}{l}
 \{ {\Omega _1}(t), \cdots ,{\Omega _L}(t)\}  \\
 \;\;\;\;\; = \arg \mathop {\min }\limits_{\sum\nolimits_{i = 1}^L {{\Omega _i}(t) = I} } {\rm{E}}\{ {[x(t) - {\rm{\hat x}}(t)]^{\rm{T}}}[x(t) - {\rm{\hat x}}(t)]\}. \\
 \end{array}
\label {eq:27}
\end{eqnarray}

2) Find stability conditions, which are dependent on the communication delay $d_i$ in (\ref{eq:9}) and the selection probability $\pi _{{\hbar _i}}^i$ in (\ref{eq:20}), such that the estimation error covariance matrix of the DKFE converges to a unique positive matrix, i.e.,
\begin{eqnarray}
\mathop {\lim }\limits_{t \to \infty } {\rm{E}}\{ [x(t) - {\rm{\hat x}}(t)]{[x(t) - {\rm{\hat x}}(t)]^{\rm{T}}}\}  = P,
\label {eq:28}
\end{eqnarray}
and $P$ is independent of the initial values.

\textbf{Remark 3}. When the $i{\rm{th}}$ sink node knows the selection probability ${\varsigma _i}$ in advance, the binary variables $\sigma _{{\hbar _i}}^i(t)({\hbar _i} = 1, \cdots ,{\Delta _i})$ obeying the categorical distribution will be randomly generated at each time step, and then the selected ASC ${{{\rm{\hat x}}}_{{s_i}}}(t)$ can be determined by (\ref{eq:13}) at the sink node. Under this case, one of the important issues in this paper is how to design the satisfactory probability selection criteria, which will be solved in Section IV. On the other hand, when the result (\ref{eq:28}) holds, the limit of each weighting matrix ${\Omega _i}(t)$ must exist, and will be independent of the initial values. This is because the estimation error covariance matrix of the DKFE is dependent on each time-varying matrix ${\Omega _i}(t)$. In such a case, the steady-state DKFE with simple calculation will be given in this paper.

\section{Finite-Horizon DKFE for the CPSs}
In this section, the recursive DKFE will be derived by using the optimal fusion criterion weighted by matrices in the linear minimum variance sense. Define ${\rm{\tilde x}}(t) \buildrel \Delta \over = x(t) - {\rm{\hat x}}(t)$ and ${I_a} = {\rm{col}}\{ {I_n}, \cdots ,{I_n}\}  \in {{\rm{R}}^{nL \times n}}$. Then, from the results in \cite{c8,c9}, the optimal weighting matrices ${\Omega _1}(t), \cdots ,{\Omega _L}(t)$ in (\ref{eq:27}) and the corresponding fusion estimation error covariance matrix $P(t) \buildrel \Delta \over = {\rm E}\{ {\rm{\tilde x}}(t){{\rm{\tilde x}}^{\rm{T}}}(t)\}$ can be calculated by:
\begin{eqnarray}
[{\Omega _1}(t),{\Omega _2}(t), \cdots ,{\Omega _L}(t)] = {(I_a^{\rm{T}}{\Xi ^{ - 1}}(t){I_a})^{ - 1}}I_a^{\rm{T}}{\Xi ^{ - 1}}(t)
\label {eq:29}\;\;\;\\
P(t) = {(I_a^{\rm{T}}{\Xi ^{ - 1}}(t){I_a})^{ - 1}},
\label {eq:30}\;\;\;\;\;\;\;\;\;\;\;\;\;\;\;\;\;\;\;
\end{eqnarray}
where the weighting matrices $\Omega _i(t)(i = 1,2, \cdots ,L)$ determined by (\ref{eq:29}) satisfy the constraint $\sum\nolimits_{i = 1}^L {{\Omega _i}(t)}  = {I_n}$, and
\begin{eqnarray}
\Xi (t) = {\left( {{\Xi _{ij}}(t)} \right)_{nL \times nL}},{\Xi _{ij}}(t) = {\rm E}\{ {\rm{\tilde x}}_i^{\rm{c}}(t){({\rm{\tilde x}}_j^{\rm{c}}(t))^{\rm{T}}}\}.
\label {eq:31}
\end{eqnarray}
It is concluded from (\ref{eq:29}) and (\ref{eq:31}) that if the computation procedure of $\Xi (t)$ is given, then the optimal weighting matrices $\Omega _i(t)(i = 1, \cdots ,L)$ in (\ref{eq:29}) can be thus obtained.

In what follows, six lemmas will be given before deriving the recursive form of ${\Xi _{ij}}(t)$. For notational convenience, the following indicator function is introduced:
\begin{eqnarray}
{{\rm{C}}_{\rm{o}}}({t_1},{t_2}) = \left\{ \begin{array}{l}
 1\;\;\;{\rm{if}}\;\;{t_1} > {t_2} \\
 0\;\;\;{\rm{if}}\;\;{t_1} \le {t_2} \\
 \end{array}. \right.
\label {eq:32}
\end{eqnarray}
Meanwhile, if ${\tau _1} > {\tau _2}$, it will be specified that $\prod\nolimits_{\tau  = {\tau _1}}^{{\tau _2}} {F(\tau )}  = {I_m}$ and $\sum\nolimits_{\tau  = {\tau _1}}^{{\tau _2}} {G(\tau )}  = 0$, where $F(\tau ) \in {{\rm{R}}^{m \times m}}$ and $G(\tau ) \in {{\rm{R}}^{n \times n}}$ represent different matrix functions with respect to the variable $\tau$.

\textbf{Lemma 1} \cite{c32} For stochastic matrices~$U$,~$B$,~$G$,~where
\begin{eqnarray*}
\begin{array}{c}
  U\buildrel \Delta \over = {\rm{diag\{ }}{u_1}{\rm{,}} \cdots {\rm{,}}{u_n}{\rm{\} ,}}\;B \buildrel \Delta \over = {\rm{diag}}\{ {b_1}, \cdots ,{b_n}\}  \\
 G \buildrel \Delta \over = \left[ {\begin{array}{*{20}{c}}
   {{g_{11}}} &  \cdots  & {{g_{1n}}}  \\
    \vdots  &  \ddots  &  \vdots   \\
   {{g_{n1}}} &  \cdots  & {{g_{nn}}}  \\
\end{array}} \right] \\
 \end{array}.
\end{eqnarray*}
If each random variable~${g_{ij}}$~in~$G$~is independent of any random variables of~${u_k}$~and~${b_k}(k = 1,2, \cdots ,n),$~then
\begin{eqnarray*}
{\rm E}\{ UGB\}  = {\rm E}\{ U \odot B\}  \otimes {\rm E}\{ G\},
\end{eqnarray*}
where~``$\otimes$''~is defined as~${[{G^1} \otimes {G^2}]_{ij}} = G_{ij}^1G_{ij}^2$,~and the product~``$\odot$''~for the matrices~$U$~and~$B$~is defined by
\begin{eqnarray*}
U \odot B = \left[ {\begin{array}{*{20}{c}}
   {{u_1}{b_1}} &  \cdots  & {{u_1}{b_n}}  \\
    \vdots  &  \ddots  &  \vdots   \\
   {{u_n}{b_1}} &  \cdots  & {{u_n}{b_n}}  \\
\end{array}} \right].
\end{eqnarray*}

\textbf{Lemma 2} Define
\begin{eqnarray}
\left\{ \begin{array}{l}
 {\Phi _{{{\rm{K}}_i}}}(t) \buildrel \Delta \over = {{\rm{G}}_{{{\rm{K}}_i}}}(t)A,
 \Phi _{{{\rm{x}}_i}}^w({t_1},{t_2}) \buildrel \Delta \over = {\rm{E}}\{ {{{\rm{\tilde x}}}_i}({t_1}){w^{\rm{T}}}({t_2})\}  \\
 \Phi _{{{\rm{x}}_i}}^{{{\rm{x}}_j}}({t_1},{t_2}) \buildrel \Delta \over = {\rm{E}}\{ {{{\rm{\tilde x}}}_i}({t_1}){\rm{\tilde x}}_j^{\rm{T}}({t_2})\}  \\
 \Phi _{{{\rm{x}}_i}}^{\rm{F}}({t_1},g,{t_2}) \buildrel \Delta \over = {\rm{E}}\{ {{{\rm{\tilde x}}}_i}({t_1}){\rm{F}}_w^{\rm{T}}(g,{t_2})\}  \\
 \Phi _{\rm{F}}^w(g,{t_1},{t_2}) \buildrel \Delta \over = {\rm E}\{ {{\rm{F}}_w}(g,{t_1}){w^{\rm{T}}}({t_2})\}  \\
 \end{array}, \right.
\label {eq:33}
\end{eqnarray}
where ${{\rm{G}}_{{{\rm K}_i}}}(t)$ and ${{\rm{F}}_w}(g,{t_2})$ are determined by (\ref{eq:6}) and (\ref{eq:24}), respectively. Then, $\Phi _{{{\rm{x}}_i}}^w({t_1},{t_2})$, $\Phi _{{{\rm{x}}_i}}^{{v_j}}({t_1},{t_2})$, $\Phi _{{{\rm{x}}_i}}^{{{\rm{x}}_j}}({t_1},{t_2})$ and $\Phi _{{{\rm{x}}_i}}^{\rm{F}}({t_1},g,{t_2})$ are given by:
\begin{eqnarray}
\begin{array}{l}
 \Phi _{{{\rm{x}}_i}}^w({t_1},{t_2}) = {{\rm{C}}_{\rm{o}}}({t_1},{t_2})\left( {\prod\nolimits_{{\varphi _i} = 0}^{{t_1} - {t_2} - 2} {{\Phi _{{{\rm K}_i}}}({t_1} - {\varphi _i})} } \right) \\
 \;\;\;\;\;\;\;\;\;\;\;\;\;\;\;\;\;\;\;\; \times {{\rm{G}}_{{{\rm{K}}_i}}}({t_2} + 1){Q_w} \\
 \end{array}
\label {eq:34}\;\;\;\;\;\\
\Phi _{{{\rm{x}}_i}}^{{{\rm{x}}_j}}({t_1},{t_2}) = \left\{ \begin{array}{l}
 \left( {\prod\nolimits_{{\varphi _i} = 0}^{{t_1} - {t_2} - 1} {{\Phi _{{{\rm{K}}_i}}}({t_1} - {\varphi _i})} } \right){P_{ij}}({t_2}) \\
 \;\;\;\;\;\;\;\;\;\;\;\;\;\;\;\;\;\;\;\;\;\;\;\;\;{\rm{if}}\;{t_1} \ge {t_2} \\
 {[\Phi _{{{\rm{x}}_j}}^{{{\rm{x}}_i}}({t_2},{t_1})]^{\rm{T}}}\:\:\;\;\;{\rm{if}}\:\:{t_1} < {t_2} \\
 \end{array} \right.\;\;\;\;
\label {eq:35}\\
\Phi _{{{\rm{x}}_i}}^{\rm{F}}({t_1},g,{t_2}) = \sum\nolimits_{\theta  = 1}^g {\Phi _{{{\rm{x}}_i}}^w({t_1},{t_2} - \theta ){{[{A^{\theta  - 1}}]}^{\rm{T}}}}\;\;\;\;\;\;\;\;\;\;\;
\label {eq:36}\\
\Phi _{\rm{F}}^w(g,{t_1},{t_2}) = {{\rm{C}}_{\rm{o}}}({t_1},{t_2}){{\rm{C}}_{\rm{o}}}({t_2},{t_1} - g - 1){A^{{t_2} - 1}}{Q_w},
\label {eq:37}\;
\end{eqnarray}
where ${\delta _{i,j}}$ and ${{\rm{C}}_{\rm{o}}}({t_1},{t_2})$ are determined by (\ref{eq:4}) and (\ref{eq:32}), respectively. ${P_{ij}}(t_2)$ in (\ref{eq:35}) is calculated by (\ref{eq:7}) or (\ref{eq:8}).

\begin{proof}
See A.1 in Appendix.
\end{proof}

\textbf{Lemma 3} Define
\begin{eqnarray}
\left\{ \begin{array}{l}
 {f_i}(t) \buildrel \Delta \over = t - {d_i} - 1,f_{i{\rm{o}}}^0(t) \buildrel \Delta \over = t \\
 {\chi _i}({t_1},{t_2}) \buildrel \Delta \over = \min \{ {\chi _i}({t_1},{t_2})|f_{i{\rm{o}}}^{{\chi _i}({t_1},{t_2})}({t_2}) - {t_1} \le 0\}  \\
 \Theta _{{\rm{x}}_i^{\rm{c}}}^w({t_1},{t_2}) \buildrel \Delta \over = {\rm E}\{ {\rm{\tilde x}}_i^{\rm{c}}({t_1}){w^{\rm{T}}}({t_2})\}  \\
 \Theta _{{\rm{x}}_i^{\rm{c}}}^{\rm{F}}({t_1},g,{t_2}) \buildrel \Delta \over = {\rm E}\{ {\rm{\tilde x}}_i^{\rm{c}}({t_1}){\rm{F}}_w^{\rm{T}}(g,{t_2})\}  \\
 \end{array}. \right.
\label {eq:41}
\end{eqnarray}
Then, $\Theta _{{\rm{x}}_i^{\rm{c}}}^w({t_1},{t_2})$, $\Theta _{{\rm{x}}_i^{\rm{c}}}^{{v_j}}({t_1},{t_2})$, $\Theta _{{\rm{x}}_i^{\rm{c}}}^{\rm{F}}({t_1},g,{t_2})$ are given by:
\begin{eqnarray}
\begin{array}{l}
 \Theta _{{\rm{x}}_i^{\rm{c}}}^w({t_1},{t_2}) = {{\rm{C}}_{\rm{o}}}({t_1} - {d_i},{t_2}) \\
 \;\;\;\;\;\;\;\; \times \left\{ {\sum\nolimits_{\hbar  = 0}^{{\chi _i}({t_1},{t_2}) - 1} {\{ {\delta _{f_{i{\rm{o}}}^{\hbar  + 1}({t_1}),{t_2}}}{\rm{H}}_{{\rm{A}}{{\rm{d}}_i}}^\hbar {{{\rm{\bar H}}}_{{\rm{A}}{{\rm{d}}_i}}}{Q_w}} } \right. \\
 \;\;\;\;\;\;\;\;{\rm{ + }}{{\rm{H}}_{{{\rm{d}}_i}}}\Phi _{{{\rm{x}}_i}}^w(f_{i{\rm{o}}}^\hbar ({t_1}) - {d_i},{t_2})\}  + \Phi _{\rm{F}}^w({d_i},{t_1},{t_2}) \\
 \left. {\;\;\;\;\;\;\;\; + \sum\nolimits_{\hbar  = 1}^{{\chi _i}({t_1},{t_2}) - 1} {{{\rm{H}}_{{\rm{A}}{{\rm{d}}_i}}}\Phi _{\rm{F}}^w({d_i},f_{i{\rm{o}}}^\hbar ({t_1}),{t_2})} } \right\} \\
 \end{array}
\label {eq:42}\\
\Theta _{{\rm{x}}_i^{\rm{c}}}^{\rm{F}}({t_1},g,{t_2}) = \sum\nolimits_{\theta  = 1}^g {\Theta _{{\rm{x}}_i^{\rm{c}}}^w({t_1},{t_2} - \theta ){{({A^{\theta  - 1}})}^{\rm{T}}}},
\label {eq:43}\;\;
\end{eqnarray}
where
\begin{eqnarray}
\left\{ \begin{array}{l}
 {{\rm{H}}_{{{\rm{d}}_i}}} = {A^{{d_i}}}{H_i},{{{\rm{\bar H}}}_{{\rm{A}}{{\rm{d}}_i}}} = {A^{{d_i}}} - {{\rm{H}}_{{{\rm{d}}_i}}} \\
 {{\rm{H}}_{{\rm{A}}{{\rm{d}}_i}}} = {A^{{d_i}}}[{I_n} - {H_i}]A \\
 \end{array}, \right.
\label {eq:44}
\end{eqnarray}
where $H_i$ is determined by (\ref{eq:21}), while $\Phi _{{{\rm{x}}_i}}^w({t_1},{t_2})$ and $\Phi _{\rm{F}}^w({d_i},{t_1},{t_2})$ are calculated by (\ref{eq:34}) and (\ref{eq:37}).

\begin{proof}
See A.2 in Appendix.
\end{proof}

The statistical correlation between ${{{\rm{\tilde x}}}_i}(t)$ and $w(t)$ is presented in Lemma 2, while Lemma 3 gives the statistical correlation between ${\rm{\tilde x}}_i^{\rm{c}}(t)$ and $w(t)$. Additionally, to obtain the ${\Xi _{ij}}(t)$, it is still required to know the statistical correlations between ${{{\rm{\tilde x}}}_i}({t_1})$ and ${\rm{\tilde x}}_i^{\rm{c}}({t_2})$, and between ${\rm{\tilde x}}_i^{\rm{c}}({t_1})$ and ${\rm{\tilde x}}_i^{\rm{c}}({t_2})$, which will be derived in Lemmas 4--6.

\textbf{Lemma 4} Define
\begin{eqnarray}
\left\{ \begin{array}{l}
 {\Gamma _{ij}}(t) \buildrel \Delta \over = {\rm E}\{ {{{\rm{\tilde x}}}_i}(t){[{\rm{\tilde x}}_j^{\rm{c}}(t)]^{\rm{T}}}\}  \\
 {\Gamma _{ij}}({t_1},{t_2}) \buildrel \Delta \over = {\rm E}\{ {{{\rm{\tilde x}}}_i}({t_1}){[{\rm{\tilde x}}_j^{\rm{c}}({t_2})]^{\rm{T}}}\} ({t_1} > {t_2}) \\
 \end{array}. \right.
\label {eq:50}
\end{eqnarray}
Then, ${\Gamma _{ij}}(t)$ is calculated by the following recursive form:
\begin{eqnarray}
\begin{array}{l}
 {\Gamma _{ij}}(t) = \left( {\prod\nolimits_{{\varphi _j} = 0}^{{d_j}} {{\Phi _{{{\rm K}_i}}}(t - {\varphi _i})} } \right){\Gamma _{ij}}(t - {d_j} - 1){\rm{H}}_{{\rm{A}}{{\rm{d}}_j}}^{\rm{T}} \\
 \;\;\;\;\;\;\;\;\;\;\;\; + \Phi _{{{\rm{x}}_i}}^{{{\rm{x}}_j}}(t,t - {d_j}){\rm{H}}_{{{\rm{d}}_j}}^{\rm{T}} + \Phi _{{{\rm{x}}_i}}^{\rm{F}}(t,{d_j},t) \\
 \;\;\;\;\;\;\;\;\;\;\;\; + \Phi _{{{\rm{x}}_i}}^w(t,t - {d_j} - 1){\rm{\bar H}}_{{\rm{A}}{{\rm{d}}_j}}^{\rm{T}} \\
 \end{array},
\label {eq:51}
\end{eqnarray}
where ${{\rm{H}}_{{{\rm{d}}_j}}},{{{\rm{\bar H}}}_{{\rm{A}}{{\rm{d}}_j}}},{{\rm{H}}_{{\rm{A}}{{\rm{d}}_j}}}$ are given by (\ref{eq:44}), while $\Phi _{{{\rm{x}}_i}}^{{{\rm{x}}_j}}(t,t - {d_j}),\Phi _{{{\rm{x}}_i}}^{\rm{F}}(t,{d_j},t),\Phi _{{{\rm{x}}_i}}^w(t,t - {d_j} - 1)$ are computed by (\ref{eq:34}), (\ref{eq:35}) and (\ref{eq:36}). In this case, ${\Gamma _{ij}}({t_1},{t_2})$ is calculated by:
\begin{eqnarray}
{\Gamma _{ij}}({t_1},{t_2}) = \left( {\prod\nolimits_{{\varphi _i} = 0}^{{t_1} - {t_2} - 1} {{\Phi _{{{\rm K}_i}}}({t_1} - {\varphi _i})} } \right){\Gamma _{ij}}({t_2}).
\label {eq:52}
\end{eqnarray}

\begin{proof}
See A.3 in Appendix.
\end{proof}

\textbf{Lemma 5} Define
\begin{eqnarray}
{\Psi _{ij}}(t) \buildrel \Delta \over = {\rm{E}}\{ {{{\rm{\tilde x}}}_i}(t - {d_i}){[{\rm{\tilde x}}_j^{\rm{c}}(t - {d_j} - 1)]^{\rm{T}}}\}.
\label {eq:56}
\end{eqnarray}
For $i=j$, ${\Psi _{ii}}(t)$ is calculated by
\begin{eqnarray}
{\Psi _{ii}}(t) = {\Phi _{{{\rm K}_i}}}(t - {d_i}){\Gamma _{ij}}(t - {d_i} - 1).
\label {eq:57}
\end{eqnarray}
For $i \ne j$, let ${\eta _{ij}} \buildrel \Delta \over = \min \{ {\eta _{ij}}|{\eta _{ij}}({d_j} + 1) - {d_i} \ge 0\} $. Then, ${\Psi _{ij}}(t)$ is calculated by
\begin{eqnarray}
\begin{array}{l}
 {\Psi _{ij}}(t) = \sum\nolimits_{\kappa  = 1}^{{\eta _{ij}} - 1} {\{ [\Phi _{{{\rm{x}}_i}}^{{{\rm{x}}_j}}(t - {d_i},f_{j{\rm{o}}}^\kappa (t) - {d_j})} {\rm{H}}_{{{\rm{d}}_j}}^{\rm{T}} \\
 \;\;\;\;\;\;\;\;\;\;\; + \Phi _{{{\rm{x}}_i}}^w{\rm{(}}t - {d_i},f_{j{\rm{o}}}^{\kappa  + 1}(t)){\rm{\bar H}}_{{\rm{A}}{{\rm{d}}_j}}^{\rm{T}} \\
 \;\;\;\;\;\;\;\;\;\;\; + \Phi _{{{\rm{x}}_i}}^{\rm{F}}{\rm{(}}t - {d_i},{d_j},f_{j{\rm{o}}}^\kappa (t))]{({\rm{H}}_{{\rm{A}}{{\rm{d}}_j}}^{\kappa  - 1})^{\rm{T}}}\}  \\
 \;\;\;\;\;\;\;\;\;\;\; + \left( {\prod\nolimits_{{\varphi _i} = 0}^{{\eta _{ij}}({d_j} + 1) - 1 - {d_i}} {{\Phi _{{{\rm K}_i}}}(t - {d_i} - {\varphi _i})} } \right) \\
 \;\;\;\;\;\;\;\;\;\;\; \times {\Gamma _{ij}}(f_{j{\rm{o}}}^{{\eta _{ij}}}(t)){({\rm{H}}_{{\rm{A}}{{\rm{d}}_j}}^{{\eta _{ij}} - 1})^{\rm{T}}} \\
 \end{array},
\label {eq:58}
\end{eqnarray}
where ${{\rm{H}}_{{{\rm{d}}_j}}},{{{\rm{\bar H}}}_{{\rm{A}}{{\rm{d}}_j}}},{{\rm{H}}_{{\rm{A}}{{\rm{d}}_j}}}$ are given by (\ref{eq:44});
${\Gamma _{ij}}(t)$ is computed by (\ref{eq:51}), while $\Phi _{{{\rm{x}}_i}}^w{\rm{(}}t - {d_i},f_{j{\rm{o}}}^{\kappa  + 1}(t))$, ${\Phi _{{{\rm{x}}_i}}^{{{\rm{x}}_j}}(t - {d_i},f_{j{\rm{o}}}^\kappa (t) - {d_j})}$ and $\Phi _{{{\rm{x}}_i}}^{\rm{F}}{\rm{(}}t - {d_i},{d_j},f_{j{\rm{o}}}^\kappa (t))$ are calculated by (\ref{eq:34}), (\ref{eq:35}), (\ref{eq:36}).

\begin{proof}
See A.4 in Appendix.
\end{proof}

\textbf{Remark 4}. It should be pointed out that ${\Psi _{ij}}(t)$ in Lemma 5 is not a special case of ${\Gamma _{ij}}({t_1},{t_2})$ in Lemma 4 because there may exist the case ${d_i} > {d_j} + 1$. Notice that when ${d_i} \le {d_j} + 1$, one has ${\eta _{ij}} = 1$. Then, according to the definitions of $\sum\nolimits_{\tau  = {\tau _1}}^{{\tau _2}} {G(\tau )}  = 0$ and $\prod\nolimits_{\tau  = {\tau _1}}^{{\tau _2}} {F(\tau )}  = I$ for ${\tau _1} > {\tau _2}$, ${\Psi _{ij}}(t)$ is calculated by ${\Psi _{ij}}(t) = \left( {\prod\nolimits_{{\varphi _i} = 0}^{{d_j} - {d_i}} {{\Phi _{{{\rm K}_i}}}(t - {d_i} - {\varphi _i})} } \right){\Gamma _{ij}}(t - {d_j} - 1)$.

\textbf{Lemma 6} Define
\begin{eqnarray}
\left\{ \begin{array}{l}
 {\tau _{ij}}={\tau _{ji}} \buildrel \Delta \over = {\rm{lcm}}({d_i} + 1,{d_j} + 1),{\tau _{{d_i}}} \buildrel \Delta \over = {{{\tau _{ij}}} \mathord{\left/
 {\vphantom {{{\tau _{ij}}} {({d_i} + 1}}} \right.
 \kern-\nulldelimiterspace} {({d_i} + 1}}) \\
 {\Upsilon _{ij}}(t) \buildrel \Delta \over = {\rm E}\{ {\rm{\tilde x}}_i^{\rm{c}}(t - {d_i} - 1){[{\rm{\tilde x}}_j^{\rm{c}}(t - {d_j} - 1)]^{\rm{T}}}\}  \\
 \Upsilon _{ij}^{\rm{x}}(t) \buildrel \Delta \over = {\rm E}\{ {\rm{\bar x}}_{{f_i}}^w(t){[{\rm{\bar x}}_{{f_j}}^w(t)]^{\rm{T}}}\}  \\
 \Upsilon _{ij}^{{\rm{c}}}(t) \buildrel \Delta \over = {\rm E}\{ {\rm{\bar x}}_{{f_i}}^w(t){[{\rm{\tilde x}}_j^{\rm{c}}(t - {\tau _{ij}})]^{\rm{T}}}\}  \\
 \end{array}, \right.
\label {eq:61}
\end{eqnarray}
where ${\rm{\bar x}}_f^w(t)$ is defined as follows:
\begin{eqnarray}
\left\{ \begin{array}{l}
 {{{\rm{\tilde x}}}_{{f_i}}}(t) \buildrel \Delta \over = {\rm{col\{ }}{{{\rm{\tilde x}}}_i}(f_{i{\rm{o}}}^1(t) - {d_i}), \cdots ,{{{\rm{\tilde x}}}_i}(f_{i{\rm{o}}}^{{\tau _{{d_i}}} - 1}(t) - {d_i}){\rm{\} }} \\
 {w_{{f_i}}}(t) \buildrel \Delta \over = {\rm{col\{ }}w(f_{i{\rm{o}}}^2(t)), \cdots ,w(f_{i{\rm{o}}}^{{\tau _{{d_i}}}}(t)){\rm{\} }} \\
 {{\rm{F}}_{{f_i}}}(t) \buildrel \Delta \over = {\rm{col\{ }}{{\rm{F}}_w}({d_i},f_{i{\rm{o}}}^1(t)), \cdots ,{{\rm{F}}_w}({d_i},f_{i{\rm{o}}}^{{\tau _{{d_i}}} - 1}(t)){\rm{\} }} \\
 {\rm{\bar x}}_f^w(t) \buildrel \Delta \over = {\rm{col}}\{ {{{\rm{\tilde x}}}_{{f_i}}}(t),{w_{{f_i}}}(t),{{\rm{F}}_{{f_i}}}(t)\}  \\
 \end{array}. \right.
\label {eq:62}
\end{eqnarray}
Then, $\Upsilon _{ij}^{\rm{x}}(t)$ can be calculated by (\ref{eq:34}--\ref{eq:36}) (see Lemma 2), while $\Upsilon _{ij}^{{\rm{c}}}(t)$ can be calculated by (\ref{eq:42}--\ref{eq:43}) (see Lemma 3) and (\ref{eq:52}). In this case, ${\Upsilon _{ij}}(t)$ is calculated by:
\begin{eqnarray}
\left\{ \begin{array}{l}
 {\Upsilon _{ij}}(t) = {\rm{H}}_{{\rm{A}}{{\rm{d}}_i}}^{{\tau _{{d_i}}} - 1}{\Xi _{ij}}(t - {\tau _{ij}}){[{\rm{H}}_{{\rm{A}}{{\rm{d}}_j}}^{{\tau _{{d_j}}} - 1}]^{\rm{T}}} + {{\hat \Upsilon }_{ij}}(t) \\
 {{\hat \Upsilon }_{ij}}(t) = (1 - {\delta _{1,{\tau _{{d_i}}}}})(1 - {\delta _{1,{\tau _{{d_j}}}}}){\Sigma _i}\Upsilon _{ij}^{\rm{x}}(t)\Sigma _j^{\rm{T}} \\
 \;\;\;\;\;\;\;\;\;\;\;\;\; + (1 - {\delta _{1,{\tau _{{d_i}}}}}){\Sigma _i}\Upsilon _{ij}^{\rm{c}}(t){[{\rm{H}}_{{\rm{A}}{{\rm{d}}_j}}^{{\tau _{{d_j}}} - 1}]^{\rm{T}}} \\
 \;\;\;\;\;\;\;\;\;\;\;\;\; + (1 - {\delta _{1,{\tau _{{d_j}}}}}){\rm{H}}_{{\rm{A}}{{\rm{d}}_i}}^{{\tau _{{d_i}}} - 1}{[\Upsilon _{ji}^{\rm{c}}(t)]^{\rm{T}}}\Sigma _j^{\rm{T}} \\
 \end{array}, \right.
\label {eq:63}
\end{eqnarray}
where ${\delta _{1,{\tau _{{d_i}}}}}$ is defined in (\ref{eq:4}), and
\begin{eqnarray}
\left\{ \begin{array}{l}
 {\Sigma _{1i}} \buildrel \Delta \over = [{{\rm{H}}_{{{\rm{d}}_i}}},{{\rm{H}}_{{\rm{A}}{{\rm{d}}_i}}}{{\rm{H}}_{{{\rm{d}}_i}}}, \cdots ,{\rm{H}}_{{\rm{A}}{{\rm{d}}_i}}^{{\tau _{{d_i}}} - 2}{{\rm{H}}_{{{\rm{d}}_i}}}] \\
 {\Sigma _{2i}} \buildrel \Delta \over = [{{{\rm{\bar H}}}_{{\rm{A}}{{\rm{d}}_i}}},{{\rm{H}}_{{\rm{A}}{{\rm{d}}_i}}}{{{\rm{\bar H}}}_{{\rm{A}}{{\rm{d}}_i}}}, \cdots ,{\rm{H}}_{{\rm{A}}{{\rm{d}}_i}}^{{\tau _{{d_i}}} - 2}{{{\rm{\bar H}}}_{{\rm{A}}{{\rm{d}}_i}}}] \\
 {\Sigma _{3i}} \buildrel \Delta \over = [{I_n},{{\rm{H}}_{{\rm{A}}{{\rm{d}}_i}}}, \cdots ,{\rm{H}}_{{\rm{A}}{{\rm{d}}_i}}^{{\tau _{{d_i}}} - 2}] \\
 {\Sigma _i} \buildrel \Delta \over = [{\Sigma _{1i}}\;{\Sigma _{2i}}\;{\Sigma _{3i}}] \\
 \end{array}. \right.
\label {eq:64}
\end{eqnarray}

\begin{proof}
See A.5 in Appendix.
\end{proof}

\textbf{Remark 5}. Notice that the structure of $\Upsilon _{ij}^{\rm{x}}(t)$ consists of $\Phi _{{{\rm{x}}_i}}^w({t_1},{t_2})$, $\Phi _{{{\rm{x}}_i}}^{{{\rm{x}}_j}}({t_1},{t_2})$, $\Phi _{{{\rm{x}}_i}}^{\rm{F}}({t_1},g,{t_2})$ and $\Phi _{{{\rm{x}}_i}}^{\rm{F}}({t_1},g,{t_2})$, and thus $\Upsilon _{ij}^{\rm{x}}(t)$ can be calculated by Lemma 2. Meanwhile, $\Upsilon _{ij}^{{\rm{c}}}(t)$ can be calculated by Lemma 3 and (\ref{eq:52}), because its structure consists of ${\Gamma _{ij}}({t_1},{t_2})({t_1} > {t_2})$, $\Theta _{{\rm{x}}_i^{\rm{c}}}^w({t_1},{t_2})$ and $\Theta _{{\rm{x}}_i^{\rm{c}}}^{\rm{F}}({t_1},g,{t_2})$. On the other hand, in most cases, the delay $d_i$ is not equal to $d_j$. Thus, to design the recursive form of ${\Xi _{ij}}(t)$, \emph{one of the key issues is} how to obtain the relationship between ${\rm E}\{ {\rm{\tilde x}}_i^{\rm{c}}(t - {d_i} - 1){[{\rm{\tilde x}}_j^{\rm{c}}(t - {d_j} - 1)]^{\rm{T}}}\}$ and ${\Xi _{ij}}(t)$, which has been solved by Lemma 6.

According to the results in Lemmas 1--6, the recursive form of ${\Xi _{ij}}(t)$ in (\ref{eq:31}) will be given by Theorem 1.

\textbf{Theorem 1} Define
\begin{eqnarray}
\left\{ \begin{array}{l}
 {\Lambda _i} \buildrel \Delta \over = {\rm E}\{ {H_i}(t) \odot {H_i}(t)\}  \\
 {{\rm{V}}_i} \buildrel \Delta \over = {\rm E}\{ {H_i}(t) \odot [{I_n} - {H_i}(t)]\}  \\
 {{\rm{W}}_i} \buildrel \Delta \over = {\rm E}\{ [{I_n} - {H_i}(t)] \odot [{I_n} - {H_i}(t)]\}  \\
 \end{array}. \right.
\label {eq:69}
\end{eqnarray}
Then, the local estimation error covariance matrix ${\Xi _{ii}}(t)\buildrel \Delta \over = {\rm E}\{ {\rm{\tilde x}}_i^{{\rm{c}}}(t){[{\rm{\tilde x}}_i^{{\rm{c}}}(t)]^{\rm{T}}}\}$ for each CSE ${\rm{\hat x}}_i^{{\rm{c}}}(t)$ is given by:
\begin{eqnarray}
\begin{array}{l}
 {\Xi _{ii}}(t) = {A^{{d_i}}}[{{\rm{W}}_i} \otimes (A{\Xi _{ii}}(t - {d_i} - 1){A^{\rm{T}}})]{({A^{{d_i}}})^{\rm{T}}} \\
 \;\;\;\;\;\;\;\;\;\;\;\;\; + {A^{{d_i}}}[{\Lambda _i} \otimes {P_{ii}}(t - {d_i}) + {{\rm{W}}_i} \otimes {Q_w}]{({A^{{d_i}}})^{\rm{T}}} \\
 \;\;\;\;\;\;\;\;\;\;\;\;\; + {A^{{d_i}}}{\rm{[}}{{\rm{V}}_i} \otimes ({\Phi _{{{\rm K}_i}}}(t - {d_i}){\Gamma _{ii}}(t - {d_i} - 1){A^{\rm{T}}} \\
 \;\;\;\;\;\;\;\;\;\;\;\;\; + {{\rm{G}}_{{{\rm K}_i}}}(t - {d_i}){Q_w})]{({A^{{d_i}}})^{\rm{T}}} \\
 \;\;\;\;\;\;\;\;\;\;\;\;\; + {A^{{d_i}}}[{\rm{V}}_i^{\rm{T}} \otimes (A\Gamma _{ii}^{\rm{T}}(t - {d_i} - 1)\Phi _{{{\rm K}_i}}^{\rm{T}}(t - {d_i}) \\
 \;\;\;\;\;\;\;\;\;\;\;\;\; + {Q_w}{\rm{G}}_{{{\rm K}_i}}^{\rm{T}}(t - {d_i}))]{({A^{{d_i}}})^{\rm{T}}} \\
 \;\;\;\;\;\;\;\;\;\;\;\;\; + \sum\nolimits_{\theta  = 1}^{{d_i}} {{A^{\theta  - 1}}{Q_w}{{[{A^{\theta  - 1}}]}^{\rm{T}}}}  \\
 \end{array},
\label {eq:70}
\end{eqnarray}
where ${P_{ii}}(t - {d_i})$ and ${\Gamma _{ii}}(t - {d_i} - 1)$ are calculated by (\ref{eq:7}) and (\ref{eq:51}) (see Lemma 4). On the other hand, the estimation error cross-covariance matrix ${\Xi _{ij}}(t) \buildrel \Delta \over = {\rm E}\{ {\rm{\tilde x}}_i^{{\rm{c}}}(t){[{\rm{\tilde x}}_j^{{\rm{c}}}(t)]^{\rm{T}}}\}$ is given by:
\vspace{-8pt}
\begin{eqnarray}
\left\{ \begin{array}{l}
 {\Xi _{ij}}(t) = {\rm{H}}_{{\rm{A}}{{\rm{d}}_i}}^{{\tau _{{d_i}}}}{\Xi _{ij}}{\rm{(}}t - {\tau _{ij}}{\rm{)(H}}_{{\rm{A}}{{\rm{d}}_j}}^{{\tau _{{d_j}}}}{)^{\rm{T}}} + {{\hat \Xi }_{ij}}(t) \\
 {{\hat \Xi }_{ij}}(t) = {{\rm{H}}_{{\rm{A}}{{\rm{d}}_i}}}{{\hat \Upsilon }_{ij}}(t){\rm{H}}_{{\rm{A}}{{\rm{d}}_j}}^{\rm{T}} + \sum\nolimits_{\theta  = 1}^{\min \{ {d_i},{d_j}\} } {{A^\theta }{Q_w}{{({A^\theta })}^{\rm{T}}}}  \\
 \;\;\;\;\;\;\;\;\;\;\;\;\;{\rm{ + }}{{\rm{H}}_{{{\rm{d}}_i}}}\{ \Phi _{{{\rm{x}}_i}}^{{{\rm{x}}_j}}(t - {d_i},t - {d_j}){\rm{H}}_{{d_j}}^{\rm{T}} \\
 \;\;\;\;\;\;\;\;\;\;\;\;\; + {\Psi _{ij}}(t){\rm{H}}_{{\rm{A}}{{\rm{d}}_j}}^{\rm{T}}  + \Phi _{{{\rm{x}}_i}}^{\rm{F}}(t - {d_i},{d_j},t) \\
 \;\;\;\;\;\;\;\;\;\;\;\;\; + \Phi _{{{\rm{x}}_i}}^w(t - {d_i},t - {d_j} - 1){\rm{\bar H}}_{{\rm{A}}{{\rm{d}}_j}}^{\rm{T}}\}  \\
 \;\;\;\;\;\;\;\;\;\;\;\;\; + {{\rm{H}}_{{\rm{A}}{{\rm{d}}_i}}}\{ \Theta _{{\rm{x}}_i^{\rm{c}}}^w(t - {d_i} - 1,t - {d_j} - 1){\rm{H}}_{{\rm{A}}{{\rm{d}}_j}}^{\rm{T}} \\
 \;\;\;\;\;\;\;\;\;\;\;\;\; + \Psi _{ji}^{\rm{T}}(t){\rm{H}}_{{{\rm{d}}_j}}^{\rm{T}} + \Theta _{{\rm{x}}_i^{\rm{c}}}^{\rm{F}}(t - {d_i} - 1,{d_j},t)\}  \\
 \;\;\;\;\;\;\;\;\;\;\;\;\; + {{{\rm{\bar H}}}_{{\rm{A}}{{\rm{d}}_i}}}\{ {\delta _{{d_i},{d_j}}}{Q_w}{\rm{\bar H}}_{{\rm{A}}{{\rm{d}}_j}}^{\rm{T}} + {{\rm{C}}_{\rm{o}}}({d_j},{d_i}){Q_w}{({A^{{d_i}}})^{\rm{T}}} \\
 \;\;\;\;\;\;\;\;\;\;\;\;\; + {(\Phi _{{{\rm{x}}_j}}^w(t - {d_j},t - {d_i} - 1))^{\rm{T}}}{\rm{H}}_{{d_j}}^{\rm{T}} \\
 \;\;\;\;\;\;\;\;\;\;\;\;\; + {(\Theta _{{\rm{x}}_i^{\rm{c}}}^w(t - {d_j} - 1,t - {d_i} - 1))^{\rm{T}}}{\rm{H}}_{{\rm{A}}{{\rm{d}}_j}}^{\rm{T}}\}  \\
 \;\;\;\;\;\;\;\;\;\;\;\;\; + {(\Phi _{{{\rm{x}}_j}}^{\rm{F}}(t - {d_j},{d_i},t))^{\rm{T}}}{\rm{H}}_{{d_j}}^{\rm{T}} \\
 \;\;\;\;\;\;\;\;\;\;\;\;\; + {(\Theta _{{\rm{x}}_j^{\rm{c}}}^{\rm{F}}(t - {d_j} - 1,{d_i},t))^{\rm{T}}}{\rm{H}}_{{\rm{A}}{{\rm{d}}_j}}^{\rm{T}} \\
 \;\;\;\;\;\;\;\;\;\;\;\;\; + {{\rm{C}}_{\rm{o}}}({d_i},{d_j}){A^{{d_j}}}{Q_w}{\rm{\bar H}}_{{\rm{A}}{{\rm{d}}_j}}^{\rm{T}} \\
 \end{array} \right.
\label {eq:71}
\end{eqnarray}
where ${\delta _{{d_i},{d_j}}}$ is determined by (\ref{eq:4}), and ${{\rm{C}}_{\rm{o}}}({d_i},{d_j})$ is determined by (\ref{eq:32}); ${{\rm{H}}_{{{\rm{d}}_i}}},{{{\rm{\bar H}}}_{{\rm{A}}{{\rm{d}}_i}}},{{\rm{H}}_{{\rm{A}}{{\rm{d}}_i}}}$ are defined by (\ref{eq:44}), while ${\tau _{ij}}$, ${\tau _{{d_i}}}$ and ${\tau _{{d_j}}}$ are defined in (\ref{eq:61}). ${{\hat \Upsilon }_{ij}}(t)$ is calculated by (\ref{eq:63}) (see Lemma 6), and ${\Psi _{ij}}(t)$ is calculated by (\ref{eq:58}) (see Lemma 5); $\Theta _{{\rm{x}}_i^{\rm{c}}}^w({t_1},{t_2}),\;\Theta _{{\rm{x}}_i^{\rm{c}}}^{\rm{F}}({t_1},{t_2})$ are calculated by (\ref{eq:42}--\ref{eq:43}) (see Lemma 3), while $\Phi _{{{\rm{x}}_i}}^{{{\rm{x}}_j}}({t_1},{t_2}),\Phi _{{{\rm{x}}_i}}^w({t_1},{t_2}),\Phi _{{{\rm{x}}_i}}^{\rm{F}}({t_1},{t_2})$ are calculated by (\ref{eq:34}--\ref{eq:36}) (see Lemma 2). Moreover, the relationship between the CSE ${\rm{\hat x}}_i^{{\rm{c}}}(t)$ and the DKFE ${{{\rm{\hat x}}}}(t)$ is
\begin{eqnarray}
{\rm{Tr}}\{P(t)\}  \le {\rm{Tr\{ }}{\Xi _{ii}}(t){\rm{\} (}}i \in {\rm{\{ 1,2,}} \cdots {\rm{,L\} )}}.
\label {eq:72}
\end{eqnarray}
\begin{proof}
See A.6 in Appendix.
\end{proof}

From Theorem 1, $\Xi (t)$ can be calculated by (\ref{eq:70}--\ref{eq:71}), then the optimal weighting matrices ${\Omega _1}(t), \cdots ,{\Omega _L}(t)$ are obtained by (\ref{eq:29}). Moreover, the computation procedures for the DKFE ${\rm{\hat x}}(t)$ can be summarized by Algorithm 1.

\begin{algorithm}
\caption{For the given selection probabilities $\pi _{{\hbar _i}}^i({\hbar _i} = 1, \cdots ,{\Delta _i};i = 1, \cdots ,L)$ satisfying (\ref{eq:20})}
\begin{algorithmic}[1]\label{algo:1}
\STATE At each sink node:
\FOR{$i := 1$ \TO $L$}
\STATE Calculate ${{\rm{G}}_{{{\rm K}_i}}}(t)$ and ${{\rm K}_i}(t)$ by (\ref{eq:6}) and (\ref{eq:7});
\STATE Calculate the LOE ${{{\rm{\hat x}}}_i}(t)$ by (\ref{eq:5});
\STATE Generate the binary variables $\sigma _{{\hbar _i}}^i(t)({\hbar _i} = 1, \cdots ,{\Delta _i})$ satisfying the categorical distribution, then determine the selected ASC ${{{\rm{\hat x}}}_{{s_i}}}(t)$ by (\ref{eq:13});
\ENDFOR
\STATE At the FC:
\FOR{$i := 1$ \TO $L$}
\STATE Calculate ${{\rm{G}}_{{{\rm K}_i}}}(t)$ by (\ref{eq:6});
\STATE Calculate the CSE ${\rm{\hat x}}_i^{\rm{c}}(t)$ by (\ref{eq:15});
\FOR{$j := i$ \TO $L$}
\STATE Calculate ${P_{ij}}(t)$ by (\ref{eq:6}--\ref{eq:8});
\STATE Calculate ${\Gamma _{ij}}(t)$,${\Psi _{ij}}(t)$,${\Upsilon _{ij}}(t)$ by (\ref{eq:51}), (\ref{eq:58}), (\ref{eq:63});
\STATE Calculate ${\Xi _{ij}}(t)$ by (\ref{eq:70}--\ref{eq:71});
\ENDFOR
\ENDFOR
\STATE Calculate $\Omega _1(t),\Omega _2(t), \cdots ,\Omega _L(t)$ by (\ref{eq:29});
\STATE Calculate the DKFE ${\rm{\hat x}}(t)$ by (\ref{eq:26}).
\end{algorithmic}
\end{algorithm}

\textbf{Remark 6.} According to (\ref{eq:70}--\ref{eq:71}), each covariance matrix ${\Xi _{ij}}(t)$ is independent of the measurement $y_i(t)$ and the LOE ${{{\rm{\hat x}}}_i}(t)$. Thus, ${\Xi _{ij}}(t)$ can be calculated at the FC when the initial values are given. In this case, only if each selected ASC ${{{\rm{\hat x}}}_{{s_i}}}(t) \in {{\rm{R}}^{{r_i}}}$ is sent to the FC, Algorithm 1 will be implemented in practical applications. On the other hand, when the communication delays and the number of local estimates increase slightly, the computational complexity of Algorithm 1 will be high. In this case, the steady-sate DKFE with time-invariant weighting matrices can reduce the amount of computation. Therefore, to obtain the steady-sate DKFE, we should find the stability conditions satisfying the following two points: i) The covariance matrix of the recursive DKFE converges to a positive-definite matrix; ii) The limit of the covariance matrix $P(t)$ is independent of the initial values. Following this idea, the steady-state DKFE will be derived in the next section.

\vspace{-8pt}
\section{Stability Analysis for the DKFE}
\subsection{Stability Condition of Each Local CSE}
The estimation performance of each CSE ${\rm{\hat x}}_i^{\rm{c}}(t)$ will be discussed in this subsection. First, it is considered that the $i{\rm{th}}$ subsystem satisfies
\begin{eqnarray}
(A,\sqrt {{Q_w}} )\;\;{\rm{is}}\;{\rm{stabilizable\;and}}\;{\rm{ (}}A,{C_i}{\rm{)\;is\;detectable}}.
\label {eq:78}
\end{eqnarray}
When the condition (\ref{eq:78}) holds, it is well known that the estimation error covariance matrix ${P_{ii}}(t)$ in (\ref{eq:7}) will converge from any initial conditions ${P_{ii}}(0)>0$ to the unique positive semi-definite solution $P_{ii}$. This means that
\begin{eqnarray}
\mathop {\lim }\limits_{t \to \infty } {P_{ii}}(t) = {P_{ii}},\mathop {\lim }\limits_{t \to \infty } {\Phi _{{{\rm K}_i}}}(t) = {\Phi _{{{\rm K}_i}}},\mathop {\lim }\limits_{t \to \infty } {{\rm K}_i}(t) = {{\rm K}_i}
\label {eq:79}\;\;\;\;
\end{eqnarray}
where the limits ${P_{ii}}$, ${\Phi _{{{\rm K}_i}}}$ and ${{\rm K}_i}$ are independent of the initial values. Moreover, ${\Phi _{{{\rm K}_i}}}$ is a stable matrix. Thus, there must exist an integer ${N_{{P_i}}} > 0$ such that, for $t>{N_{{P_i}}}$, the estimation error system (\ref{eq:38}) reduces to:
\begin{eqnarray}
{{{\rm{\tilde x}}}_i}(t) = {\Phi _{{{\rm K}_i}}}{{{\rm{\tilde x}}}_i}(t - 1) + {{\rm{G}}_{{{\rm K}_i}}}w(t - 1) - {{\rm K}_i}{v_i}(t).
\label {eq:80}
\end{eqnarray}
Then, it follows from (\ref{eq:80}) that
\begin{eqnarray}
{{{\rm{\tilde x}}}_i}(t + 1) = \Phi _{{{\rm K}_i}}^{{d_i} + 1}{{{\rm{\tilde x}}}_i}(t - {d_i}) + \zeta _i^{\rm{o}}(t),
\label {eq:81}
\end{eqnarray}
where
\begin{eqnarray}
\begin{array}{l}
 \zeta _i^{\rm{o}}(t) = \sum\nolimits_{{\alpha _i} = 1}^{{d_i} + 1} {\Phi _{{{\rm K}_i}}^{{\alpha _i} - 1}{{\rm{G}}_{{{\rm K}_i}}}w(t - {\alpha _i} + 1)}  \\
 \;\;\;\;\;\;\;\;\; - \sum\nolimits_{{\alpha _i} = 0}^{{d_i}} {\Phi _{{{\rm K}_i}}^{{\alpha _i} - 1}{{\rm{G}}_{{{\rm K}_i}}}{{\rm K}_i}{v_i}(t - {\alpha _i} + 1)}.  \\
 \end{array}
\label {eq:82}
\end{eqnarray}
Meanwhile, it is derived from (\ref{eq:23}) and (\ref{eq:80}) that
\begin{eqnarray}
{\rm{\tilde x}}_i^{\rm{c}}(t + 1) = {{\rm{A}}_{i1}}(t){\rm{\tilde x}}_i^{\rm{c}}(t - {d_i}) + {{\rm{A}}_{i2}}(t){{{\rm{\tilde x}}}_i}(t - {d_i}) + \zeta _i^{\rm{c}}(t)
\label {eq:83}\;\;\;\;
\end{eqnarray}
where
\begin{eqnarray}
\left\{ \begin{array}{l}
 {{\rm{A}}_{i1}}(t) = {A^{{d_i}}}[{I_n} - {H_i}(t - {d_i} + 1)]A \\
 {{\rm{A}}_{i2}}(t) = {A^{{d_i}}}{H_i}(t - {d_i} + 1){\Phi _{{{\rm K}_i}}} \\
 \zeta _i^{\rm{c}}(t) = {A^{{d_i}}}{H_i}(t - {d_i} + 1){{\rm{G}}_{{{\rm K}_i}}}w(t - {d_i}) \\
 \;\;\;\;\;\;\; + {A^{{d_i}}}{H_i}(t - {d_i} + 1)w(t - {d_i}) + {{\rm{F}}_w}({d_i},t + 1) \\
 \;\;\;\;\;\;\; - {A^{{d_i}}}{H_i}(t - {d_i} + 1){v_i}(t - {d_i} + 1) \\
 \end{array}. \right.
\label {eq:84}
\end{eqnarray}
Define ${{\tilde \xi }_i}(t) \buildrel \Delta \over = {\rm{col}}\{ {\rm{\tilde x}}_i^{\rm{c}}(t), {{{\rm{\tilde x}}}_i}(t)\} $, ${\zeta _i}(t) \buildrel \Delta \over = {\rm{col\{ }}\zeta _i^{\rm{c}}(t),\zeta _i^{\rm{o}}(t){\rm{\} }}$. Then, combining (\ref{eq:81}) and (\ref{eq:83}) yields that
\begin{eqnarray}
{{\tilde \xi }_i}(t + 1) = {{\rm{A}}_i}(t){{\tilde \xi }_i}(t - {d_i}) + {\zeta _i}(t),
\label {eq:85}
\end{eqnarray}
where
\begin{eqnarray}
{{\rm{A}}_i}(t) = \left[ {\begin{array}{*{20}{c}}
   {{{\rm{A}}_{i1}}(t)} & {{{\rm{A}}_{i2}}(t)}  \\
   0 & {\Phi _{{{\rm K}_i}}^{{d_i} + 1}}  \\
\end{array}} \right].
\label {eq:86}
\end{eqnarray}
It is concluded from the definition of ${{\tilde \xi }_i}(t)$ that if the covariance matrix ${\tilde \Xi _{{\xi _i}}}(t) \buildrel \Delta \over = {\rm E}\{ {{\tilde \xi }_i}(t){[{{\tilde \xi }_i}(t)]^{\rm{T}}}\} $ converges to the unique matrix ${\tilde \Xi _{{\xi _i}}}$, there must exist the unique limit of the estimation error covariance matrix ${\Xi _{ii}}(t)$ for the $i{\rm{th}}$ CSE.

\textbf{Lemma 7} Define
\begin{eqnarray}
f(B) \buildrel \Delta \over = {\rm E}\{ {\rm{A}}_i^{\rm{T}}(t)B{{\rm{A}}_i}(t)\},
\label {eq:87}
\end{eqnarray}
where $B = \left[ {\begin{array}{*{20}{c}}
   {{B_{11}}} & {{B_{12}}}  \\
   {{B_{21}}} & {{B_{22}}}  \\
\end{array}} \right]$. Then, $f(B)$ is calculated by:
\begin{eqnarray}
f(B) = \left[ {\begin{array}{*{20}{c}}
   {{{{\rm{\tilde B}}}_{11}}} & {{{{\rm{\tilde B}}}_{12}}}  \\
   {{{{\rm{\tilde B}}}_{21}}} & {{{{\rm{\tilde B}}}_{22}}}  \\
\end{array}} \right],
\label {eq:88}
\end{eqnarray}
where
\begin{eqnarray}
\left\{ \begin{array}{l}
 {{{\rm{\tilde B}}}_{11}} = {A^{\rm{T}}}{\rm{\{ }}{{\rm{W}}_i} \otimes [{({A^{{d_i}}})^{\rm{T}}}{B_{11}}{A^{{d_i}}}]\} A \\
 {{{\rm{\tilde B}}}_{12}} = {A^T}\{ {\rm{V}}_i^{\rm{T}} \otimes [{({A^{{d_i}}})^{\rm{T}}}{B_{11}}{A^{{d_i}}}]\} {\Phi _{{{\rm K}_i}}} \\
 \;\;\;\;\;\;\;\; + {A^{\rm{T}}}[{I_n} - {H_i}]{({A^{{d_i}}})^{\rm{T}}}{B_{12}}\Phi _{{{\rm K}_i}}^{{d_i} + 1} \\
 {{{\rm{\tilde B}}}_{21}} = \Phi _{{{\rm K}_i}}^{\rm{T}}{\rm{\{ }}{{\rm{V}}_i} \otimes [{({A^{{d_i}}})^{\rm{T}}}{B_{11}}{A^{{d_i}}}]\} A \\
 \;\;\;\;\;\;\;\; + {[\Phi _{{{\rm K}_i}}^{{d_i} + 1}]^{\rm{T}}}{B_{21}}{A^{{d_i}}}[{I_n} - {H_i}]A \\
 {{{\rm{\tilde B}}}_{22}} = \Phi _{{{\rm K}_i}}^{\rm{T}}\{ {\Lambda _i} \otimes [{({A^{{d_i}}})^{\rm{T}}}{B_{11}}{A^{{d_i}}}]\} {\Phi _{{{\rm K}_i}}} \\
 \;\;\;\;\;\;\;\; + {[\Phi _{{{\rm K}_i}}^{{d_i} + 1}]^{\rm{T}}}{B_{21}}{A^{{d_i}}}{H_i}{\Phi _{{{\rm K}_i}}} + {[\Phi _{{{\rm K}_i}}^{{d_i} + 1}]^{\rm{T}}}{B_{22}}\Phi _{{{\rm K}_i}}^{{d_i} + 1} \\
 \;\;\;\;\;\;\;\; + \Phi _{{{\rm K}_i}}^{\rm{T}}{H_i}{({A^{{d_i}}})^{\rm{T}}}{B_{12}}\Phi _{{{\rm K}_i}}^{{d_i} + 1} \\
 \end{array} \right.
\label {eq:89}
\end{eqnarray}
where the probability selection matrix $H_i$ is given by (\ref{eq:21}), and ${{\rm{W}}_i}$, ${\Lambda _i}$, ${{\rm{V}}_i}$ are given by (\ref{eq:69}). Moreover, for any matrices ${B_1}$,${B_2}$, there will be:
\begin{eqnarray}
f({B_1} + {B_2}) = f({B_1}) + f({B_2}).
\label {eq:90}
\end{eqnarray}
\begin{proof}
(\ref{eq:88}) can be obtained from (\ref{eq:16}), (\ref{eq:19}), Lemma 1 and the definition of $f(B)$, while (\ref{eq:90}) is derived from (\ref{eq:88}). This completes the proof.
\end{proof}

Based on Lemma 7, the delay-dependent stability condition of the CSE ${\rm{\hat x}}_i^{\rm{c}}(t)$ will be given in Theorem 2.

\textbf{Theorem 2} For the communication delay $d_i$ and selection probabilities $\pi _{{\hbar _i}}^i({\hbar _{i}} = 1, \cdots ,{\Delta _i})$ in (\ref{eq:88}), if there exist $D_i>0$, ${{X_i}}$, $Y_i$, $Z_i$ and $S_i>0$ such that
\begin{eqnarray}
\left[ {\begin{array}{*{20}{c}}
   {{X_i}} & {{Y_i}}  \\
   {Y_i^{\rm{T}}} & {{Z_i}}  \\
\end{array}} \right] \ge 0,
\label {eq:91}\;\;\;\;\;\;\;\;\;\;\;\;\;\;\;\;\;\;\;\;\;\;\;\;\;\;\;\;\;\;\\
{{\rm M}_i} = \left[ {\begin{array}{*{20}{c}}
   {{{\rm M}_i}(1,1)} & { - {Y_i} - {d_i}{Z_i}{{\rm{A}}_i}}  \\
   { - Y_i^{\rm{T}} -{d_i}{\rm{A}}_i^{\rm{T}}Z_i} & {f({D_i}) + {d_i}f({Z_i}) - {S_i}}  \\
\end{array}} \right] < 0,
\label {eq:92}\;\;
\end{eqnarray}
where ${{\rm M}_i}(1,1) =  - {D_i} + {X_i} + Y_i^{\rm{T}} + {Y_i} + {d_i}{Z_i} + {S_i}$ and ${{\rm{A}}_i} = {\rm E}\{ {{\rm{A}}_i}(t)\} $, while ${f({D_i})}$ and ${f({Z_i})}$ are calculated by (\ref{eq:88}) in Lemma 7, then the covariance matrix ${\Xi _{ii}}(t)$ (\ref{eq:70}) converges to the unique matrix, i.e.,
\begin{eqnarray}
\mathop {\lim }\limits_{t \to \infty } {\Xi _{ii}}(t) = {\Xi _{ii}},
\label {eq:93}
\end{eqnarray}
and the limit ${\Xi _{ii}}$ is independent of the initial values.

\begin{proof}
See A.7 in Appendix.
\end{proof}

\textbf{Remark 7}. When ${d_i} \ne 0$, it is calculated from (\ref{eq:69}) and (\ref{eq:70}) that ${\rm{Tr}}\{ {A^{{d_i}}}{\rm{[}}{{\rm{V}}_i} \otimes ({\Phi _{{{\rm K}_i}}}(t - {d_i}){\Gamma _{ii}}(t - {d_i} - 1){A^{\rm{T}}} + {{\rm{G}}_{{{\rm K}_i}}}(t - {d_i}){Q_w})]{({A^{{d_i}}})^{\rm{T}}}\}  \ne 0$ and ${\rm{Tr\{ }}{A^{{d_i}}}[{\rm{V}}_i^{\rm{T}} \otimes (A\Gamma _{ii}^{\rm{T}}(t - {d_i} - 1)\Phi _{{{\rm K}_i}}^{\rm{T}}(t - {d_i}) + {Q_w}{\rm{G}}_{{{\rm K}_i}}^{\rm{T}}(t - {d_i}))]{({A^{{d_i}}})^{\rm{T}}}\}  \ne 0$, which are different from the results (101) and (102) in \cite{c32}. This implies that the stability condition for each CSE is difficult to be obtained by the derivation method based on the property of the operator ${\rm{Tr}}\{  \bullet \}$ in \cite{c32}. In contrast, by adopting a novel derivation idea in this paper, the stability conditions (\ref{eq:91}) and (\ref{eq:92}) are linear matrix inequalities (LMIs), and thus they can be verified by resorting to Matlab LMI Toolbox \cite{c42}. On the other hand, according to the similar derivation of stability conditions in \cite{o24}, if the following conditions are satisfied
\begin{eqnarray}
\left\{ \begin{array}{l}
 {\zeta _{\infty ,i}} \buildrel \Delta \over = ||{A^{{d_i}}}|{|_1}||{({I_n} - {H_i})^{{\textstyle{1 \over 2}}}}A|{|_1} \\
 \;\;\;\;\;\;\;\;\;\;\; \times ||{A^{{d_i}}}|{|_\infty }||{({I_n} - {H_i})^{{\textstyle{1 \over 2}}}}A|{|_\infty } < 1 \\
 {\zeta _{2,i}} \buildrel \Delta \over = ||\Phi _{{{\rm K}_i}}^{{d_i} + 1}|{|_2}||{A^{{d_i}}}({I_n} - {H_i})A|{|_2} < 1 \\
 \end{array} \right.
\label {eq:c105}
\end{eqnarray}
where ${\Phi _{{{\rm K}_i}}} \buildrel \Delta \over = \mathop {\lim }\limits_{t \to \infty } {\Phi _{{{\rm K}_i}}}(t)$, while $|| \bullet |{|_1}$ and $|| \bullet |{|_\infty }$ represent the 1-norm and $\infty  - {\rm{norm}}$ of matrices, respectively.Then, the covariance matrix ${\Xi _{ii}}(t)$ (\ref{eq:70}) will converge. However, the stability conditions in Theorem 2 are derived from the perspective of approaching the spectral radius when using Laypunov stability theory, while the condition (\ref{eq:c105}) is directly given by using the relaxation technique of matrix norms. Notice that, for any matrix ``${\rm o}$'', there must be $\rho ({\rm o}) \le ||{\rm o}|{|_i}(i \in \{ 1,2,\infty \} )$, where $\rho ( \bullet )$ denotes the spectral radius. In this sense, the stability conditions in Theorem 2 has less conservative than (\ref{eq:c105}) derived by \cite{o24}. Moreover, according to the definitions of 1-norm and $\infty  - {\rm{norm}}$, the condition (\ref{eq:c105}) will become more conservatism because of the product of $|| \bullet |{|_1}$ and $|| \bullet |{|_\infty }$. The above-mentioned result has been illustrated by Examples 1-2.

\textbf{Remark 8.} When $d_i=0$, the Lyapunov function candidate can be chosen as ${V_{{\xi _i}}}(t) = {\rm E}\{ \xi _i^{\rm{T}}(t){D_i}{\xi _i}(t)\}$. Then it is concluded from the similar derivation of Theorem 2 that if there exists ${D_i} > 0$ such that
\begin{eqnarray}
\hat f({D_i}) - {D_i} < 0,
\label {eq:a105}
\end{eqnarray}
where $\hat f({D_i})$ is calculated by $f({D_i})$ (i.e., (\ref{eq:88}) in Lemma 7) for $d_i=0$, then the corresponding covariance matrix ${\Xi _{ii}}(t)$ will converge to the unique steady-state value. Meanwhile, it is concluded from the result in \cite{c32} that if
\begin{eqnarray}
{\lambda _{\max }}({A^{\rm{T}}}({I_n} - {H_i})A) < 1,
\label {eq:b105}
\end{eqnarray}
where $H_i$ is determined by (\ref{eq:21}), then the $\mathop {{\rm{lim}}}\limits_{t \to \infty } {\rm{Tr}}\{ {\Xi _{ii}}(t)\} $ will exist for $d_i=0$. It should be pointed out that, under the condition (\ref{eq:b105}), one cannot prove the following results: a) the limit of the covariance matrix ${\Xi _{ii}}(t)$ exists, and b) the limit of ${\Xi _{ii}}(t)$ is independent of the initial conditions. Since the results (a) and (b) are necessary before deriving the steady-sate DKFE, the steady-state fusion estimator cannot be given under the condition (\ref{eq:b105}). Moreover, when only considering the convergence of the sequence ${\rm{\{ Tr}}\{ {\Xi _{ii}}(t)\} \}$, the condition (\ref{eq:a105}) has less conservatism than the condition (\ref{eq:b105}). This is because the relaxation technique of matrix trace inequality is introduced to derive (\ref{eq:b105}), but the condition (\ref{eq:a105}) is derived by the stability theory without any relaxation. This result has been demonstrated by Example 1.

\subsection{Steady-State DKFE Design}
According to (\ref{eq:30}) and (\ref{eq:31}), the stability of the DKFE is also dependent on each estimation error cross-covariance matrix ${\Xi _{ij}}(t)$ (see (\ref{eq:71})). Thus, the convergence of the sequence $\{ {\Xi _{ij}}(t)\}$ will first be discussed in this subsection, and then combining Theorem 2 leads to the delay-dependent stability condition of the DKFE and the steady-state DKFE (SDKFE). The main result will be presented in Theorem 3.

\textbf{Theorem 3} Consider the CPSs (\ref{eq:1}--\ref{eq:2}) under the condition (\ref{eq:78}), if the selection probabilities $\pi _{{\hbar _i}}^i({\hbar _i} = 1,2, \cdots ,{\Delta _i},\;i = 1,2, \cdots ,L)$ and the communication delays ${d_i}(i = 1,2, \cdots ,L)$ satisfy (\ref{eq:91}), (\ref{eq:92}), and
\begin{eqnarray}
\rho ({A^{{d_i}}}({I_n} - {H_i})A) < 1 (i = 1,2, \cdots ,L),
\label {eq:106}
\end{eqnarray}
where $H_i$ is given by (\ref{eq:21}), then the fusion estimation error covariance matrix $P(t)$ (\ref{eq:30}) will converge to the unique matrix $P$, i.e.,
\begin{eqnarray}
\mathop {\lim }\limits_{t \to \infty } P(t) = P
\label {eq:107}
\end{eqnarray}
with $P$ independent of the initial values. Moreover, the steady-state weighing matrices ${\Omega _i}(i = 1,2, \cdots ,L)$ are calculated by
\begin{eqnarray}
[{\Omega _1},{\Omega _2}, \cdots ,{\Omega _L}] = {P^{ - 1}}I_a^{\rm{T}}{\Xi ^{ - 1}},
\label {eq:108}
\end{eqnarray}
where $\mathop {\lim }\limits_{t \to \infty } \Xi (t) = \Xi$, and the limit $\Xi$ is independent of the initial values. In this case, the SDKFE ${{{\rm{\hat x}}}_s}(t)$ at the FC side is given by:
\begin{eqnarray}
{{{\rm{\hat x}}}_s}(t) = \sum\nolimits_{i = 1}^L {{\Omega _i}{\rm{\hat x}}_i^{\rm{c}}(t)},
\label {eq:109}
\end{eqnarray}
where ${{\rm{\hat x}}_i^{\rm{c}}(t)}$ is calculated by (\ref{eq:15}).

\begin{proof}
See A.8 in Appendix.
\end{proof}

According to Theorem 3, the computation procedures for the SDKFE ${{{\rm{\hat x}}}_s}(t)$ can be summarized by Algorithm 2.

\begin{algorithm}
\caption{For the given selection probabilities $\pi _{{\hbar _i}}^i({\hbar _i} = 1, \cdots ,{\Delta _i};i = 1, \cdots ,L)$ satisfying (\ref{eq:20})}
\begin{algorithmic}[1]\label{algo:2}
\STATE Determine the weighting matrices ${\Omega _i}(i = 1, \cdots ,L)$ by (\ref{eq:108});
\STATE At each sink node:
\FOR{$i := 1$ \TO $L$}
\STATE Calculate the LOE ${{{\rm{\hat x}}}_i}(t)$ by (\ref{eq:5});
\STATE Generate the binary variables $\sigma _{{\hbar _i}}^i(t)({\hbar _i} = 1, \cdots ,{\Delta _i})$ satisfying the categorical distribution, then determine the selected ASC ${{{\rm{\hat x}}}_{{s_i}}}(t)$ by (\ref{eq:13});
\ENDFOR
\STATE At the FC:
\FOR{$i := 1$ \TO $L$}
\STATE Calculate the CSE ${\rm{\hat x}}_i^{\rm{c}}(t)$ by (\ref{eq:15});
\ENDFOR
\STATE Calculate the SDKFE ${{{\rm{\hat x}}}_s}(t)$ by (\ref{eq:109}).
\end{algorithmic}
\end{algorithm}

\textbf{Remark 9}. It has been proved in Theorem 3 that when the conditions (\ref{eq:91}), (\ref{eq:92}) and (\ref{eq:106}) hold, the estimation error covariance matrix $P(t)$ can converge to the unique steady-state values for any initial conditions. Thus, the steady-state weighting matrices (\ref{eq:108}) can be obtained off-line by implementing the Step 7--Step 16 of Algorithm 1. It is noted that the computational complexity of the SDKFE obtained by Algorithm 2 is much lower than that of the DKFE obtained by Algorithm 1.

Notice that the stability condition in Theorems 2--3 are dependent on the communication delays and the selection probabilities of dimensionality reduction. Since each communication delay is determined by the property of communication channel, it is difficult to adjust the parameter $d_i$ to satisfy the stability condition. In this case, from the result (\ref{eq:72}) in Theorem 1 and Theorems 2--3, how to determine the selection probabilities (\ref{eq:20}) such that the MSE of the DKFE is bounded or convergent will be presented in Theorem 4.

\textbf{Theorem 4} For the CPSs (\ref{eq:1}--\ref{eq:2}), when each communication delay $d_i$ is known in prior, two probability criteria to determine the dimensionality reduction strategy are presented as follows:

\textbf{(C.1):} To guarantee that the MSE of the DKFE is bounded, one needs to determine one group of the selection probability ${\varsigma _i} \buildrel \Delta \over = {\rm{col}}\{ \pi _1^i, \cdots ,\pi _{{\Delta _i}}^i\}$ in (\ref{eq:22}) by (\ref{eq:91}--\ref{eq:92}).

\textbf{(C.2):} To guarantee the existence of the SDKFE, one needs to determine the $L$ selection probabilities ${\varsigma _i}(i = 1, \cdots ,L)$ in (\ref{eq:22}) by (\ref{eq:91}--\ref{eq:92}) and (\ref{eq:106}).

%

\section{Numerical Examples}
In this section, two illustrative examples are presented to show the advantage and effectiveness of the proposed dimensionality reduction fusion estimation methods.

\textbf{Example 1:} Consider a CPS (\ref{eq:1}) with the following system parameters \cite{c25}:
\begin{eqnarray}
A = \left[ {\begin{array}{*{20}{c}}
   {1.25} & 0  \\
   1 & {1.1}  \\
\end{array}} \right],{Q_w} = \left[ {\begin{array}{*{20}{c}}
   {20} & 0  \\
   0 & {20}  \\
\end{array}} \right],
\label {eq:125}
\end{eqnarray}
where the parameters of the first measurement equation in (\ref{eq:2}) are taken as:
\begin{eqnarray}
{C_1} = [0\;\;\;1],{Q_{{v_1}}} = 2.5.
\label {eq:126}
\end{eqnarray}
Then, it is calculated from (\ref{eq:125}) and (\ref{eq:126}) that ${\rm{rank}}\{ [\sqrt {{Q_w}} \;A\sqrt {{Q_w}} ]\}  = 2$
and ${\rm{rank}}\{ {\rm{col\{ }}{C_1}{\rm{,}}{C_1}A{\rm{\} }}\}  = 2$, which means that (\ref{eq:78}) holds. In this case, one has by (\ref{eq:6}), (\ref{eq:7}) and (\ref{eq:79}) that
\begin{eqnarray}
{{\rm{G}}_{{{\rm K}_i}}} = \left[ {\begin{array}{*{20}{c}}
   {0.4760} & { - 0.8573}  \\
   {0.0314} & {0.0376}  \\
\end{array}} \right].
\label {eq:127}
\end{eqnarray}
According to the dimensionality reduction strategy, it is considered in this example that only one component of ${{{\rm{\hat x}}}_1}(t)$  is allowed to be transmitted to the FC at each time step. In this case, it is calculated from (\ref{eq:69}) that
\begin{eqnarray}
\left\{ \begin{array}{l}
 {\Lambda _1} = {\rm{diag}}\{ {\gamma _{11}},1 - {\gamma _{11}}\} ,{{\rm{W}}_1} = {\rm{diag}}\{ 1 - {\gamma _{11}},{\gamma _{11}}\}  \\
 {{\rm{V}}_1} = \left[ {\begin{array}{*{20}{c}}
   0 & {{\gamma _{11}}}  \\
   {1 - {\gamma _{11}}} & 0  \\
\end{array}} \right],{H_1} = {\rm{diag}}\{ {\gamma _{11}},1 - {\gamma _{11}}\}  \\
 \end{array} \right.
\label {eq:128}
\end{eqnarray}
where $0 \le {\gamma _{11}} \le 1$.

\begin{figure}[thpb]
\centering
\includegraphics[height=5.5cm, width=8.5cm]{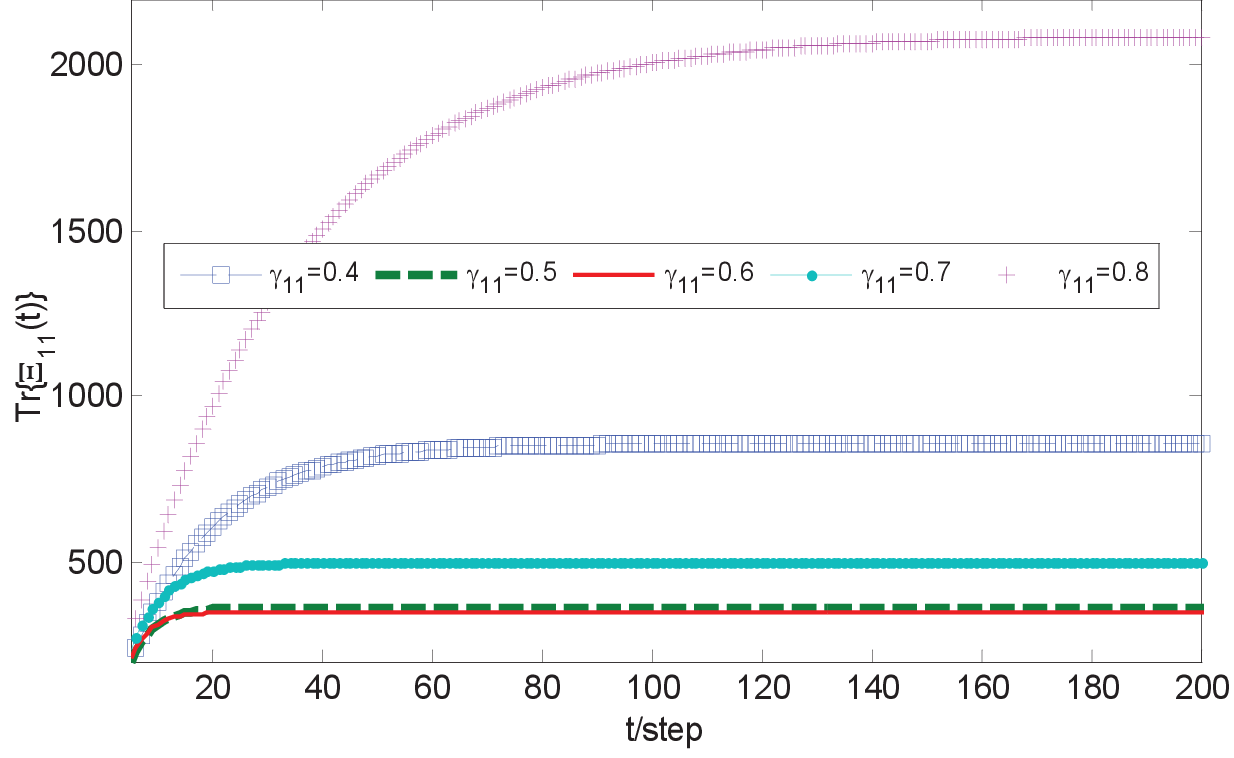}
\caption{The estimation performances of the CSE (i.e., ${\rm{Tr}}\{ {\Xi _{11}}(t)\} $) with different selection probabilities ${\gamma _{11}}$.}
\label{fig2}
\end{figure}

\begin{figure}[thpb]
\centering
\includegraphics[height=5.5cm, width=8.5cm]{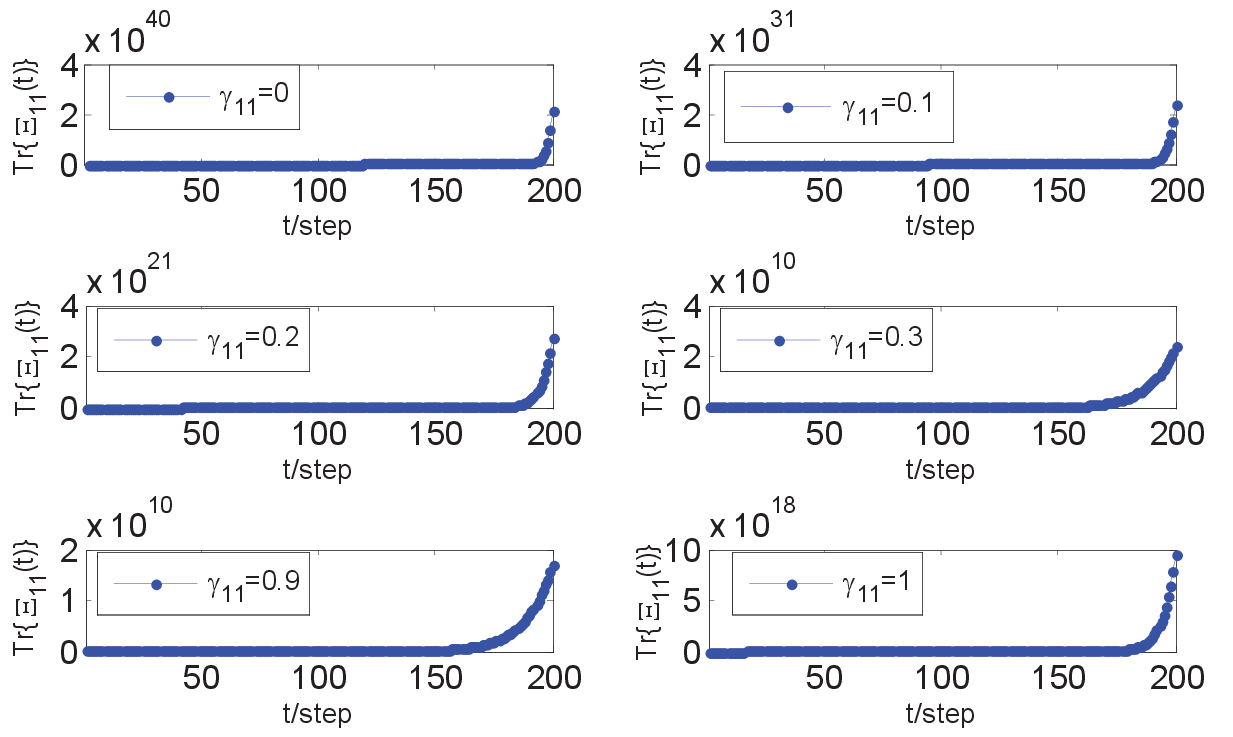}
\caption{The estimation performances of the CSE (i.e., ${\rm{Tr}}\{ {\Xi _{11}}(t)\} $) with different selection probabilities ${\gamma _{11}}$.}
\label{fig3}
\end{figure}

\emph{To demonstrate the advantage of the designed stability condition, it is assumed that there is no communication delay for this example}, and the selection probability ${\gamma _{11}}$ is taken as ${\gamma _{11}} = 0.5$. Then, by using LMI Toolbox in Matlab to solve the inequality matrix (\ref{eq:a105}), one has
\begin{eqnarray*}
{D_1} = \left[ {\begin{array}{*{20}{c}}
   {{\rm{1}}{\rm{.3284}}} & {{\rm{0}}{\rm{.1730}}} & { - 0.0731} & { - {\rm{0}}{\rm{.0395}}}  \\
   {{\rm{0}}{\rm{.1730}}} & {{\rm{0}}{\rm{.3727}}} & {{\rm{0}}{\rm{.0009}}} & { - {\rm{0}}{\rm{.0782}}}  \\
   { - {\rm{0}}{\rm{.0731}}} & {{\rm{0}}{\rm{.0009}}} & {{\rm{1}}{\rm{.0315}}} & { - {\rm{0}}{\rm{.5204}}}  \\
   { - {\rm{0}}{\rm{.0395}}} & { - {\rm{0}}{\rm{.0782}}} & { - {\rm{0}}{\rm{.5204}}} & {{\rm{1}}{\rm{.9647}}}  \\
\end{array}} \right],
\end{eqnarray*}
while it is calculated from (\ref{eq:125}) and (\ref{eq:128}) that
\begin{eqnarray*}
\left\{ \begin{array}{l}
 {\lambda _{\max }}({A^{\rm{T}}}({I_2} - {H_1})A) = 1.5887 > 1 \\
 {\xi _{\infty ,1}} = ||{({I_2} - {H_1})^{{\textstyle{1 \over 2}}}}A|{|_1}||{({I_2} - {H_1})^{{\textstyle{1 \over 2}}}}A|{|_\infty } = 2.3625 > 1 \\
 \end{array} \right.
\end{eqnarray*}
Therefore, it is concluded from Theorem 2 that the limit of ${\rm{Tr}}\{ {\Xi _{11}}(t)\}$ exists, however, the condition (\ref{eq:c105}) derived by \cite{o24} and the condition (\ref{eq:b105}) derived by \cite{c32} do not hold for this example. Moreover, when choosing $\gamma _{11}$ from 0 to 1, the effectiveness of the conditions (\ref{eq:a105}) and (\ref{eq:b105}) is shown in Table I, Fig.\ref{fig2} and Fig.\ref{fig3}. It can be seen from Fig.\ref{fig2} and Fig.\ref{fig3} that the sequence of ${\rm{Tr}}\{ {\Xi _{11}}(t)\} $ is convergent when ${\gamma _{11}} \in \{ 0.4,0.5,0.6,0.7,0.8\} $. This result can be directly obtained by the judgement condition (\ref{eq:a105}), however, the judgement condition (\ref{eq:c105}) derived by \cite{o24} and the judgment condition (\ref{eq:b105}) derived by \cite{c32} are invalid for the above cases. Therefore, the judgement condition (\ref{eq:a105}) in this paper has much less conservatism than the result in \cite{c32}, and thus is applicable to more fusion systems under the dimensionality reduction.
\begin{table}[htb]
\renewcommand{\arraystretch}{1.3}
\caption{Comparison of the result in Theorem 2 and the results in \cite{o24} and \cite{c32}}
\label{table_example}
\centering
\begin{tabular}{|c|c|c|c|}
\hline
${\gamma _{11}}$                 & $(75)$    &    ${\rm{(76)}\;derived\;by\;[32]}$ &${\rm{(74)}\;derived\;by\;[26]}$    \\
\hline
$0$     &  ${\rm{False}}$ & ${\rm{False}}$  & ${\rm{False}}$  \\
\hline
$0.1$     &  ${\rm{False}}$ &  ${\rm{False}}$  & ${\rm{False}}$  \\
\hline
$0.2$     &  ${\rm{False}}$ &  ${\rm{False}}$  & ${\rm{False}}$  \\
\hline
$0.3$     &  ${\rm{False}}$ &  ${\rm{False}}$  & ${\rm{False}}$  \\
\hline
$0.4$     &  ${\rm{\emph{True}}}$ &  ${\rm{False}}$  & ${\rm{False}}$  \\
\hline
$0.5$     &  ${\rm{\emph{True}}}$ &  ${\rm{False}}$  & ${\rm{False}}$  \\
\hline
$0.6$     &  ${\rm{\emph{True}}}$ &  ${\rm{False}}$  & ${\rm{False}}$  \\
\hline
$0.7$     &  ${\rm{\emph{True}}}$ &  ${\rm{False}}$  & ${\rm{False}}$  \\
\hline
$0.8$     &  ${\rm{\emph{True}}}$ &  ${\rm{False}}$ & ${\rm{False}}$  \\
\hline
$0.9$     &  ${\rm{False}}$ &  ${\rm{False}}$ & ${\rm{False}}$  \\
\hline
$1.0$     &  ${\rm{False}}$ &  ${\rm{False}}$ & ${\rm{False}}$  \\
\hline
\end{tabular}
\end{table}

\textbf{Example 2:} Consider a CPS described by power grid with a 4-bus model for the distribution test feeders \cite{c43}. To monitor the work status of the grid, two sink nodes collect their sensor measurements, and the local estimates computed by the sink nodes are transmitted to the FC (e.g., monitoring center or control center). According to the continuous-time smart grid system in \cite{c16}, and setting the sampling time $T_0={10^{ - 4}}{\rm{s}}$, the discretized system matrix in (\ref{eq:1}) is given by:
\begin{eqnarray}
A = \left[ {\begin{array}{*{20}{c}}
   {1.0156} & {0.0139} & {0.0457} & {0.0971}  \\
   { - 0.0353} & {0.9997} & { - 0.0008} & { - 0.0017}  \\
   { - 0.0526} & { - 0.0448} & {0.9625} & { - 0.0797}  \\
   { - 0.008} & { - 0.0505} & { - 0.0903} & {0.9011}  \\
\end{array}} \right],
\label {eq:129}
\end{eqnarray}
where ${\lambda _{\max }}(A) = 1.0441 > 1$ means that this 4-bus smart grid system is unstable, and the covariance of the process noise is taken as:
\begin{eqnarray}
{Q_w} = \left[ {\begin{array}{*{20}{c}}
   {0.04} & {0.1} & {0.06} & {0.08}  \\
   {0.1} & {0.25} & {0.15} & {0.2}  \\
   {0.06} & {0.15} & {0.09} & {0.12}  \\
   {0.08} & {0.2} & {0.12} & {0.16}  \\
\end{array}} \right].
\label {eq:130}
\end{eqnarray}
Then, the measurement matrices in (\ref{eq:2}) are given by
\begin{eqnarray}
 {C_1} = \left[ {\begin{array}{*{20}{c}}
   1 & 0 & 1 & 0  \\
   0 & 1 & 0 & 0  \\
   0 & 1 & 1 & 0  \\
   1 & 0 & 1 & 0  \\
\end{array}} \right],{C_2} = \left[ {\begin{array}{*{20}{c}}
   1 & 0 & 0 & 1  \\
   1 & 0 & 1 & 0  \\
   0 & 0 & 0 & 1  \\
   1 & 0 & 1 & 0  \\
\end{array}} \right],
\label {eq:131}
\end{eqnarray}
which means that the measurement information on the fourth component of ``x(t)'' cannot be obtained by the first sink node, while the measurement information on the second component of~``x(t)''~cannot be obtained by the second sink node. The covariances of $v_i(t)(i=1,2)$ in (\ref{eq:2}) are taken as ${Q_{{v_1}}} = {\rm{diag}}\{ 0.9,0.6,0.9,0.4\}$ and ${Q_{{v_2}}} = {\rm{diag}}\{ 0.3,0.4,0.5,0.2\}$, respectively. Then it is calculated from (\ref{eq:129}--\ref{eq:131}) that ${\rm{rank}}({\rm{col}}\{ {C_i},{C_i}A,{C_i}{A^2},{C_i}{A^3}\} ) = 4\;(i=1,2)$ and ${\rm{rank}}([\sqrt {{Q_w}} ,A\sqrt {{Q_w}} ,{A^2}\sqrt {{Q_w}} ,{A^3}\sqrt {{Q_w}} ]) = 4$, which means that the condition (\ref{eq:78}) holds. Thus, the limits in (\ref{eq:79}) exist, and one has
\begin{eqnarray}
\left\{ \begin{array}{l}
 {\Phi _{{{\rm K}_1}}} = \left[ {\begin{array}{*{20}{c}}
   {0.7915} & { - 0.0778} & { - 0.1333} & {0.0886}  \\
   { - 0.3288} & {0.6887} & { - 0.4510} & {0.0035}  \\
   { - 0.1573} & { - 0.2237} & {0.6459} & { - 0.0659}  \\
   { - 0.2790} & { - 0.2490} & { - 0.4294} & {0.9001}  \\
\end{array}} \right] \\
 {\Phi _{{{\rm K}_2}}} = \left[ {\begin{array}{*{20}{c}}
   {0.7324} & {0.025} & { - 0.0683} & { - 0.0919}  \\
   { - 0.6848} & {1.0244} & { - 0.2209} & { - 0.4579}  \\
   { - 0.4547} & { - 0.0287} & {0.5831} & { - 0.1738}  \\
   { - 0.5093} & { - 0.0291} & { - 0.3937} & {0.6284}  \\
\end{array}} \right] \\
 \end{array}. \right.
\label {eq:132}
\end{eqnarray}
\begin{figure}[thpb]
\centering
\includegraphics[height=5.5cm, width=8.5cm]{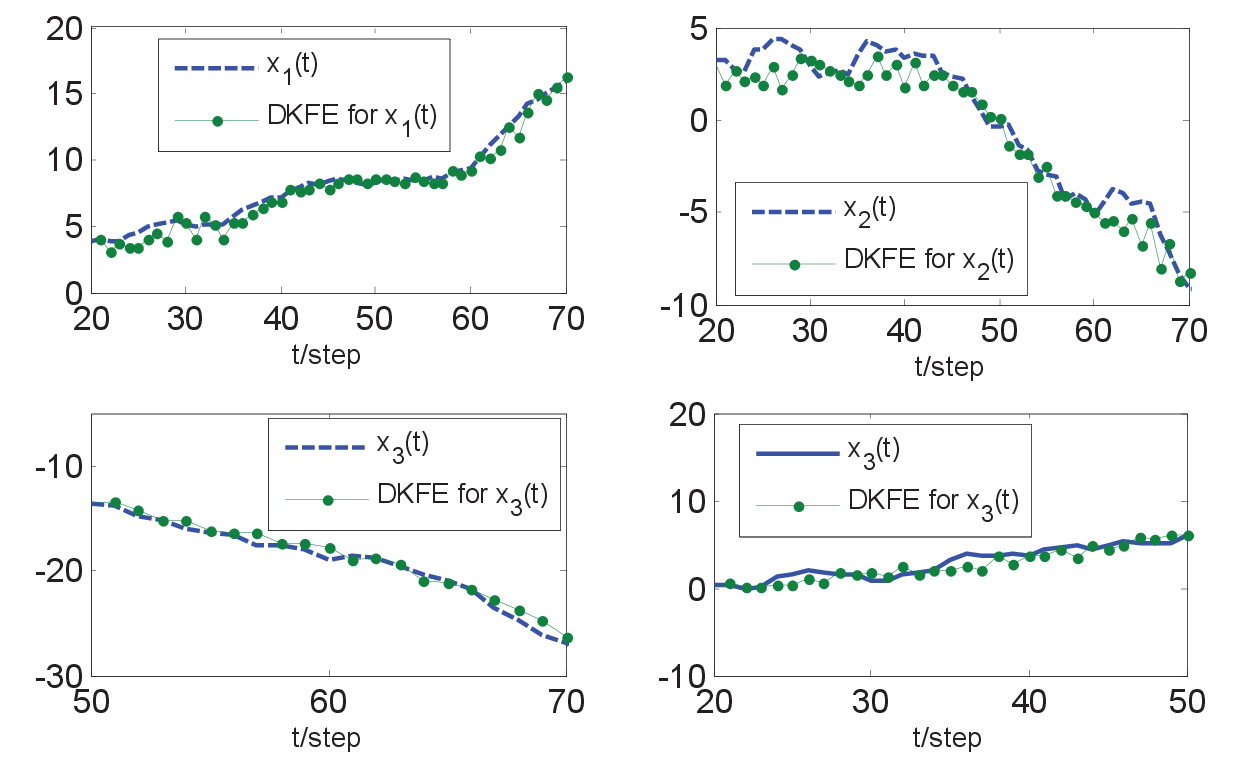}
\caption{The trajectories of the DKFE ${\rm{\hat x}}(t)$ and the state ``x(t)''.}
\label{fig4}
\end{figure}
\begin{figure}[thpb]
\centering
\includegraphics[height=5.5cm, width=8.0cm]{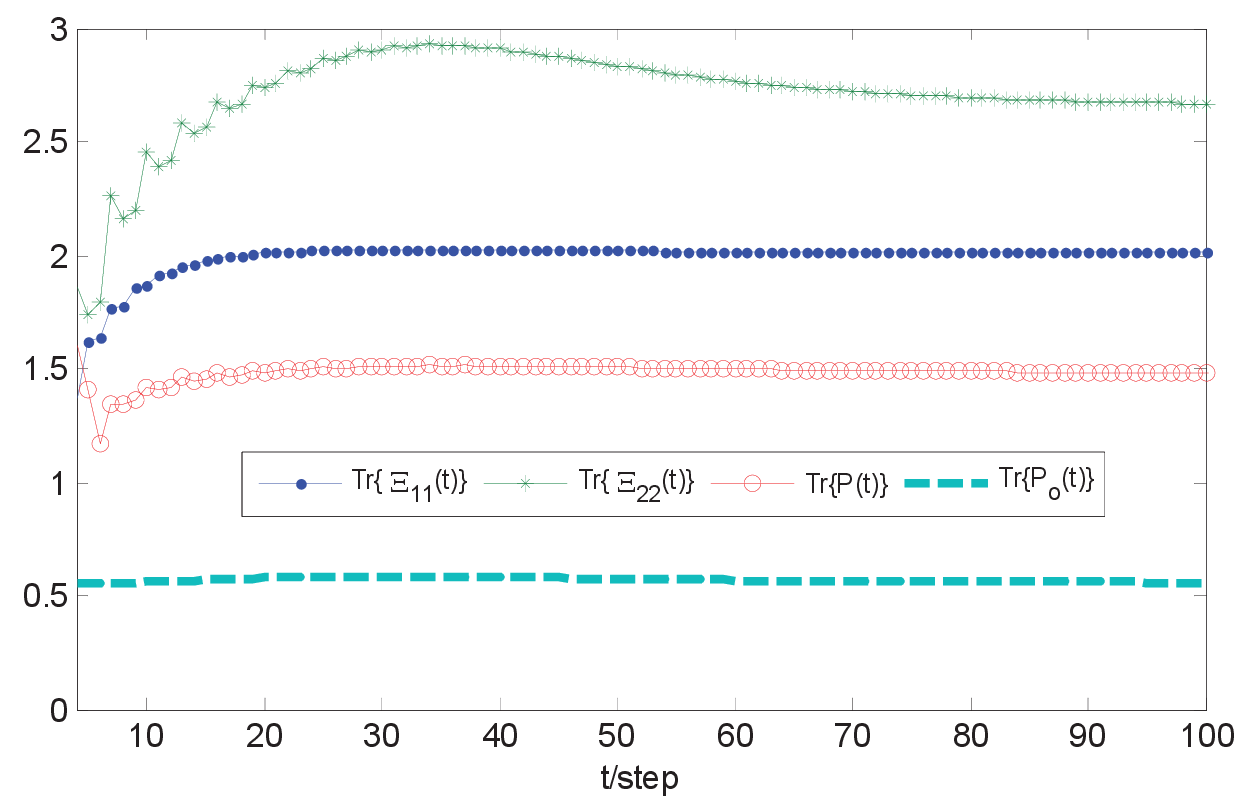}
\caption{Comparison of estimation performance for the CSEs, DKFE and ODKFE.}
\label{fig5}
\end{figure}

For this example, according to the dimensionality reduction strategy, it is considered that only two components of ${{\hat x}_i}(t)$ are allowed to be transmitted to the FC for satisfying the finite bandwidth, and thus ${r_1} = {r_2} = 2$ and ${\Delta _1} = {\Delta _2} = 6$. In this case, the diagonal matrices $H_{{\hbar _i}}^i(t)(i = 1,2;{\hbar _i} = 1,2,3,4,5,6)$ in (\ref{eq:16}) are given by
\begin{eqnarray}
\left\{ \begin{array}{l}
 H_1^i = {\rm{diag}}\{ 1,1,0,0\} ,H_2^i = {\rm{diag}}\{ 1,0,1,0\}  \\
 H_3^i = {\rm{diag}}\{ 1,0,0,1\} ,H_4^i = {\rm{diag}}\{ 0,1,1,0\}  \\
 H_5^i = {\rm{diag}}\{ 0,1,0,1\} ,H_6^i = {\rm{diag}}\{ 0,0,1,1\}  \\
 \end{array}. \right.
\label {eq:133}
\end{eqnarray}
Then it follows from (\ref{eq:16}) and (\ref{eq:133}) that
\begin{eqnarray}
\begin{array}{l}
 {H_i}(t) = {\rm{diag}}\{ \sigma _1^i(t) + \sigma _2^i(t) + \sigma _3^i(t),\sigma _1^i(t) + \sigma _4^i(t) \\
  + \sigma _5^i(t),\sigma _2^i(t) + \sigma _4^i(t) + \sigma _6^i(t),\sigma _3^i(t) + \sigma _5^i(t) + \sigma _6^i(t)\}  \\
 \end{array},
\label {eq:134}\;\;\;
\end{eqnarray}
where $\sigma _{{\hbar _i}}^i(t)({\hbar _i} = 1,2,3,4,5,6)$ are determined by (\ref{eq:11}--\ref{eq:12}), and each stochastic process $\{ \sigma _{{\hbar _i}}^i(t)\}$ obeys the categorical distribution. To determine the signal ${{{\rm{\hat x}}}_{{s_i}}}(t)$ (see (\ref{eq:13})), the selection probabilities in (\ref{eq:20}) are taken as follows:
\begin{eqnarray}
\left\{ \begin{array}{l}
 \pi _1^1 = 0.3,\pi _2^1 = 0.2,\pi _3^1 = 0.1,\pi _4^1 = 0.1 \\
 \pi _5^1 = 0.1,\pi _6^1 = 0.2,\pi _1^2 = 0.2,\pi _2^2 = 0.1 \\
 \pi _3^2 = 0.2,\pi _4^2 = 0.1,\pi _5^2 = 0.3,\pi _6^2 = 0.1 \\
 \end{array}. \right.
\label {eq:135}
\end{eqnarray}
Thus, the selection probability matrices $H_1$ and $H_2$ (see (\ref{eq:21})) are given by:
\begin{eqnarray}
\left\{ \begin{array}{l}
 {H_1} = {\rm{diag}}\{ 0.6,0.5,0.5,0.4\}  \\
 {H_2} = {\rm{diag}}\{ 0.5,0.6,0.3,0.6\}  \\
 \end{array}. \right.
\label {eq:136}
\end{eqnarray}
When each selected ASC ${{{\rm{\hat x}}}_{{s_i}}}(t)$ is transmitted to the FC, the communication delays are taken as $d_1=1$ and $d_2=2$. In this case, it is calculated from (\ref{eq:129}) and (\ref{eq:136}) that
\begin{eqnarray}
\left\{ \begin{array}{l}
 \rho (A({I_4} - {H_1})A) = 0.5759 < 1 \\
 \rho ({A^2}({I_4} - {H_2})A) = 0.6631 < 1 \\
 \end{array}, \right.
\label {eq:137}
\end{eqnarray}
which means that the condition (\ref{eq:106}) holds in Theorem 3. Meanwhile, by using LMI Toolbox in Matlab, the variables $D_i$, $X_i$, $Y_i$, $Z_i$ and $S_i(i=1,2)$ are obtained by solving the matrix inequalities (\ref{eq:91}) and (\ref{eq:92}), i.e., the conditions (\ref{eq:91}--\ref{eq:92}) hold for the two local CSEs with different selection probabilities and communication delays. Under this case, it is concluded from Theorem 3 that the fusion estimation covariance matrix $P(t)$ for this example converges to a unique matrix, and the SDKFE exists. Then, implementing Algorithm 1 obtains the steady-state weighting matrices as follows:
\begin{eqnarray}
\left\{ \begin{array}{l}
 {\Omega _1} = \left[ {\begin{array}{*{20}{c}}
   {0.6254} & {0.0921} & {0.3294} & {0.044}  \\
   {0.0585} & {0.7874} & {{\rm{0}}{\rm{.2587}}} & {{\rm{0}}{\rm{.2107}}}  \\
   {{\rm{0}}{\rm{.1654}}} & {{\rm{0}}{\rm{.0271}}} & {{\rm{0}}{\rm{.6857}}} & {{\rm{0}}{\rm{.2670}}}  \\
   {{\rm{0}}{\rm{.0065}}} & {{\rm{0}}{\rm{.0765}}} & {{\rm{0}}{\rm{.2257}}} & {{\rm{0}}{\rm{.6729}}}  \\
\end{array}} \right] \\
 {\Omega _2} = \left[ {\begin{array}{*{20}{c}}
   {{\rm{0}}{\rm{.3746}}} & { - 0.0921} & { - 0.3294} & { - 0.044}  \\
   { - 0.0585} & {{\rm{0}}{\rm{.2126}}} & { - {\rm{0}}{\rm{.2587}}} & { - {\rm{0}}{\rm{.2107}}}  \\
   { - {\rm{0}}{\rm{.1654}}} & { - {\rm{0}}{\rm{.0271}}} & {{\rm{0}}{\rm{.3143}}} & { - {\rm{0}}{\rm{.2670}}}  \\
   { - {\rm{0}}{\rm{.0065}}} & { - {\rm{0}}{\rm{.0765}}} & { - {\rm{0}}{\rm{.2257}}} & {{\rm{0}}{\rm{.3271}}}  \\
\end{array}} \right] \\
 \end{array}. \right.
\label {eq:138}
\end{eqnarray}
Thus, the SDKFE ${{{\rm{\hat x}}}_s}(t)$ for this example is obtained by substituting (\ref{eq:138}) into (\ref{eq:109}). Notice that one also has by (\ref{eq:129}) and (\ref{eq:136}) that
\begin{eqnarray}
\left\{ \begin{array}{l}
 {\zeta _{\infty ,1}} = ||A|{|_1}||{({I_n} - {H_1})^{{\textstyle{1 \over 2}}}}A|{|_1} \\
 \;\;\;\;\;\;\;\; \times ||A|{|_\infty }||{({I_n} - {H_1})^{{\textstyle{1 \over 2}}}}A|{|_\infty } = 0.8657 < 1 \\
 {\zeta _{\infty ,2}} = ||{A^2}|{|_1}||{({I_n} - {H_1})^{{\textstyle{1 \over 2}}}}A|{|_1} \\
 \;\;\;\;\;\;\;\;\; \times ||{A^2}|{|_\infty }||{({I_n} - {H_1})^{{\textstyle{1 \over 2}}}}A|{|_\infty } = 1.3554 > 1 \\
 \end{array} \right.
\label {eq:a137}
\end{eqnarray}
It is obvious that the judgement condition (\ref{eq:c105}) derived by \cite{o24} is invalid for the second sink node in this example, which implies that the stability condition in this paper has less conservatism than the one in \cite{o24}.

By using Algorithm 1, the trajectories of the DKFE ``${\rm{\hat x}}(t)$'' and the state ``x(t)'' are plotted in Fig.\ref{fig4}, which shows that the designed DKFE is able to estimate the original state ``x(t)'' well. Meanwhile, let $P_o(t)$ denote the original DKFE (ODKFE) under the dimensionality reduction when there are no communication delays between the sink nodes and the FC. Then, the estimation performances (assessed by the trace of the estimation error covariance matrix) of the local CSEs, DKFE and ODKFE are shown in Fig.\ref{fig5}. It is seen from this figure that the estimation performance of the DKFE is better than that of each CSE at each time-step, which is in line with the result (\ref{eq:72}). However, the estimation performance of the DKFE is worse than that of the ODKFE, which implies that the communication delays can affect the fusion estimation performance. Moreover, it is known from the this figure that the MSEs of the DKFE and CSEs all converge to the steady-sate values, which accords with the results (\ref{eq:93}) and (\ref{eq:107}).

\begin{figure}[thpb]
\centering
\includegraphics[height=5.5cm, width=8.0cm]{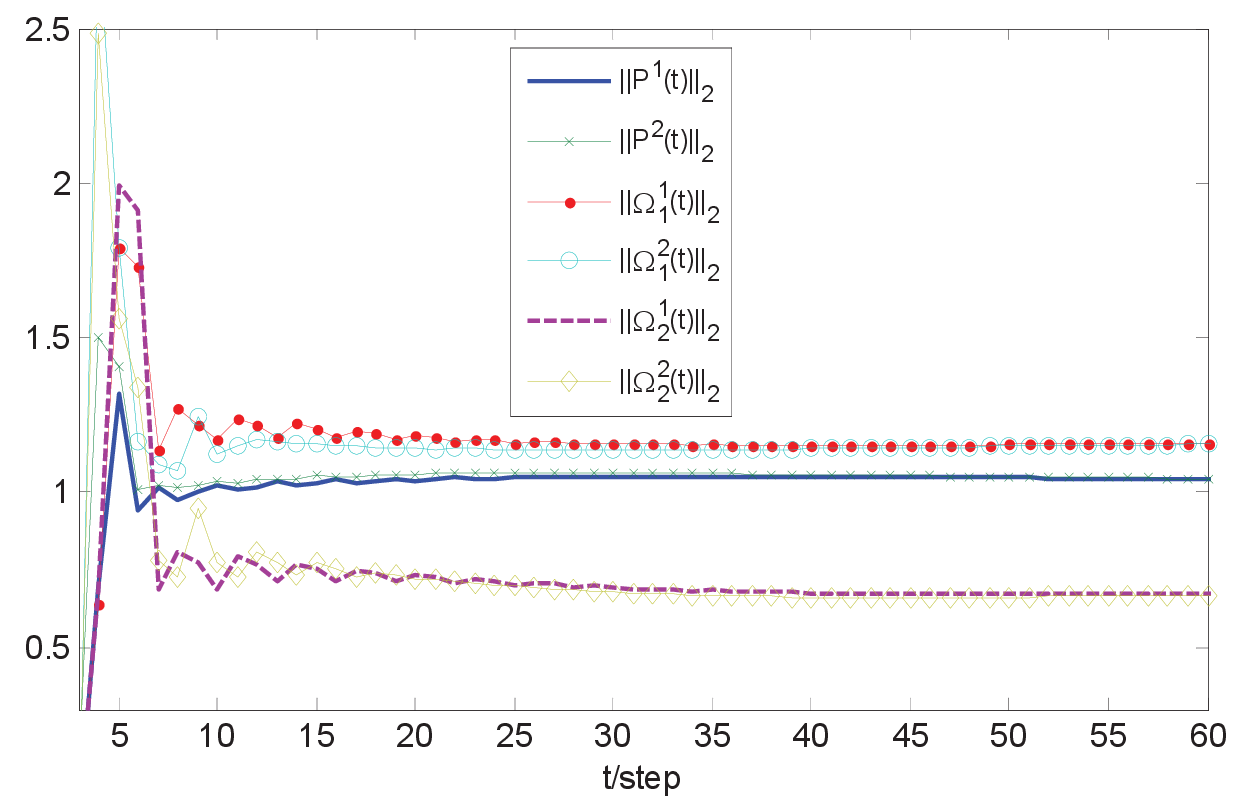}
\caption{The trajectories of $||{P^i}(t)|{|_2}$, $||\Omega _1^i(t)|{|_2}$ and $||\Omega _2^i(t)|{|_2}(i = 1,2)$, where ${P^i}(t)$, $\Omega _1^i(t)$ and $\Omega _2^i(t)$ represents the covariance matrix $P(t)$ and the weighting matrices ${\Omega _1}(t),{\Omega _2}(t)$ under different initial values for $i \ne j$.}
\label{fig6}
\end{figure}

\begin{figure}[thpb]
\centering
\includegraphics[height=5.5cm, width=8.0cm]{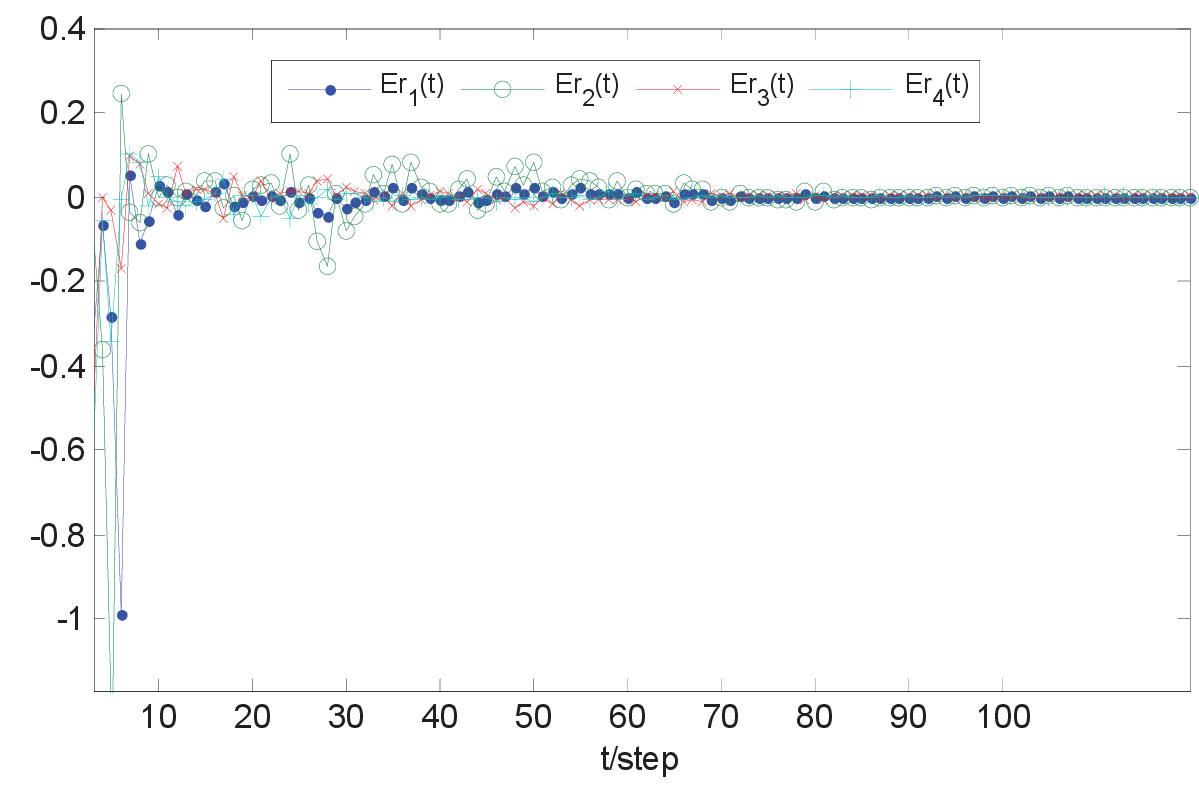}
\caption{The trajectories of ${\rm{E}}{{\rm{r}}_i}(t)\;(i=1,2,3,4)$.}
\label{fig7}
\end{figure}

To demonstrate the effectiveness of the SDKFE for this example, the matrix 2-norms of $P(t)$ and ${\Omega _i}(t)(i = 1,2)$ are shown in Fig.\ref{fig6} under different initial values. It is seen from this figure that $||P(t)||_2$ and $||{\Omega _i}(t)||_2(i \in \{ 1,2\} )$ can converge to the unique steady-state values under different initial values, which is in line with the result in Theorem 3. On the other hand, let ${\rm{E}}{{\rm{r}}_i}(t) \buildrel \Delta \over = {\rm{\hat x}}(i,t) - {{{\rm{\hat x}}}_s}(i,t)$, where ${\rm{\hat x}}(i,t)$ represents the $i{\rm{th}}$ component of the DKFE ${\rm{\hat x}}(t)$, and the meaning of ${{{\rm{\hat x}}}_s}(i,t)$ is the same as that of ${\rm{\hat x}}(i,t)$. Then, implementing Algorithms 1--2, the trajectories of ${\rm{E}}{{\rm{r}}_i}(t)(i = 1,2,3,4)$ are depicted in Fig.\ref{fig7}, where the measurement sequences $\{ {y_i}(t)(i = 1,2)\}$ are the same when computing the DKFE ${\rm{\hat x}}(t)$ and the SDKFE ${{{\rm{\hat x}}}_s}(t)$. It is shown from this figure that the errors between the DKFE and SDKFE will converge to zero as $t$ increases, which is in line with the property of the SDKFE. It should be pointed out that the SDKFE is much easier to implement as compared with the DKFE in practical applications.


\section{CONCLUSIONS}
CPSs are being widely integrated in various critical infrastructures and running on wired or wireless communication networks, however, bandwidth constraints and communication delays are usually unavoidable. State estimation plays an essential role in the monitoring and supervision of CPSs, and its importance has made the robustness and estimation performance a major concern. Therefore, to guarantee the satisfactory estimation performance in CPSs, the distributed dimensionality reduction fusion estimation problem with communication delays has been studied in this paper. Based on the stochastic dimensionality reduction strategy, a mathematical model was proposed to establish the relationship between the dimensionality reduction and communication delays, and then the recursive DKFE was obtained by resorting to the optimal weighted fusion criterion. A delay-dependent and probability-dependent condition, which can be easily judged by using Matlab LMI Toolbox, was derived for the DKFE such that the fusion estimation error covariance matrix $P(t)$ converges to a unique steady-state matrix. This result is very important to derive the SDKFE, and the computational complexity of the SDKFE is much lower than that of the DKFE. Meanwhile, when the communication delay $d_i$ is known in advance, the selection probability criterion to determine the dimensionality reduction strategy has also been presented. Moreover, it has been shown that the stability condition in this paper has less conservatism than the exiting ones. Finally, two examples were given to demonstrate the advantage and effectiveness of the proposed methods.


\begin{center}
APPENDIX
\end{center}

\begin{center}
    \textbf{A.1: The proof of Lemma 2}
\end{center}
\begin{proof}
It follows from (\ref{eq:1}) and (\ref{eq:5}) that
\vspace{-3pt}
\begin{eqnarray}
\begin{array}{l}
 {{{\rm{\tilde x}}}_i}(t) = {\Phi _{{{\rm{K}}_i}}}(t){{{\rm{\tilde x}}}_i}(t - 1) \\
 \;\;\;\;\;\;\;\;\;\;\; + {{\rm{G}}_{{{\rm{K}}_i}}}(t)w(t - 1) - {{\rm{K}}_i}(t){v_i}(t) \\
 \end{array}
\label {eq:38}
\end{eqnarray}
For ${t_1} \ge {t_2}$, it is derived from (\ref{eq:38}) that
\vspace{-3pt}
\begin{eqnarray}
\begin{array}{l}
 {{{\rm{\tilde x}}}_i}({t_1}) = \left( {\prod\nolimits_{{\varphi _i} = 0}^{{t_1} - {t_2} - 1} {{\Phi _{{{\rm K}_i}}}({t_1} - {\varphi _i})} } \right){{{\rm{\tilde x}}}_i}({t_2}) \\
 \;\;\;\;\;\;\;\;\; + \sum\nolimits_{{\alpha _i} = 1}^{{t_1} - {t_2}} {\left\{ {\left( {\prod\nolimits_{{\varphi _i} = 0}^{{\alpha _i} - 2} {{\Phi _{{{\rm K}_i}}}({t_1} - {\varphi _i})} } \right)} \right.}  \\
 \;\;\;\;\;\;\;\;\;\;\;\;\;\;\;\;\;\;\; \times {{\rm{G}}_{{{\rm K}_i}}}({t_1} - {\alpha _i} + 1)w({t_1} - {\alpha _i})\}  \\
 \;\;\;\;\;\;\;\;\; - \sum\nolimits_{{\alpha _i} = 0}^{{t_1} - {t_2} - 1} {\left\{ {\left( {\prod\nolimits_{{\varphi _i} = 0}^{{\alpha _i} - 1} {{\Phi _{{{\rm K}_i}}}({t_1} - {\varphi _i})} } \right)} \right.}  \\
 \;\;\;\;\;\;\;\;\;\;\;\;\;\;\;\;\;\;\; \times {{\rm K}_i}({t_1} - {\alpha _i}){v_i}({t_1} - {\alpha _i})\}  \\
 \end{array}
\label {eq:39}
\end{eqnarray}
\vspace{-3pt}
On the other hand, it is concluded from (\ref{eq:3}) and the geometric meaning of ${{{\rm{\tilde x}}}_i}(t)$ that
\vspace{-6pt}
\begin{eqnarray}
\left\{ \begin{array}{l}
 {{{\rm{\tilde x}}}_i}({t_1}) \bot w({t_2})({t_2} \ge {t_1}) \\
 {{{\rm{\tilde x}}}_i}({t_1}) \bot {v_i}({t_2})({t_2} > {t_1}) \\
 {{{\rm{\tilde x}}}_i}({t_1}) \bot {v_j}({t_2})(i \ne j,\forall {t_1},{t_2}) \\
 w({t_1}) \bot {v_i}({t_2})(\forall i,{t_1},{t_2}) \\
 \end{array} \right.
\label {eq:40}
\end{eqnarray}
One has by (\ref{eq:32}) that ${{\rm{C}}_o}({t_1},{t_2} - 1) = 1$ for ${t_1} \ge {t_2}$, and thus the results (\ref{eq:34})-(\ref{eq:35}) are obtained by (\ref{eq:39}) and (\ref{eq:40}). Moreover, the result (\ref{eq:36}) is directly obtained by the definitions of $\Phi _{{{\rm{x}}_i}}^w({t_1},{t_2})$ and $\Phi _{{{\rm{x}}_i}}^{\rm{F}}({t_1},g,{t_2})$ in (\ref{eq:33}), while the result (\ref{eq:37}) is obtained from (\ref{eq:3}) and (\ref{eq:24}).
\end{proof}
\begin{center}
   \textbf{A.2: The proof of Lemma 3}
\end{center}

\begin{proof}
Let us define
\begin{eqnarray}
\left\{ \begin{array}{l}
 {{\rm{H}}_{{\rm{A}}{{\rm{d}}_i}}}(t) \buildrel \Delta \over = {A^{{d_i}}}[{I_n} - {H_i}(t - {d_i})]A \\
 {{\rm{H}}_{{{\rm{d}}_i}}}(t) \buildrel \Delta \over = {A^{{d_i}}}{H_i}(t - {d_i}),{{{\rm{\bar H}}}_{{\rm{A}}{{\rm{d}}_i}}}(t) = {A^{{d_i}}} - {{\rm{H}}_{{{\rm{d}}_i}}}(t) \\
 \end{array} \right.
\label {eq:45}
\end{eqnarray}
For ${t_1} \ge {t_2}$, it follows from (\ref{eq:23}) that
\begin{eqnarray}
\begin{array}{l}
 {\rm{\tilde x}}_i^{\rm{c}}({t_1}) = \sum\nolimits_{\hbar  = 0}^{{\chi _i}({t_1},{t_2}) - 1} {\{ {{\rm{H}}_{{{\rm{d}}_i}}}(f_i^\hbar ({t_1})){{\tilde x}_i}(f_{i{\rm{o}}}^\hbar ({t_1}) - {d_i})}  \\
  \;\;\;+ \prod\nolimits_{\mu  = 0}^{\hbar  - 1} {{{\rm{H}}_{{\rm{A}}{{\rm{d}}_i}}}(f_{i^\circ }^\mu ({t_1}))} {{{\rm{\bar H}}}_{{\rm{A}}{{\rm{d}}_i}}}(f_{i{\rm{o}}}^\hbar ({t_1}))w(f_{i{\rm{o}}}^{\hbar  + 1}({t_1}))\}  \\
  \;\;\;+ \sum\nolimits_{\hbar  = 1}^{{\chi _i}({t_1},{t_2}) - 1} {\{ {{\rm{H}}_{{\rm{A}}{{\rm{d}}_i}}}(f_{i{\rm{o}}}^{\hbar  - 1}({t_1})){{\rm{F}}_w}({d_i},f_{i{\rm{o}}}^\hbar ({t_1}))\} }  \\
  \;\;\;+ \prod\nolimits_{\mu  = 0}^{{\chi _i}({t_1},{t_2}) - 1} {{{\rm{H}}_{{\rm{A}}{{\rm{d}}_i}}}(f_{i{\rm{o}}}^\mu ({t_1}))} {\rm{\tilde x}}_i^{\rm{c}}(f_{i{\rm{o}}}^{{\chi _i}({t_1},{t_2})}({t_1})) \\
  \;\;\;+ (1 - {\delta _{{t_1},{t_2}}}){{\rm{F}}_w}({d_i},f_{i{\rm{o}}}^0({t_1})) \\
 \end{array}
\label {eq:46}
\end{eqnarray}
where $f_i(t)$ and ${\chi _i}({t_1},{t_2})$ are defined in (\ref{eq:41}). Notice that
\begin{eqnarray}
f_{i \circ }^{{\chi _i}({t_1},{t_2})}({t_1}) < f_{i \circ }^{{\chi _i}({t_1},{t_2}) - 1}({t_1}) <  \cdots  < f_{i \circ }^0({t_1})
\label {eq:47}
\end{eqnarray}
Then taking the statistical property of $\gamma _\ell ^i(t)$ into account yields:
\begin{eqnarray}
{\rm E}\left\{ {\left( {\prod\limits_{\mu  = 0}^{\hbar  - 1} {{{\rm{H}}_{{\rm{A}}{{\rm{d}}_i}}}(f_{i \circ }^\mu ({t_1})){{{\rm{\bar H}}}_{{\rm{A}}{{\rm{d}}_i}}}(f_{i \circ }^\hbar ({t_1}))} } \right)} \right\} = {\rm{H}}_{{\rm{A}}{{\rm{d}}_i}}^\hbar {{{\rm{\bar H}}}_{{\rm{A}}{{\rm{d}}_i}}}
\label {eq:48}\;\;\;\;\;\;
\end{eqnarray}
where  ${{\rm{H}}_{{\rm{A}}{{\rm{d}}_i}}}$ and ${{{\rm{\bar H}}}_{{\rm{A}}{{\rm{d}}_i}}}$ are given by (\ref{eq:44}). Moreover, it is concluded from (\ref{eq:3}), (\ref{eq:23}) and (\ref{eq:40}) that
\begin{eqnarray}
\left\{ \begin{array}{l}
 {\rm{\tilde x}}_i^{{\rm{c}}}({t_1}) \bot w({t_2})({t_2} \ge {t_1} - {d_1}) \\
 {\rm{\tilde x}}_i^{{\rm{c}}}({t_1}) \bot {v_i}({t_2})({t_2} > {t_1} - {d_1}) \\
 {\rm{\tilde x}}_i^{{\rm{c}}}({t_1}) \bot {v_j}({t_2})(i \ne j,\forall {t_1},{t_2}) \\
 \end{array} \right.
\label {eq:49}
\end{eqnarray}
Thus, the result (\ref{eq:42}) is derived from (\ref{eq:46}--\ref{eq:49}). On the other hand, (\ref{eq:43}) is directly obtained from the definitions of $\Theta _{{\rm{x}}_i^{\rm{c}}}^w({t_1},{t_2})$ and ${{\rm{F}}_w}(g,t)$.
\end{proof}

\begin{center}
\textbf{A.3: The proof of Lemma 4}
\end{center}

\begin{proof}
It follows from (\ref{eq:39}) that
\begin{eqnarray}
\begin{array}{l}
 {{{\rm{\tilde x}}}_i}(t) = \left( {\prod\nolimits_{{\varphi _i} = 0}^{{d_j}} {{\Phi _{{{\rm K}_i}}}(t - {\varphi _i})} } \right){{{\rm{\tilde x}}}_i}(t - {d_j} - 1) + \sum\nolimits_{{\alpha _i} = 1}^{{d_j} + 1} {}  \\
 \;\left\{ {\left( {\prod\nolimits_{{\varphi _i} = 0}^{{\alpha _i} - 2} {{\Phi _{{{\rm K}_i}}}(t - {\varphi _i})} } \right){{\rm{G}}_{{{\rm K}_i}}}(t - {\alpha _i} + 1)w(t - {\alpha _i})} \right\} \\
  - \sum\nolimits_{{\alpha _i} = 0}^{{d_j}} {\left( {\prod\nolimits_{{\varphi _i} = 0}^{{\alpha _i} - 1} {{\Phi _{{{\rm K}_i}}}(t - {\varphi _i})} } \right){{\rm K}_i}(t - {\alpha _i}){v_i}(t - {\alpha _i})}  \\
 \end{array}\;\;
\label {eq:53}
\end{eqnarray}
Then one has by (\ref{eq:49}) and (\ref{eq:53}) that
\begin{eqnarray}
\begin{array}{l}
 {\rm E}\{ {{{\rm{\tilde x}}}_i}(t){[{\rm{\tilde x}}_j^{\rm{c}}(t - {d_j} - 1)]^{\rm{T}}}\}  \\
 \;\;\;\;\;\;\;\;\;\;\;\;\;\; = \left( {\prod\nolimits_{{\varphi _j} = 0}^{{d_j}} {{\Phi _{{{\rm K}_i}}}(t - {\varphi _i})} } \right){\Gamma _{ij}}(t - {d_j} - 1) \\
 \end{array}
\label {eq:54}
\end{eqnarray}
Meanwhile, it follows from (\ref{eq:23}) that
\begin{eqnarray}
\begin{array}{l}
 {\Gamma _{ij}}(t) = {\rm E}\{ {{{\rm{\tilde x}}}_i}(t){[{\rm{\tilde x}}_j^{\rm{c}}(t - {d_i} - 1)]^{\rm{T}}}\} {\rm{H}}_{{\rm{A}}{{\rm{d}}_j}}^{\rm{T}} \\
 \;\;\;\; + {\rm E}\{ {{{\rm{\tilde x}}}_i}(t){\rm{\tilde x}}_j^{\rm{T}}(t - {d_j})\} {\rm{H}}_{{{\rm{d}}_j}}^{\rm{T}} + {\rm E}\{ {{{\rm{\tilde x}}}_i}(t){\rm{F}}_w^{\rm{T}}({d_j},t)\}  \\
 \;\;\;\; + {\rm E}\{ {{{\rm{\tilde x}}}_i}(t){w^{\rm{T}}}(t - {d_j} - 1)\} {\rm{\bar H}}_{{\rm{A}}{{\rm{d}}_j}}^{\rm{T}} \\
 \end{array}
\label {eq:55}
\end{eqnarray}
Therefore, the result (\ref{eq:51}) is derived from (\ref{eq:54}--\ref{eq:55}) and Lemma 2. Meanwhile, according to (\ref{eq:39}), (\ref{eq:52}) can be derived from the similar derivation of (\ref{eq:51}).
\end{proof}

\begin{center}
    \textbf{A.4: The proof of Lemma 5}
\end{center}
\begin{proof}
(\ref{eq:57}) can be derived from (\ref{eq:54}). On the other hand, it follows from (\ref{eq:23}) that
\begin{eqnarray}
\begin{array}{l}
 {\rm{\tilde x}}_j^{\rm{c}}(t - {d_j} - 1) = {\rm{\tilde x}}_j^{\rm{c}}({f_j}(t)) \\
  = \sum\nolimits_{\kappa  = 1}^{{\eta _{ij}} - 1} {\left\{ {\left( {\prod\nolimits_{\upsilon  = 1}^{\kappa  - 1} {{{\rm{H}}_{{\rm{A}}{{\rm{d}}_j}}}(f_{j{\rm{o}}}^\upsilon (t))} } \right){{\rm{H}}_{{{\rm{d}}_j}}}(f_{j{\rm{o}}}^\kappa (t))} \right.}  \\
  \;\times {{{\rm{\tilde x}}}_j}(f_{j{\rm{o}}}^\kappa (t) - {d_j})\}  + \sum\nolimits_{\kappa  = 1}^{{\eta _{ij}} - 1} {\left\{ {\left( {\prod\nolimits_{\upsilon  = 1}^{\kappa  - 1} {{{\rm{H}}_{{\rm{A}}{{\rm{d}}_j}}}(f_{j{\rm{o}}}^\upsilon (t))} } \right)} \right.}  \\
  \;\times {{{\rm{\bar H}}}_{{\rm{A}}{{\rm{d}}_j}}}(f_{j{\rm{o}}}^\kappa (t))w(f_{j^\circ }^{\kappa  + 1}(t))\}  \\
  \;+ \sum\nolimits_{\kappa  = 1}^{{\eta _{ij}} - 1} {\left( {\prod\nolimits_{\upsilon  = 1}^{\kappa  - 1} {{{\rm{H}}_{{\rm{A}}{{\rm{d}}_j}}}(f_{j{\rm{o}}}^\upsilon (t))} } \right){{\rm{F}}_w}({d_j},f_{j{\rm{o}}}^\kappa (t))}  \\
  \;+ \left( {\prod\nolimits_{\kappa  = 1}^{{\eta _{ij}} - 1} {{{\rm{H}}_{{\rm{A}}{{\rm{d}}_j}}}(f_{j{\rm{o}}}^\kappa (t))} } \right){\rm{\tilde x}}_j^{\rm{c}}(f_{j{\rm{o}}}^{{\eta _{ij}}}(t)) \\
 \end{array}
\label {eq:59}
\end{eqnarray}
where ${{\rm{H}}_{{{\rm{d}}_j}}}(t)$, ${{\rm{H}}_{{\rm{A}}{{\rm{d}}_j}}}(t)$ and ${{{\rm{\bar H}}}_{{\rm{A}}{{\rm{d}}_j}}}(t)$ are defined in (\ref{eq:45}). Meanwhile, it follows from the similar derivation of (\ref{eq:54}) that
\begin{eqnarray}
\begin{array}{l}
 {\rm E}\{ {{{\rm{\tilde x}}}_i}(t - {d_i}){[{\rm{\tilde x}}_j^{\rm{c}}(f_{j{\rm{o}}}^{{\eta _{ij}}}(t))]^{\rm{T}}}\}  \\
 \;\;\;\;= \left( {\prod\nolimits_{{\varphi _i} = 0}^{{\eta _{ij}}({d_j} + 1) - 1 - {d_i}} {{\Phi _{{{\rm K}_i}}}(t - {d_i} - {\varphi _i})} } \right)\Gamma_{ij} (f_{j{\rm{o}}}^{{\eta _{ij}}}(t)) \\
 \end{array}
\label {eq:60}
\end{eqnarray}
Therefore, (\ref{eq:58}) is obtained from (\ref{eq:48}), (\ref{eq:59}), (\ref{eq:60}) and Lemma 2.
\end{proof}

\begin{center}
    \textbf{A.5: The proof of Lemma 6}
\end{center}
\begin{proof}
To establish the relationship between ${\Upsilon _{ij}}(t)$ and ${\Xi _{ij}}(t)$, the least common multiple of~$d_i+1$~and~$d_j+1$~is introduced, and thus one has
\begin{eqnarray}
f_{i\circ}^{{\tau _{{d_i}}}}  (t) = f_{i\circ}^{{\tau _{{d_j}}}}(t) = t - {\tau _{ij}}
\label {eq:65}
\end{eqnarray}
where ${f_i}(t)$ is defined in (\ref{eq:41}). On the other hand, for $i \ne j$, it is concluded from the statistical property of $H_i(t)$ that
\begin{eqnarray}
\begin{array}{l}
 {\Upsilon _{ij}}(t) = {\rm E}\{ {\rm{\tilde x}}_i^{\rm{c}}({f_i}(t)){[{\rm{\tilde x}}_j^{\rm{c}}({f_j}(t))]^{\rm{T}}}\}  \\
 \;\;\;\;\;\;\;\;\;\;\; = {\rm E}\{ {\rm{\bar x}}_i^{\rm{c}}({f_i}(t)){[{\rm{\bar x}}_j^{\rm{c}}({f_j}(t))]^{\rm{T}}}\}  \\
 \end{array}
\label {eq:66}
\end{eqnarray}
where
\begin{eqnarray}
\begin{array}{l}
 {\rm{\bar x}}_i^{\rm{c}}({f_i}(t)) = \sum\nolimits_{\kappa  = 1}^{{\tau _{{d_i}}} - 1} {{\rm{H}}_{{\rm{A}}{{\rm{d}}_i}}^{\kappa  - 1}{{\rm{H}}_{{{\rm{d}}_i}}}{{{\rm{\tilde x}}}_i}(f_{i{\rm{o}}}^\kappa (t) - {d_i})}  \\
 \;\;\;\;\;\;\;\;\;\;\;\;\;\;\;\;\;\; + \sum\nolimits_{\kappa  = 1}^{{\tau _{{d_i}}} - 1} {{\rm{H}}_{{\rm{A}}{{\rm{d}}_i}}^{\kappa  - 1}{{{\rm{\bar H}}}_{{\rm{A}}{{\rm{d}}_i}}}w(f_{i{\rm{o}}}^{\kappa  + 1}(t))}  \\
 \;\;\;\;\;\;\;\;\;\;\;\;\;\;\;\;\;\; + \sum\nolimits_{\kappa  = 1}^{{\tau _{{d_i}}} - 1} {{\rm{H}}_{{\rm{A}}{{\rm{d}}_i}}^{\kappa  - 1}{{\rm{F}}_w}({d_i},f_{i{\rm{o}}}^\kappa (t))}  \\
 \;\;\;\;\;\;\;\;\;\;\;\;\;\;\;\;\;\; + {\rm{H}}_{{\rm{A}}{{\rm{d}}_i}}^{{\tau _{{d_i}}} - 1}{\rm{\tilde x}}_i^{\rm{c}}(t - {\tau _{ij}}) \\
 \end{array}
\label {eq:67}
\end{eqnarray}
Notice that when ${\tau _{{d_i}}} = 1$, ${\rm{\bar x}}_i^{\rm{c}}({f_i}(t))={\rm{\tilde x}}_i^{\rm{c}}(t - {\tau _{ij}})$. Then, (\ref{eq:67}) can be written as:
\begin{eqnarray}
{\rm{\bar x}}_i^{\rm{c}}({f_i}(t)) = (1 - {\delta _{1,{\tau _{{d_i}}}}}){\Sigma _i}{\rm{\bar x}}_{{f_i}}^w(t) + {\rm{H}}_{{\rm{A}}{{\rm{d}}_i}}^{{\tau _{{d_i}}} - 1}{\rm{\tilde x}}_i^{\rm{c}}(t - {\tau _{ij}})
\label {eq:68}
\end{eqnarray}
where ${\rm{\bar x}}_{{f_i}}^w(t)$ and ${\Sigma _i}$ are given by (\ref{eq:62}) and (\ref{eq:64}), respectively. Therefore, (\ref{eq:63}) is derived from (\ref{eq:66}) and (\ref{eq:68}).
\end{proof}

\begin{center}
    \textbf{A.6: The proof of Theorem 1}
\end{center}
\begin{proof}
It is concluded from (\ref{eq:3}), (\ref{eq:40}) and (\ref{eq:49}) that
\vspace{-5pt}
\begin{eqnarray}
\begin{array}{l}
 {{{\rm{\tilde x}}}_i}(t - {d_i}) \bot {{\rm{F}}_w}({d_i},t),
 {\rm{\tilde x}}_i^{\rm{c}}(t - {d_i}) \bot {{\rm{F}}_w}({d_i},t) \\
 \end{array}
\label {eq:73}
\end{eqnarray}
where ${{\rm{F}}_w}({d_i},t)$~is defined by (\ref{eq:24}). Notice that
\vspace{-5pt}
\begin{eqnarray}
\left\{ \begin{array}{l}
 {\rm E}\{ {\rm{\tilde x}}_i^{\rm{c}}(t - {d_i} - 1){\rm{\tilde x}}_i^{\rm{T}}(t - {d_i})\}  = \Psi _{ii}^{\rm{T}}(t) \\
 {\rm E}\{ [{I_n} - {H_i}(t)] \odot {H_i}(t)\}  = {\rm{V}}_i^{\rm{T}} \\
 \end{array} \right.
\label {eq:74}
\end{eqnarray}
Then, (\ref{eq:70}) is derived from Lemma 1, (\ref{eq:23}), (\ref{eq:40}), (\ref{eq:49}), (\ref{eq:51}), (\ref{eq:57}) and (\ref{eq:73}--\ref{eq:74}). On the other hand, (\ref{eq:23}) is rewritten as:
\begin{eqnarray}
\begin{array}{l}
 {\rm{\tilde x}}_i^{\rm{c}}(t) = {{\rm{H}}_{{{\rm{d}}_i}}}(t){{{\rm{\tilde x}}}_i}(t - {d_i}) + {{\rm{H}}_{{\rm{A}}{{\rm{d}}_i}}}(t){\rm{\tilde x}}_i^{\rm{c}}(t - {d_i} - 1) \\
 \;\;\;\;\;\;\;\;\;\;\; + {{{\rm{\bar H}}}_{{\rm{A}}{{\rm{d}}_i}}}(t)w(t - {d_i} - 1) + {{\rm{F}}_w}({d_i},t) \\
 \end{array}
\label {eq:75}
\end{eqnarray}
where ${{\rm{H}}_{{{\rm{d}}_i}}}(t)$, ${{\rm{H}}_{{\rm{A}}{{\rm{d}}_i}}}(t)$ and ${{{\rm{\bar H}}}_{{\rm{A}}{{\rm{d}}_i}}}(t)$ are defined in (\ref{eq:45}). Meanwhile, for $i \ne j$, one has by (\ref{eq:16}) and (\ref{eq:19}) that
\begin{eqnarray}
{\rm E}\{ {H_i}(t){H_j}(t)\}  = {\rm E}\{ {H_i}(t)\} {\rm E}\{ {H_j}(t)\} \}
\label {eq:76}
\end{eqnarray}
Moreover, it is derived from (\ref{eq:3}) and (\ref{eq:24}) that
\begin{eqnarray}
\left\{ \begin{array}{l}
 {\rm E}{\rm{\{ }}{w_i}(t - {d_i} - 1){\rm{F}}_w^{\rm{T}}({d_j},t{\rm{)\} }}{{\rm{C}}_{\rm{o}}}({d_j},{d_i}){Q_w}{({A^{{d_i}}})^{\rm{T}}} \\
 {\rm E}\{ {{\rm{F}}_w}({d_i},t{\rm{)F}}_w^{\rm{T}}({d_j},t{\rm{)}}\}  = \sum\nolimits_{\theta  = 1}^{\min \{ {d_i},{d_j}\} } {{A^\theta }{Q_w}{{({A^\theta })}^{\rm{T}}}}  \\
 \end{array} \right.
\label {eq:77}
\end{eqnarray}
where ${{\rm{C}}_{\rm{o}}}({d_j},{d_i})$ is determined by (\ref{eq:32}). Therefore, (\ref{eq:71}) is
derived from (\ref{eq:75}--\ref{eq:77}) and the results in Lemmas 2--6. Furthermore, at a particular time, the optimal fusion estimation error covariance matrix of (\ref{eq:26}) can be calculated by (\ref{eq:30}), while each local estimation error
covariance matrix of (\ref{eq:15}) is calculated by (\ref{eq:70}), then (\ref{eq:72}) is obtained from the results in \cite{c8,c9}.
\end{proof}

\begin{center}
    \textbf{A.7: The proof of Theorem 2}
\end{center}
\begin{proof}
Consider the following stochastic system:
\begin{eqnarray}
{ \xi _i}(t + 1) = {{\rm{A}}_i}(t){ \xi _i}(t - {d_i}),
\label {eq:94}
\end{eqnarray}
where ${{\rm{A}}_i}(t)$ is defined by (\ref{eq:86}). Define ${\eta _i}(t) \buildrel \Delta \over = {{ \xi }_i}(t) - {{ \xi }_i}(t - 1)$. Then, one chooses a Lyapunov function candidate for the system (\ref{eq:94}) as follows:
\begin{eqnarray}
\begin{array}{l}
 {V_{{\xi _i}}}(t) = {\rm E}\{  \xi _i^{\rm{T}}(t){D_i}{{ \xi }_i}(t)\}  + \sum\nolimits_{\kappa  = t - {d_i}}^{t - 1} {{\rm E}\{  \xi _i^{\rm{T}}(\kappa ){S_i}{{ \xi }_i}(\kappa )\} }  \\
  \;\;\;\;\;\;\;\;\;\;+ \sum\nolimits_{\beta_i  =  - {d_i}}^{ - 1} {\sum\nolimits_{\kappa  = t + {\beta _i} + 1}^t {{\rm E}\{ \eta _i^{\rm{T}}(\kappa ){Z_i}{\eta _i}(\kappa )\} } }  \\
 \end{array}.
\label {eq:95}\;\;\;
\end{eqnarray}
Notice that
\begin{eqnarray}
\left\{ \begin{array}{l}
 {{ \xi }_i}(t - {d_i}) = {{ \xi }_i}(t) - \sum\nolimits_{\kappa  = t - {d_i} + 1}^t {{\eta _i}(\kappa )}  \\
 {{\xi }_i}(t + 1) = {{\rm{A}}_i}(t){{\xi }_i}(t) - {{\rm{A}}_i}(t)\sum\nolimits_{\kappa  = t - {d_i} + 1}^t {{\eta _i}(\kappa )}  \\
 \end{array}. \right.
\label {eq:96}
\end{eqnarray}
From the similar derivation of Theorem 1 in \cite{c41}, it can be derived from (\ref{eq:94}--\ref{eq:96}) that
\begin{eqnarray}
\begin{array}{l}
 \Delta {V_{{\xi _i}}}(t) = {V_{{\xi _i}}}(t + 1) - {V_{{\xi _i}}}(t) \\
 \;\;\;\;\;\;\;\;\;\;\;\;\; \le  \xi _i^{\rm{T}}(t)\{  - {D_i} + {X_i} + Y_i^{\rm{T}} + Y_i + {d_i}{Z_i} \\
 \;\;\;\;\;\;\;\;\;\;\;\;\; + {S_i}\} {{ \xi }_i}(t) +  \xi _i^{\rm{T}}(t - {d_i}){\rm E}\{ {\rm{A}}_i^{\rm{T}}(t){D_i}{{\rm{A}}_i}(t) \\
 \;\;\;\;\;\;\;\;\;\;\;\;\; + {d_i}{\rm E}\{ {\rm{A}}_i^{\rm{T}}(t){Z_i}{{\rm{A}}_i}(t)\}  - {S_i}\} {{ \xi }_i}(t - {d_i}) \\
 \;\;\;\;\;\;\;\;\;\;\;\;\; +  \xi _i^{\rm{T}}(t)\{  - {Y_i} - {d_i}{Z_i}{\rm E}\{ {{\rm{A}}_i}(t)\} \} {{ \xi }_i}(t - {d_i}) \\
 \;\;\;\;\;\;\;\;\;\;\;\;\; +  \xi _i^{\rm{T}}(t - {d_i})\{  - Y_i^{\rm{T}} - {d_i}{\rm E}\{ {\rm{A}}_i^{\rm{T}}(t)\} {Z_i}\} {{ \xi }_i}(t), \\
 \end{array}
\label {eq:97}
\end{eqnarray}
where $X_i$, $Y_i$ and $Z_i$ are required to satisfy (\ref{eq:91}). Under this case, according to Lyapunov stability theory (see \cite{c42}: p.131), when the condition (\ref{eq:92}) holds, the system (\ref{eq:94}) is mean-square stable. This means that the state covariance matrix ${\Xi _{{\xi _i}}}(t) \buildrel \Delta \over = {\rm E}\{ {\xi _i}(t)\xi _i^{\rm{T}}(t)\}$ of (\ref{eq:94}) converges to zero under the conditions (\ref{eq:91}) and (\ref{eq:92}) (i.e., $\mathop {\lim }\limits_{t \to \infty } {\Xi _{{\xi _i}}}(t) = 0$), where
\begin{eqnarray}
\begin{array}{l}
 {\Xi _{{\xi _i}}}(t + 1) = {\rm E}\{ {{\rm{A}}_i}(t){\xi _i}(t - {d_i})\xi _i^{\rm{T}}(t - {d_i}){\rm{A}}_i^{\rm{T}}(t)\}  \\
 \;\;\;\;\;\;\;\;\;\;\;\;\;\;\;\; = f({\Xi _{{\xi _i}}}(t - {d_i})). \\
 \end{array}
\label {eq:98}
\end{eqnarray}
On the other hand, one has by (\ref{eq:16}),(\ref{eq:19}),(\ref{eq:40}),(\ref{eq:49}) that
\begin{eqnarray}
\left\{ \begin{array}{l}
 {{\tilde \xi }_i}(t - {d_i}) \bot {\zeta _i}(t) \\
 {\rm E}\{ {{\rm{A}}_i}(t){{\tilde \xi }_i}(t - {d_i})\tilde \xi _i^{\rm{T}}(t - {d_i}){\rm{A}}_i^{\rm{T}}(t)\}  = f({{\tilde \Xi }_{{\xi _i}}}(t - {d_i})) \\
 \end{array} \right.
\label {eq:99},\;\;\;
\end{eqnarray}
where ${{\tilde \Xi }_{{\xi _i}}}(t) = {\rm E}\{ {{\tilde \xi }_i}(t)\tilde \xi _i^{\rm{T}}(t)\}$. Then, it follows from (\ref{eq:85}) and (\ref{eq:99}) that
\begin{eqnarray}
{{\tilde \Xi }_{{\xi _i}}}(t + 1) = f({{\tilde \Xi }_{{\xi _i}}}(t - {d_i})) + {Q_{{\zeta _i}}},
\label {eq:100}
\end{eqnarray}
where ${Q_{{\zeta _i}}} = {\rm E}\{ {\zeta _i}(t)\zeta _i^{\rm{T}}(t)\}$, and $f({{\tilde \Xi }_{{\xi _i}}}(t - {d_i}))$ is calculated by (\ref{eq:88}) in Lemma 7. In what follows, we will prove that, under the conditions (\ref{eq:91}--\ref{eq:92}), the sequence $\{ {{\tilde \Xi }_{{\xi _i}}}(t)\}$ obtained from (\ref{eq:100}) is convergent, and the limit is independent of the initial values.

Define $\Delta {{\tilde \Xi }_{{\xi _i}}}(t) \buildrel \Delta \over = {{\tilde \Xi }_{{\xi _i}}}(t) - {{\tilde \Xi }_{{\xi _i}}}(t - 1)$. Then, it is derived from (\ref{eq:100}) that
\begin{eqnarray}
\Delta {{\tilde \Xi }_{{\xi _i}}}(t + 1) = f({{\tilde \Xi }_{{\xi _i}}}(t - {d_i})) - f({{\tilde \Xi }_{{\xi _i}}}(t - {d_i} - 1)).
\label {eq:101}\;\;
\end{eqnarray}
Combining (\ref{eq:90}) in Lemma 7 yields that
\begin{eqnarray}
\Delta {{\tilde \Xi }_{{\xi _i}}}(t + 1) = f(\Delta {{\tilde \Xi }_{{\xi _i}}}(t - {d_i})).
\label {eq:102}
\end{eqnarray}
Notice that the recursive form of (\ref{eq:102}) is the same as that of (\ref{eq:98}), thus one has $\mathop {\lim }\limits_{t \to \infty } \Delta {{\tilde \Xi }_{{\xi _i}}}(t) = 0$, which leads to
\begin{eqnarray}
\mathop {\lim }\limits_{t \to \infty } {{\tilde \Xi }_{{\xi _i}}}(t) = {{\tilde \Xi }_{{\xi _i}}}.
\label {eq:103}
\end{eqnarray}
For the recursive equation (\ref{eq:100}), let $\tilde \Xi _{{\xi _i}}^1$ and $\tilde \Xi _{{\xi _i}}^2(t)$ denote any matrices with different initial conditions, and define ${{\hat \Xi }_{{\xi _i}}}(t) \buildrel \Delta \over = \tilde \Xi _{{\xi _i}}^1(t) - \tilde \Xi _{{\xi _i}}^2(t)$. In this case, it is derived from (\ref{eq:100}) and (\ref{eq:102}) that ${{\hat \Xi }_{{\xi _i}}}(t + 1) = f({{\hat \Xi }_{{\xi _i}}}(t - {d_i}))$, whose recursive form is similar to (\ref{eq:98}). Then, it is concluded that $\mathop {\lim }\limits_{t \to \infty } {{\hat \Xi }_{{\xi _i}}}(t) = 0$, which implies
\begin{eqnarray}
\mathop {\lim }\limits_{t \to \infty } \tilde \Xi _{{\xi _i}}^1(t) = \mathop {\lim }\limits_{t \to \infty } \tilde \Xi _{{\xi _i}}^2(t),
\label {eq:104}
\end{eqnarray}
i.e., the limit ${{\tilde \Xi }_{{\xi _i}}}$ in (\ref{eq:103}) is unique. Moreover, for $t > {N_{{P_i}}}$, it follows from the definition of ${{\tilde \Xi }_{{\xi _i}}}(t)$ that
\begin{eqnarray}
{{\tilde \Xi }_{{\xi _i}}}(t) = \left[ {\begin{array}{*{20}{c}}
   {{\Xi _{ii}}(t)} & {\Gamma _{ii}^{\rm{T}}(t)}  \\
   {{\Gamma _{ii}}(t)} & {{P_{ii}}}  \\
\end{array}}\right],
\label {eq:105}
\end{eqnarray}
where $P_{ii}$ is given by (\ref{eq:79}), while ${{\Gamma _{ii}}(t)}$,${{\Xi _{ii}}(t)}$
are calculated by (\ref{eq:51}) and (\ref{eq:70}). In this case, it can be concluded from (\ref{eq:103}--\ref{eq:105}) that $\mathop {\lim }\limits_{t \to \infty } {\Xi _{ii}}(t) = {\Xi _{ii}}$, and the limit ${\Xi _{ii}}$ is independent of the initial values.
\end{proof}

\begin{center}
    \textbf{A.8: The proof of Theorem 3}
\end{center}
\begin{proof}
When the CPSs (\ref{eq:1}--\ref{eq:2}) satisfy the condition (\ref{eq:78}), it is concluded from the result in\cite{c9} that
\begin{eqnarray}
\mathop {\lim }\limits_{t \to \infty } {P_{ij}}(t) = {P_{ij}},
\label {eq:111}
\end{eqnarray}
where $P_{ij}(t)$ is calculated by (\ref{eq:8}), and the limit ${P_{ij}}$ is independent of the initial values. Then, it follows from Lemmas 2--3, (\ref{eq:79}) and (\ref{eq:111}) that
\begin{eqnarray}
\left\{ \begin{array}{l}
 \mathop {\lim }\limits_{t \to \infty } \Phi _{{{\rm{x}}_i}}^{{{\rm{x}}_j}}(t,t - \varepsilon ) = \Phi _{{{\rm{x}}_i}}^{{{\rm{x}}_j}}(\varepsilon ) \\
 \mathop {\lim }\limits_{t \to \infty } \Phi _{{{\rm{x}}_i}}^{\rm{F}}(t,g,t - \varepsilon ) = \Phi _{{{\rm{x}}_i}}^{\rm{F}}(\varepsilon ) \\
 \mathop {\lim }\limits_{t \to \infty } \Phi _{{{\rm{x}}_i}}^w(t,t - \varepsilon ) = \Phi _{{{\rm{x}}_i}}^w(\varepsilon ) \\
 \mathop {\lim }\limits_{t \to \infty } \Theta _{{\rm{x}}_i^{\rm{c}}}^w(t,t - \varepsilon ) = \Theta _{{\rm{x}}_i^{\rm{c}}}^w(\varepsilon ) \\
 \mathop {\lim }\limits_{t \to \infty } \Theta _{{\rm{x}}_i^{\rm{c}}}^{\rm{F}}(t,g,t - \varepsilon ) = \Theta _{{\rm{x}}_i^{\rm{c}}}^{\rm{F}}(\varepsilon ) \\
 \end{array}, \right.
\label {eq:112}
\end{eqnarray}
where these limits are independent of the initial values.

Define ${{\hat \Gamma }_{ij}}(t) \buildrel \Delta \over = \Phi _{{{\rm{x}}_i}}^{{{\rm{x}}_j}}(t,t - {d_j}){\rm{H}}_{{{\rm{d}}_j}}^{\rm{T}} + \Phi _{{{\rm{x}}_i}}^{\rm{F}}(t,{d_j},t) + \Phi _{{{\rm{x}}_i}}^w(t,t - {d_j} - 1){\rm{\bar H}}_{{\rm{A}}{{\rm{d}}_j}}^{\rm{T}}$. Then, one has by (\ref{eq:79}) and (\ref{eq:112}) that
\begin{eqnarray}
\left\{ \begin{array}{l}
 \mathop {\lim }\limits_{t \to \infty } \left( {\prod\nolimits_{{\varphi _j} = 0}^{{d_j}} {{\Phi _{{{\rm K}_i}}}(t - {\varphi _j})} } \right) = \Phi _{{{\rm K}_i}}^{{d_j} + 1} \\
 \mathop {\lim }\limits_{t \to \infty } {{\hat \Gamma }_{ij}}(t) = {{\hat \Gamma }_{ij}} \\
 \end{array}, \right.
\label {eq:113}
\end{eqnarray}
where ${{\hat \Gamma }_{ij}}$ is independent of the initial values. Thus, there must exist an integer ${N_{{\Gamma _i}}}( > {N_{{P_i}}})$ such that, for $t > {N_{{\Gamma _i}}}$, (\ref{eq:51}) reduces to:
\begin{eqnarray}
{\Gamma _{ij}}(t) = \Phi _{{{\rm K}_i}}^{{d_j} + 1}{\Gamma _{ij}}(t - {d_j} - 1){\rm{H}}_{{\rm{A}}{{\rm{d}}_j}}^{\rm{T}} + {{\hat \Gamma }_{ij}},
\label {eq:114}
\end{eqnarray}
where ${{\rm{H}}_{{\rm{A}}{{\rm{d}}_j}}}$ is defined by (\ref{eq:44}). In this case, it follows from (\ref{eq:114}) that
\vspace{-6pt}
\begin{eqnarray}
\begin{array}{l}
 {\Gamma _{ij}}(t) = {[\Phi _{{{\rm{K}}_i}}^{{d_j} + 1}]^{\hbar-\hbar_0} }{\Gamma _{ij}}(f_{j{\rm{o}}}^{\hbar-\hbar_0} (t)){[{\rm{H}}_{{\rm{A}}{{\rm{d}}_j}}^{\rm{T}}]^{\hbar-\hbar_0} }\\
  \;\;\;\;\;\;\;\;\;\;\;\;+ \sum\nolimits_{\kappa  = 0}^{\hbar-\hbar_0  - 1} {{{[\Phi _{{{\rm{K}}_i}}^{{d_j} + 1}]}^\kappa }{{\hat \Gamma }_{ij}}{{[{\rm{H}}_{{\rm{A}}{{\rm{d}}_j}}^{\rm{T}}]}^\kappa }},  \\
 \end{array}
\label {eq:115}
\end{eqnarray}
where $f_j(t)$ is defined in (\ref{eq:41}), and the variable $\hbar_0$ is determined by:
\begin{eqnarray}
\hbar_0  = \min \{ {\hbar ^*}|f_{j{\rm{o}}}^{\hbar *}(t) - {N_{{\Gamma _i}}} \ge 0\},
\label {eq:116}
\end{eqnarray}
which implies that $t \to \infty  \Leftrightarrow \hbar  \to \infty $. Meanwhile, it is known from (\ref{eq:79}) and (\ref{eq:106}) that  ${\Phi _{{{\rm K}_i}}}$ and ${{\rm{H}}_{{\rm{A}}{{\rm{d}}_j}}}$ are stable matrices, which means that
\begin{eqnarray}
\mathop {\lim }\limits_{t \to \infty } \Phi _{{{\rm K}_i}}^t = 0,\mathop {\lim }\limits_{t \to \infty } {\rm{H}}_{{\rm{A}}{{\rm{d}}_j}}^t = 0.
\label {eq:117}
\end{eqnarray}
Then, it is concluded from (\ref{eq:116}) and (\ref{eq:117}) that
\begin{eqnarray}
\left\{ \begin{array}{l}
 \mathop {\lim }\limits_{t \to \infty } {[\Phi _{{{\rm{K}}_i}}^{{d_j} + 1}]^{\hbar-\hbar_0} }{\Gamma _{ij}}(f_{j{\rm{o}}}^{\hbar_0} (t)){[{\rm{H}}_{{\rm{A}}{{\rm{d}}_j}}^{\rm{T}}]^{\hbar-\hbar_0} } = 0 \\
 \mathop {\lim }\limits_{\hbar  \to \infty } {[\Phi _{{{\rm{K}}_i}}^{{d_j} + 1}]^{\hbar-\hbar_0  - 1}}{{\hat \Gamma }_{ij}}{[{\rm{H}}_{{\rm{A}}{{\rm{d}}_j}}^{\rm{T}}]^{\hbar-\hbar_0  - 1}} = 0 \\
 \end{array}. \right.
\label {eq:118}
\end{eqnarray}
Therefore, one has by (\ref{eq:115}) and (\ref{eq:118}) that
\begin{eqnarray}
\mathop {\lim }\limits_{t \to \infty } {\Gamma _{ij}}(t) = {\Gamma _{ij}}.
\label {eq:119}
\end{eqnarray}
Let $\Gamma _{ij}^1(t) = \Phi _{{{\rm K}_i}}^{{d_j} + 1}\Gamma _{ij}^1(t - {d_j} - 1){\rm{H}}_{{\rm{A}}{{\rm{d}}_j}}^{\rm{T}} + {{\hat \Gamma }_{ij}}$ and $\Gamma _{ij}^2(t) = \Phi _{{{\rm K}_i}}^{{d_j} + 1}\Gamma _{ij}^2(t - {d_j} - 1){\rm{H}}_{{\rm{A}}{{\rm{d}}_j}}^{\rm{T}} + {{\hat \Gamma }_{ij}}$, where $\Gamma _{ij}^1(t - {d_j} - 1)$ and $\Gamma _{ij}^2(t - {d_j} - 1)$ denote the different initial values. Then, it is known from (\ref{eq:119}) that $\mathop {\lim }\limits_{t \to \infty } \Gamma _{ij}^1(t) = \Gamma _{ij}^1$ and $\mathop {\lim }\limits_{t \to \infty } \Gamma _{ij}^2(t) = \Gamma _{ij}^2$. Meanwhile, defining $\Delta {\Gamma _{ij}}(t) \buildrel \Delta \over = \Gamma _{ij}^1(t) - \Gamma _{ij}^2(t)$ yields that $\Delta {\Gamma _{ij}}(t) = {[\Phi _{{{\rm{K}}_i}}^{{d_j} + 1}]^{\hbar-\hbar_0} }\Delta {\Gamma _{ij}}(f_{j{\rm{o}}}^{\hbar_0} (t)){[{\rm{H}}_{{\rm{A}}{{\rm{d}}_j}}^{\rm{T}}]^{\hbar-\hbar_0}}$, and thus it follows from (\ref{eq:118}) that $\mathop {\lim }\limits_{t \to \infty } \Delta {\Gamma _{ij}}(t) = 0$, i.e.,
\begin{eqnarray}
\Gamma _{ij}^1 = \Gamma _{ij}^2.
\label {eq:120}
\end{eqnarray}
This implies that the limit ${\Gamma _{ij}}$ in (\ref{eq:119}) is independent of the initial values.

According to the computation formulas of ${\Psi _{ij}}(t)$ and ${{\hat \Upsilon }_{ij}}(t)$, it is obtained from (\ref{eq:112}) and (\ref{eq:119}) that
\begin{eqnarray}
\mathop {\lim }\limits_{t \to \infty } {\Psi _{ij}}(t) = {\Psi _{ij}},\mathop {\lim }\limits_{t \to \infty } {{\hat \Upsilon }_{ij}}(t) = {{\hat \Upsilon }_{ij}},
\label {eq:121}
\end{eqnarray}
where ${\Psi _{ij}}$ and ${{\hat \Upsilon }_{ij}}$ are independent of the initial values. Then, combining (\ref{eq:112}) and (\ref{eq:121}) yields that
\begin{eqnarray}
\mathop {\lim }\limits_{t \to \infty } {{\hat \Xi }_{ij}}(t) = {{\hat \Xi }_{ij}},
\label {eq:122}
\end{eqnarray}
where ${{\hat \Xi }_{ij}}$ is independent of the initial values. Moreover, from (\ref{eq:122}), there must exist an integer ${N_{{\Xi _i}}}( > {N_{{\Gamma _i}}})$ such that ${\Xi _{ij}}(t)$ (\ref{eq:71}) reduces to:
\begin{eqnarray}
{\Xi _{ij}}(t) = {\rm{H}}_{{\rm{A}}{{\rm{d}}_i}}^{{\tau _{{d_i}}}}{\Xi _{ij}}(t - {\tau _{ij}}){[{\rm{H}}_{{\rm{A}}{{\rm{d}}_j}}^{{\tau _{{d_j}}}}]^{\rm{T}}} + {{\hat \Xi }_{ij}}\;(t > {N_{{\Xi _i}}}).
\label {eq:123}
\end{eqnarray}
When the condition (\ref{eq:106}) holds, ${{\rm{H}}_{{\rm{A}}{{\rm{d}}_i}}}$ and ${{\rm{H}}_{{\rm{A}}{{\rm{d}}_j}}}$ in (\ref{eq:123}) are stable matrices. In this case, the form of (\ref{eq:123}) is the same as that of (\ref{eq:114}), and thus it is obtained from the similar derivation of (\ref{eq:119}--\ref{eq:120}) that
\begin{eqnarray}
\mathop {\lim }\limits_{t \to \infty } {\Xi _{ij}}(t) = {\Xi _{ij}},
\label {eq:124}
\end{eqnarray}
and ${\Xi _{ij}}$ is independent of the initial values.

Therefore, when the conditions (\ref{eq:91}), (\ref{eq:92}) and (\ref{eq:106}) hold, the result (\ref{eq:107}) can be obtained from (\ref{eq:93}) and (\ref{eq:124}). Moreover, the steady-state weighted matrices (\ref{eq:108}) can be derived from (\ref{eq:29}), the definition of $\Xi (t)$ (see (\ref{eq:31})) and (\ref{eq:107}). Notice that the results (\ref{eq:107}) and (\ref{eq:108}) have shown that the designed DKFE is independent of the initial values, the SDKFE (\ref{eq:109}) can be thus obtained.
\end{proof}

\vfill

\end{document}